\documentclass[11pt,a4paper]{article}
\usepackage{amsmath,amssymb,epsfig}
\allowdisplaybreaks[2]
\author{Marian  Koles\'{a}r and Ji\v{r}\'{\i} Novotn\'{y}\footnote{Institute of Particle and Nuclear Physics, Charles  University, V Hole\v{s}ovi\v{c}k\'{a}ch 2, 180 00, Prague 8, Czech Republic}}

\title{\textbf{\boldmath $\pi\eta$ scattering and the
resummation of vacuum fluctuation  in three-flavour $\chi
PT$}}

\begin{document}

\maketitle

\abstract{We discuss various aspects of  resummed chiral
perturbation theory, which was developed recently in order to
consistently include the possibility of large vacuum fluctuations
of the $\bar{s}s$-pairs and the scenario with smaller value of the
$\bar{q}q$ condensate for $N_f=3$. The subtleties of this approach
are illustrated using a concrete example of observables connected
with $\pi\eta$ scattering. This process seems to be a suitable
theoretical laboratory for this purpose due to its sensitivity to
the values of the $O(p^4)$ LEC's, namely to the values of the
fluctuation parameters $L_4$ and $L_6$. We discuss several issues
in detail, namely the choice of ``good'' observables and
properties of their bare expansions, the ``safe''
reparametrization in terms of  physical observables, the
implementation of exact perturbative
 unitarity and exact renormalization scale independence, the
role  of  higher order remainders and their estimates. We make 
a detailed comparison with standard chiral perturbation theory 
and use generalized $\chi PT$  as well as resonance chiral theory 
to estimate the higher order remainders.}  

\newpage
\tableofcontents

\section{ Introduction}

As it is well known, at the energy scales $E\ll \Lambda _H\sim
1\mathrm{GeV}$ the physics of QCD is nonperturbative and governed by chiral
symmetry ($\chi S$) $SU(N_f)_L$$\times$$SU(N_f)_R$. This global
symmetry is present on the classical level within the QCD with
$N_f$ {massless} quarks (in the chiral limit of QCD){\ and } on
the quantum level there exist strong theoretical (for $N_f\geq 3$)
and phenomenological arguments for spontaneous
symmetry breakdown (SSB) of $\chi S$ according to the pattern 
$SU(N_f)_L$$\times$$SU(N_f)_R\rightarrow SU(N_f)_V$. Due to the
confinement, quark and gluon fields do not represent appropriate
low energy degrees of freedom within the above mentioned energy
range; the relevant degrees of freedom correspond to the lightest
colourless hadrons in the QCD spectrum. As far as the Green
functions of quark currents are concerned, it is possible to
obtain a general solution of the chiral Ward identities in terms
of the low energy expansion. This expansion can be organized most
efficiently using the methods of effective field theory
corresponding to the low-energy limit of QCD with $N_f$ light
quark flavours which is known as chiral perturbation
theory $(\chi PT)$ \cite{Weinberg:1978kz, Gasser:1983yg, Gasser:1984gg}. $%
\chi PT$ describes the low energy QCD dynamics in terms of the lightest $%
(N_f^2-1)$-plet of the pseudoscalar mesons identified  with the
Goldstone bosons (GB) of the spontaneously broken chiral symmetry
which appear in the particle spectrum of the theory as a consequence
of the Goldstone theorem. In the chiral limit these pseudoscalars
are massless  and
dominate the low energy dynamics of QCD. They interact weekly at low energies $%
E\ll \Lambda _H$, where $\Lambda _H\sim 1\mathrm{GeV}$ is the hadronic
scale corresponding to the masses of the lightest nongoldstone
hadrons. This feature of the GB dynamics enables systematic
perturbative treatment with the expansion parameter $(E/\Lambda
_H)$. Within the real QCD the quark mass term
$\mathcal{L}_{f,mass}^{QCD}\,$breaks $\chi S$ explicitly and the
Goldstone bosons become pseudogoldstone bosons (PGB) with nonzero
masses.
Though $m_f\neq 0$, for $m_f\ll \Lambda _H $ the mass term $\mathcal{L}%
_{f,mass}^{QCD}$ can be treated as a perturbation. As a
consequence, PGB
correspond to the lightest hadrons in the QCD spectrum\footnote{%
The PGB masses $M_P$ can be expanded in the powers (and
logarithms) of the quark masses starting from the linear term and
therefore vanish in the chiral limit.} (identified with $\pi
^0,\pi ^{\pm }\,$ for $N_f=2$ and $\pi ^0,\pi ^{\pm
},K^0,\overline{K^0},K^{\pm },\eta $ for $N_f=3$) and the
interaction of PGB at the energy scale $E\ll \Lambda _H$ continues to
be weak. Because $M_P<\Lambda _H$, the QCD dynamics at $E\ll
\Lambda _H$ is still dominated by these particles and the
effective theory provides us with a simultaneous expansion in
powers of $(E/\Lambda _H)$ and $(m_f/\Lambda _H)$. The Lagrangian
of $\chi PT$ can be constructed on the basis of symmetry arguments
only; the unknown information about the nonperturbative properties
of QCD are hidden in the parameters known as low energy constants
(LEC)\cite {Gasser:1983yg, Gasser:1984gg}. These are related to
the (generally nonlocal) order parameters of the SSB of $\chi S$,
the most prominent of
them are the Goldstone boson decay constant $F_0$ and the chiral condensate%
\footnote{%
The parameter $F_0$ is however more fundamental in the sense that
$F_0\neq 0$
is both necessary and sufficient condition for SSB, while $\langle \overline{%
q_f}q_f\rangle _0\neq 0$ corresponds to the sufficient condition
only. (The
lower index zero means here the chiral limit.)} $B_0=\Sigma /F_0^2$ where $%
\Sigma =-\langle \overline{u}u\rangle _0$ .

To be more precise, $N_f$-flavour $\chi PT$ is in fact an
expansion in $m_i$, around the $SU(N_f)_L\times SU(N_f)_R$ chiral
limit $m_i=0$, $i\leq N_f$, while keeping all the other quark
masses for $i > N_f$ at their physical values. Because $m_{u,d}$
are much smaller not only in comparison with the hadronic
scale $\Lambda _H$, but also in comparison with the intrinsic QCD scale $%
\Lambda _{QCD}$, the two-flavour $\chi PT$ is expected to produce
well-behaved expansion corresponding to small corrections to the $%
SU(2)_L\times SU(2)_R$ chiral limit.

The strange quark mass on the other hand, though still small
enough with respect to $\Lambda _H$ to be treated as an expansion
parameter within the three-flavour $\chi PT$ (relating real QCD
with its $SU(3)_L\times SU(3)_R$ chiral limit), is of comparable
size with respect to $\Lambda _{QCD}$. This fact, besides the
expected worse convergence of the three-flavour $\chi PT$, might
also have interesting consequences for the possible difference
between the $N_f=2$ and $N_f=3$ chiral dynamics. As discussed
intensively in a series of papers \cite{Descotes-Genon:1999uh,
Descotes-Genon:1999zj, Descotes-Genon:2000di,
Descotes-Genon:2002yv, Descotes-Genon:2003cg,
Descotes-Genon:2007ta}, $m_s\lesssim \Lambda _{QCD}$ suggest, that
the loop effects of the vacuum $\overline{s}s$ pairs are not
suppressed as strongly as it is for the heavy quarks and might
enhance the magnitude of the $N_f=2$ chiral order parameters
relatively to their $N_f=3$ chiral limits. This applies mainly to
$F_0(N_f)$ and $\Sigma (N_f)=F_0^2(N_f)B_0(N_f)$, which should satisfy 
paramagnetic inequalities \cite
{Descotes-Genon:1999uh}
\begin{eqnarray}
\Sigma (2) &>&\Sigma (3)=\lim_{m_s\rightarrow 0}\Sigma (2) \nonumber \\
F_0(2) &>&F_0(3)=\lim_{m_s\rightarrow 0}F_0(2).
\end{eqnarray}
The leading order difference between the two-flavour and
three-flavour values is proportional to $m_s$, with coefficients
measuring the violation of the OZI rule in the $0^{++}$ channel,
\emph{e.g}.
\begin{equation}
\Sigma (2)=\Sigma (3)+m_s\overline{Z}_1^s+\ldots
\label{induced_condensate}
\end{equation}
(see \cite{Descotes-Genon:1999uh} for details) where
\begin{equation}
\overline{Z}_1^s=\lim_{m_s\rightarrow 0}\int \mathrm{d}^4x\langle \overline{u%
}u(x)\overline{s}s(0)\rangle _c
\end{equation}
and analogously for $F_0$. The fluctuation parameter
$\overline{Z}_1^s$ is related to the LEC $L_6^r(\mu )$ (and
$L_4^r(\mu )$ for $F_0$) of the three-flavour $\chi PT$ . As
discussed in \cite{Descotes-Genon:1999uh, Descotes-Genon:1999zj,
Descotes-Genon:2000di, Descotes-Genon:2002yv,
Descotes-Genon:2003cg,Descotes-Genon:2007ta}, these parameters
might be
larger than their estimate based on the large $N_c$ expansion, provided $%
N_f=3$ is close to the critical number of light quark flavours $N_f^{crit}$,
for which the chiral symmetry is restored.
Available estimates vary widely, some indicate a larger number $N_f^{crit}\sim10-12$ for $N_c=3$ \cite{Appelquist1,Appelquist2,Gardi}, while other approaches \cite{Velkovsky,Fischer} and lattice calculations \cite{Mawhinney1,Mawhinney2,Sui,Iwasaki1,Iwasaki2} discuss a possibly much lower value $N_f^{crit}\leq6$. Provided
the scenario of large vacuum fluctuations takes place, the second
term in (\ref{induced_condensate}) (called the induced condensate
in \cite{Descotes-Genon:1999zj,Descotes-Genon:2003cg, Girlanda:2001pc}) can be
numerically comparable with the first term and the three-flavour condensate $%
\Sigma (3)$ could be substantially smaller than the two-flavour
one, the value of which is experimentally accessible in the recent
experiments. Analogous reasonings apply to the relationship of
$F_0(2)$ and $F_0(3)$.

These effects could possibly have strong consequences for the
organization of the chiral expansion in the $N_f=3$ case
\cite{Descotes-Genon:1999uh, Descotes-Genon:2000di,
Descotes-Genon:2002yv, Descotes-Genon:2003cg}. Let us remind that
the general form of the Lagrangian of $\chi PT$ is
\begin{equation}
\mathcal{L}=\sum_{m,n}\mathcal{L}^{(m,n)}
\end{equation}
where
\begin{equation}
\mathcal{L}^{(m,n)}=\sum_kC_k^{(m,n)}O_k^{(m,n)}.
\end{equation}
with LEC's $C_k^{(m,n)}$ and independent set of the operators $%
O_k^{(m,n)}=O(\partial ^mm_f^n)$.

In order to be able to treat the double expansion consistently, it
is necessary to assign a single integer parameter called
\emph{chiral order} to each term $\mathcal{L}^{(m,n)}=O(\partial
^mm_f^n)$ of the effective Lagrangian. The terms $\mathcal{L}_k$
with chiral order $k$ are then called $O(p^k)$ terms. Obviously,
$\partial =O(p).$ The matter of discussion might be, however, the
question concerning the chiral power of $m_f$. This
question is intimately connected to the scenario according which the SSB of $%
\chi S$ is realized.

The standard scenario \cite{Gasser:1983yg, Gasser:1984gg}
corresponds to the assumption, that the SSB order parameters
$\Sigma (N_f)$ and $F_0(N_f)$ are large in the sense, that the ratios
\begin{equation}
X(N_f)=\frac{2\widehat{m}\Sigma (N_f)}{F_\pi ^2M_\pi ^2}
\label{X}
\end{equation}
(where and $\widehat{m}=(m_u+m_d)/2$) and
\begin{equation}
Z(N_f)=\frac{F_0^2(N_f)}{F_\pi ^2}  \label{Z}
\end{equation}
are close to one. Because $M_\pi ^2=O(p^2)$, it is then natural to take $%
m_f=O(p^2)$, i.e. $k=m+2n$. This results in the \emph{standard} $\chi PT$
($S\chi PT$ in what follows). This scenario seems to be
experimentally confirmed \cite{Pislak:2001bf} for $N_f=2$; the
recent analysis of the data yields \cite{Descotes-Genon:2001tn}

\begin{equation}
X(2) = 0.81\pm 0.07,\quad
Z(2) = 0.89\pm 0.03.
\end{equation}

\noindent The $O(p^2)$ Lagrangian \cite{Gasser:1983yg, Gasser:1984gg}
\begin{equation}
\mathcal{L}_2=\frac{F_0^2}4\left( \langle \partial _\mu
U^{+}\partial ^\mu U\rangle +2B_0\langle
U^{+}\mathcal{M}+\mathcal{M}^{+}U\rangle \right)
\end{equation}
gives $\Sigma (2)_{LO}=\Sigma (3)=B_0F_0^2$ at the leading order, thus postponing the
difference $\Sigma
(2)-$ $\Sigma (3)$ to higher orders. The same is true for the parameters $%
F_0(N_f)$. Let us also note that the quark mass ratio
$r=m_s/\widehat{m}$ is not a free parameter here\footnote{On the contrary,
the value of $r$ is usually taken as an input in standard $O(p^6)$ fits, see
{\em{e.g.}} \cite {Bijnens:2006zp} and references therein.}, at
the leading order one has
\begin{equation}
r=2\frac{M_K^2}{M_\pi ^2}-1.
\end{equation}

An alternative way of chiral power counting for $N_f=3$ is the \emph{generalized} $\chi PT$ ($G\chi PT$) \cite{Fuchs:1991cq, Stern:1993rg, Knecht:1994ug, Knecht:1994wf, Knecht:1994zb, Knecht:1995tr}, originally designed to treat the scenario with small quark
condensate $X(3)$$\,\ll$1 and to take the quark mass ratio $r$
as a free parameter. In the case $X(3)\ll 1$ it is natural to take
$m_f=O(p)$ and $B_0=O(p),$ this means $k=m+n $. In contrast to
$S\chi PT$, there are also odd chiral orders and the $O(p^2)$
Lagrangian contains additional terms which are $O(p^4)$ within the
standard
chiral counting\footnote{%
Effectively the generalized chiral power-counting means partial
resummation of these terms.}(see \emph{e.g.} \cite{Fuchs:1991cq,
Stern:1993rg, Knecht:1994zb}):
\begin{eqnarray}
\mathcal{L}_2 &=&\frac{F_0^2}4\left( \langle \partial _\mu
U^{+}\partial ^\mu U\rangle +2B_0\langle
U^{+}\mathcal{M}+\mathcal{M}^{+}U\rangle +A_0\langle
(U^{+}\mathcal{M})^2+(\mathcal{M}^{+}U)^2\rangle \right.
\nonumber \\
&&+\left. Z_0^P\langle U^{+}\mathcal{M}-\mathcal{M}^{+}U\rangle
^2+Z_0^S\langle U^{+}\mathcal{M}+\mathcal{M}^{+}U\rangle ^2\right)
. \label{L2GCHPT}
\end{eqnarray}
For the condensate $\Sigma =-\langle \overline{u}u\rangle $ we get
 at the leading order for $N_f=3$
\begin{equation}
\Sigma _{LO}=B_0F_0^2+Z_0^S(2\widehat{m}+m_s)=\Sigma (3)+Z_0^S(2\widehat{m}%
+m_s)
\end{equation}
and therefore
\begin{equation}
\Sigma _{LO}(2)=\Sigma (3)+Z_0^Sm_s.
\end{equation}
This allows  the difference $\Sigma (2)-$ $\Sigma (3)$ to appear already
at the leading order, consistently with the small $\Sigma (3)$
scenario. The
next-to-leading order Lagrangian $O(p^3)$%
\begin{equation}
\mathcal{L}_3=\frac{F_0^2}4\left( \xi \langle \partial _\mu
U^{+}\partial ^\mu UU^{+}\mathcal{M}+\mathcal{M}^{+}U\rangle
+\widetilde{\xi }\langle
\partial _\mu U^{+}\partial ^\mu U\rangle \langle U^{+}\mathcal{M}+\mathcal{M%
}^{+}U\rangle +\ldots \right)
\end{equation}
(where the ellipses stand for the additional terms which are of the order $%
O(p^6)$ in $S\chi PT$) gives rise to the $N_f=3$ relation 
\begin{equation}
F_{\pi ,NLO}^2=F_0^2(3)(1+2\widetilde{\xi }(m_s+2\widehat{m})+2\widehat{m}%
\xi ),
\end{equation}
which implies that the difference $F_0^2(2)-F_0^2(3)$ is treated
as an effect of the next-to-leading order
\begin{equation}
F_0^2(2)=F_0^2(3)(1+2\widetilde{\xi }m_s).
\end{equation}
Therefore, neither $S\chi PT$ nor $G\chi PT$ can accumulate the
case of  large fluctuation parameter $\widetilde{\xi }$ and the
ratio $Z(3)\ll 1$ at the leading order.

Quite recently, a consistent method of handling the case $X(3)$,
$Z(3)\ll 1$ was proposed \cite{Descotes-Genon:1999uh,
Descotes-Genon:2000di, Descotes-Genon:2002yv,
Descotes-Genon:2003cg, Descotes-Genon:2007ta}. Instead of changing
the chiral power counting, it is based on a more careful
manipulations with the chiral expansion. As it was discussed in
the above references, the case $X(3)$, $Z(3)\ll 1$ could
significantly influence the properties of the chiral expansion
inducing instabilities of the perturbative series corresponding to
the observables, which cannot be linearly related to the $QCD$
correlators (such as the \emph{ratios} like PGB masses, scattering
amplitudes etc.). For such  quantities, one should not perform a
perturbative chiral expansion of the denominators but rather keep
the ratios in a nonperturbative ``resummed'' form. The possibly
large
vacuum $\overline{s}s$ pair fluctuations are then parameterized in terms of $%
X(3)$, $Z(3)$ and $r$ and treated as free parameters. We return to
the detailed formulation of this recipe in the next section.

The aim of this paper is to illustrate the ``resummed`` form of
the chiral expansion with special attention to its formal
properties and to the details and subtleties of the general
procedure. Motivated by our preliminary results on $\pi \eta $
scattering within the $G\chi PT$ \cite{Novotny:2002ed}, we have
chosen the observables connected with this process as a concrete
example which seems to be sensitive to the deviations from the standard assumption $X(3)$, $%
Z(3)\sim 1$ (note that some recent phenomenological studies
suggest a possibility of $X(3)\sim 0.5$, cf.
\cite{Moussallam:1999aq, Moussallam:2000zf, Ananthanarayan:2001uy}
and \cite {Descotes-Genon:2000di,Descotes-Genon:2000ct}). Also,
from the phenomenological point of view, the off-shell
$\pi\eta\pi\eta^*$ vertex is a necessary building block for the
non-resonant part of the amplitude
 for the rare decay $%
\eta \rightarrow \pi ^0\pi ^0\gamma \gamma $. Preliminary
estimates within $G\chi PT$ \cite {Kolesar:2003zh, Kolesar:2004ra}
suggest, that the effect of deviation of this off-shell vertex
from the standard case might be at least in principle observed.
The details will be presented elsewhere \cite{Kolesar:200x}.

The amplitude of $\pi \eta $ scattering was already calculated
within $S\chi PT$ to $O(p^4)$ (and within the extended $S\chi PT$
with explicit resonance fields) in the paper
\cite{Bernard:1991xb}, where the authors presented
prediction for the scattering lengths and phase shifts of the $S$, $P$ and $%
D $ partial waves. We quote here their $O(p^4)$ results for the $S$- and $P$%
-wave scattering lengths (in the units of the pion Compton wavelength): $%
a_0^{S\chi PT}=7.2\times 10^{-3}$ and $a_1^{S\chi PT}=-5.2\times
10^{-4}$.

The paper is organized as follows. In Section 2 we
recapitulate the motivation for the resummed version of $\chi PT$
and the construction of the bare expansion of ``good'' observables. We make a detailed general discussion, connected with the four-meson amplitude,
of  the strict chiral expansion, the dispersive representation and
matching of both approaches with stress to the reconciling of
exact perturbative unitarity and exact renormalization scale
independence in Section 3. Section 4 is devoted to the general properties of
the $\pi\eta$ scattering amplitude. We  discuss the kinematics,
the definition of suitable ``good'' observables, the dispersion
representation of the amplitude and the construction of the bare
expansion.  Various possibilities of its reparametrization are
described in a detailed way in Section 5. The numerical illustration of the particular
variants is made in Section 6, where we also numerically illustrate the
subtleties of the construction of the bare expansion. We recapitulate the results of
the standard variant of $\chi PT$ and compare them with the
resummed approach. We concentrate on the
dependence on the LEC's as well as on the sensitivity to the higher
order reminders and make an attempt to estimate their values using a matching with $G\chi PT$ 
and a simple version of resonance chiral theory. Section
\ref{summary} contains the summary and conclusions.
Some technical details are postponed to the appendices.

\section{Resummation of the vacuum fluctuations - motivation and basic notation\label{motivation and notation}}

As we have mentioned in the Introduction, the potentially large
vacuum fluctuations of the $\overline{s}s$ pairs might result in
the instabilities of the chiral expansion, which originate in the
possibility that for some observables the next-to-leading order
correction could be numerically comparable with the leading order
one. As discussed in \cite {Descotes-Genon:2003cg,
Descotes-Genon:2007ta}, this could generally cause problems with
the convergence of the formal chiral expansion. Nevertheless, at
least for some carefully defined ``good'' observables, it is
natural to presume some sort of satisfactory convergence
properties. Such ``good'' observables are assumed to be those
which can be obtained directly from the low-energy correlation
functions in the domain of their analyticity far away from
singularities and which are related to the corresponding
correlator \emph{linearly }\cite{Descotes-Genon:2003cg,
Descotes-Genon:2007ta}. Typical
examples are the squares of the PGB decay constants $F_P^2$ , the products $%
F_P^2M_P^2$ where $M_P$ are the PGB masses and also the subthreshold parameters which can be derived from the products $%
A\prod_{i=1}^4F_{P_i} $ where $A$ is the PGB scattering amplitude
$1+2\rightarrow 3+4$. Let us write the expansion  of such a
``good'' observable $G$ in the form of a (carefully defined)
\emph{bare expansion} \cite{Descotes-Genon:2003cg} as\footnote{Here we tacitly
assume the standard chiral power counting. Analogous expansion
could be written also for the generalized case.}
\begin{equation}
G=G^{(2)}+G^{(4)}+G\delta _G,  \label{Gexp}
\end{equation}
where $G^{(2)}=g^{(2)}(F_0,B_0,m_q)$ and
$G^{(4)}=g^{(4)}(F_0,B_0,m_q,L_i,M_P^2)$ correspond to the sum of
the leading and next-to-leading order terms respectively and the
renormalization scale independent quantity $\delta _G$
accommodates the higher order remainders. 

As a terminological note, in what follows we use the term {\em{strict}} chiral expansion for an
unmodified expansion in terms of the LEC's  strictly respecting the
chiral orders. The {\em{bare}} expansion, though still expressed in terms of LEC's,
accumulates  some modifications dictated by physical requirements.
It is the bare expansion which is assumed to be globally
convergent.

For a ``good'' observable it is then assumed
\begin{equation}
|\delta _G|\ll 1  \label{global_convergence}
\end{equation}
as a natural assumption. This property of the bare chiral
expansion (\ref {Gexp}) is called global convergence in
\cite{Descotes-Genon:2003cg, Descotes-Genon:2007ta}. Note however,
that the validity of the inequality (\ref {global_convergence})
might depend on the definition of the reminder $\delta _G$ which
is not fixed unambiguously and might differ according to the
calculation scheme in use. We will comment on this point later on.

The above mentioned possible  instability in (\ref{Gexp}) appears when  $%
G^{(2)}\sim G^{(4)}$, {\em i.e.}  $X_G\lnsim 1$, where
\begin{equation}
X_G=\frac{G^{(2)}}G.
\end{equation}
Such an instability manifests itself in the expansion of the
observables depending on $G$ nonlinearly
\cite{Descotes-Genon:2003cg}. For instance, for a ratio of two
``good'' observables $G$ and $G^{^{\prime }}$ formally expanded in
the form (\ref{Gexp})
\begin{equation}
\frac G{G^{^{\prime }}}=\left( \frac{G^{(2)}}{G^{^{\prime
}(2)}}\right)
+\left( \frac{G^{(2)}}{G^{^{\prime }(2)}}\right) \left( \frac{G^{(4)}}{%
G^{(2)}}-\frac{G^{^{\prime }(4)}}{G^{^{\prime }(2)}}\right) +\frac
G{G^{^{\prime }}}\delta _{G/G^{^{\prime }}},
\end{equation}
we get for the remainder $\delta _{G/G^{^{\prime }}}$%
\begin{equation}
\delta _{G/G^{^{\prime }}}=\frac{(1-X_{G^{^{\prime
}}})(X_G-X_{G^{^{\prime
}}})}{X_G^{^{\prime }2}}+\frac{\delta _G}{X_{G^{^{\prime }}}}-\frac{%
X_G\delta _{G^{^{\prime }}}}{X_G^{^{\prime }2}}.
\end{equation}
For $X_{G^{^{\prime }}}\lnsim 1$ this might be numerically large
even if both $|\delta _G|$, $|\delta _{G^{^{\prime }}}|$ were
reasonably small. In this sense, a ratio of two globally
convergent observables need not to be necessarily globally
convergent too. It should be therefore much safer not to expand
such ``dangerous'' observables and rather write the ratio in the
``resummed'' form
\begin{equation}
\frac G{G^{^{\prime }}}=\frac{G^{(2)}+G^{(4)}}{G^{^{\prime
}(2)}+G^{^{\prime }(4)}}+\frac G{G^{^{\prime }}}\widetilde{\delta
}_{G/G^{^{\prime }}}. \label{dG/G}
\end{equation}
The relation (\ref{dG/G}) is an exact algebraic identity provided
we keep explicitly the remainder
\begin{equation}
\widetilde{\delta }_{G/G^{^{\prime }}}=\frac{\delta _G-\delta
_{G^{^{\prime }}}}{1-\delta _{G^{^{\prime }}}}.
\end{equation}
In this case $\widetilde{\delta }_{G/G^{^{\prime }}}$remains for
$|\delta _G|,|\delta _{G^{^{\prime }}}|\ll 1$ under numerical
control.

Of course,  only the fact that the bare expansion of some
observable is not globally convergent does not necessarily
correspond to the collapse of the convergence, because the
next-to-next-to-leading order $G^{(6)}$ can saturate the series in
such a way that the next-to-next-to-leading remainder
\begin{equation}
G\delta _G^{NNLO}=G-G^{(2)}-G^{(4)}-G^{(6)}
\end{equation}
is reasonably small. Namely this is the usual assumption behind
the $O(p^6)$ calculations. Violation of the global convergence
property here means merely
that the $O(p^6)$ contribution have unnatural size, \emph{i.e.} $%
G^{(6)}\lesssim G^{(2)}+G^{(4)}$. This could, however, destabilize the $%
O(p^6)$ chiral expansion of ratios in the way similar to that
discussed above.

Provided we allow  the expansion of the ``good'' observables only,
we are also pressed to modify the next step leading from the bare
expansion to the usual output of the $\chi PT$, consisting of a
reparametrization of the expansion by expressing some of the LEC's
in terms of the physical observables such as masses and PGB decay
constants. This step converts the series into an expansion in
powers and logs of the (squared) PGB masses instead of quark
masses. To achieve this, it is either necessary to invert a bare
chiral expansion of some observable (in the case of the $O(p^2)$ LEC's)
or to use an observable which might be generally a ``dangerous''
one. Let us briefly discuss the first case. Schematically, suppose
that some $O(p^2)$ LEC $G_0$ ({\em{e.g.}} $F_0^2$) just corresponds
 to the leading term $G^{(2)}$ of the expansion of the
observable $G$. Then we can write an algebraic identity
\begin{equation} G_0=G-G^{(4)}(G_0)-G\delta _G,
\end{equation}
where we explicitly point out the dependence of the next-to-leading term on $G_0$%
. To convert this expansion and express $G_0$ by means of the series in $G$ one substitutes $G$ for $%
G_0$ on the right hand side. This defines a new remainder $\delta _{G_0}$%
\begin{equation}
G_0=G-G^{(4)}(G)+G_0\delta _{G_0},  \label{inverted}
\end{equation}
for which we get
\begin{equation}
\delta _{G_0}=-\frac{1-X_G}{X_G}+\frac 1{X_G}\frac{G^{(4)}(G)}G.
\end{equation}
This could cause an instability of the converted expansion for $G_0$
in terms of $G$ for $X_G\lnsim 1$ even if the relative size of the
next-to-leading order $G^{(4)}(G)/G$ is reasonably small,
irrespective of the condition for  global convergence $|\delta
_G|\ll 1$.

On the other hand, suppose that some $O(p^4)$ constant $G_1$
coincides with the next-to-leading term $G^{(4)}$. In this case we
have an algebraic identity for $G_1$
\begin{equation}
G_1=G-G^{(2)}-G\delta _G
\end{equation}
and the remainder here is perfectly under control, provided $G$
has a globally convergent bare expansion and we \emph{do not}
re-express $G^{(2)}$ in terms of physical observables (i.e.
provided we treat the $O(p^2)$ LEC's as free parameters).

From the above simple considerations follows that in order to
avoid potential problems with the instabilities of the chiral
expansion, which might be present in the three-flavor $\chi PT$ in
the case of small $X(3)$ and $Z(3)$ (cf. (\ref{X}, \ref{Z})), we
should \cite{Descotes-Genon:2003cg, Descotes-Genon:2007ta}

\begin{itemize}
\item  carefully define the bare expansion

\item  confine ourselves (as far as the bare chiral expansion is
concerned) to the linear space of ``good'' observables and keep
the ``dangerous'' observables in the nonperturbative ``resummed''
form

\item  use rather $\Sigma (3)$, $F_0(3)$ (or $X(3)$ and $Z(3)$) and $%
r\,=2m_s/(m_u+m_d)$ as free parameters\footnote{%
Note, that $r$ is related to the ``dangerous'' observable
\begin{equation*}
2\frac{F_K^2M_K^2}{F_\pi ^2M_\pi ^2}-1=r+\ldots
\end{equation*}
} instead of expressing them in the form of the series in PGB
masses and decay constants

\item  eliminate the $O(p^4)$ LEC's algebraically, using bare expansions of ``good'' observables 
such as $F_P ^2$, $F_P ^2M_P ^2$ \footnote{We will do it for $L_4-L_8$ but leave $L_1-L_3$ free, 
also $L_7$ is a special case, see in what follows.}.
\end{itemize}

In the next section we shall illustrate the possible subtleties of
the first
step of this general recipe on the concrete example\footnote{%
We shall tacitly assume the case of three light flavours in what
follows.} of the PGB scattering amplitude $P_1P_2\rightarrow
P_3P_4$

\section{\boldmath Construction of the bare expansion for the scattering $%
P_1P_2\rightarrow P_3P_4$}

\subsection{Chiral expansion of the ``good'' observable}

Let us assume a scattering of pseudoscalar mesons
$P_1P_2\rightarrow P_3P_4$ with masses $M_{P_i}$. The amplitude
$S(s,t;u)$ is defined as
\begin{equation}
\langle P_3(k_3)P_4(k_4)_{out}|P_1(k_1)P_2(k_2)_{in}\rangle
=\mathrm{i}(2\pi )^4\delta ^{(4)}(k_3+k_4-k_1-k_2)S(s,t;u),
\end{equation}
where $s$, $t$ and $u$ are the usual Mandelstam variables. The
amplitude is
related to the ``good'' observable\footnote{%
Strictly speaking, the ``good'' observables correspond to the
subthreshold parameters derived form $G(t,s;u)$ in an unphysical
point away from  singularities.}

\begin{equation}
G(s,t;u)=\prod_{i=1}^4F_{P_i}S(s,t;u),  \label{G}
\end{equation}
(where $F_{P_i}$ are the decay constants) which can be directly
obtained
from the (cut) four-point function of the axial currents. Let us write for $%
G(s,t;u)$ the following {\em{strict}} chiral expansion in terms of
the low energy constants
\begin{equation}
G=G^{(2)}+G_{ct}^{(4)}+G_{tad}^{(4)}+G_{unit}^{(4)}+G\delta _G.
\label{G_strict}
\end{equation}
$G\delta _G$ accommodates the higher order remainders. Using the
functional method, $G$ can be obtained from the generating
functional
\begin{eqnarray}
F_0^4Z[U,v,p,a,s] &=&F_0^4\int d^4x\left( \mathcal{L}^{(2)}(U,v,p,a,s)+%
\mathcal{L}^{(4)}(U,v,p,a,s)\right)  \nonumber \\
&&+F_0^4Z_{loop}^{(4)}[U,v,p,a,s]+\ldots  \label{Zvaps}
\end{eqnarray}
by setting $v=s=p=0$, $s=2B_0{\cal M}$ and expanding in the fields $\Phi $ where $%
U=\exp \left( i\Phi /F_0\right) $. Following the notation in
\cite {Gasser:1984gg}, we have
\begin{eqnarray}
Z_{loop}^{(4)}[U,v,p,a,s]
&=&Z_{tad}^{(4)}[U,v,p,a,s]+Z_{unit}^{(4)}[U,v,p,a,s]  \nonumber \\
&=&\frac i2\ln \det D_0+\frac i4\mathrm{Tr}(D_0^{-1}\delta )-\frac i4\mathrm{%
Tr}(D_0^{-1}\delta D_0^{-1}\delta )+\ldots  \label{Zvaps4}.
\end{eqnarray}
In the above formulae,
\begin{equation}
D_0^{ab}=\delta ^{ab}\Box +\frac 12B_0tr(\{\lambda ^a,\lambda ^b\}\mathcal{M}%
)
\end{equation}
and $\mathcal{M}$ is the quark mass matrix. Note this
representation
of $Z_{loop}^{(4)}$ assumes that the masses running in the loops are the $%
O(p^2)$ masses rather than the physical masses. Or, in more
detail, provided we start with the chiral expansion of the squared
product of the masses and decay constants
\begin{equation}
F_P^2M_P^2=(F_P^2M_P^2)^{(2)}+(F_P^2M_P^2)^{(4)}+F_P^2M_P^2\delta
_{FM_P},
\end{equation}
the masses in the loops are defined as
\begin{equation}
\stackrel{o}{M}_P^2=\frac{(F_P^2M_P^2)^{(2)}}{F_0^2}.
\label{Op2M}
\end{equation}
Note, however, that this is the first term in a potentially
``dangerous'' expansion of the ratio
\begin{equation}
M_P^2=\frac{F_P^2M_P^2}{F_P^2}=\frac{%
(F_P^2M_P^2)^{(2)}+(F_P^2M_P^2)^{(4)}+F_P^2M_P^2\delta _P}{%
F_0^2+(F_P^2)^{(4)}+F_P^2\delta _{F_P}}=\stackrel{o}{M}_P^2+\ldots
\end{equation}
From this definition of $Z_{loop}^{(4)}$ we obtain $G^{(4)}=$ $%
G_{ct}^{(4)}+G_{tad}^{(4)}+G_{unit}^{(4)}$ which is exactly
renormalization scale independent even for the external momenta
off-shell. This meets the
requirement of the renormalization scale independence of the remainder $%
\delta _G$.

The first two terms of the above strict chiral expansion for
$G(s,t;u)$ have a serious drawback in the sense that the
singularities in the complex $stu$ planes required by unitarity
are not placed at the physical thresholds but rather at points
given by the leading order terms $\stackrel{o}{M}_P$ of
the chiral expansion of the PGB masses. Straightforward substitution $%
\stackrel{o}{M}_P\rightarrow M_P$ in the propagators of the loops,
which apparently means merely a redefinition of the remainder
$\delta _G$, could, however, in general spoil its exact
renormalization scale independence. It is therefore desirable to
use the freedom in the definition of the remainder more carefully
in order to reconcile both scale independence of $G^{(4)}$ and
unitarity. For this purpose, a useful tool is the matching with a
dispersive representation \cite{Descotes-Genon:2003cg} of the amplitude $%
S(s,t;u)$ based on the reconstruction theorem \cite{Stern:1993rg,
Knecht:1995tr}.

\subsection{Dispersive representation for $G(s,t;u)$}

The above mentioned reconstruction theorem for the PGB scattering
amplitude is based on the basic properties of unitarity,
analyticity and crossing symmetry and provides us with the most
general form of the PGB scattering amplitude up to the order
$O(p^6)$ in terms of dispersive
integrals with known discontinuities. It was first proved for the case of $%
\pi \pi $ scattering in \cite{Stern:1993rg, Knecht:1995tr} and for
$\pi K$ scattering in \cite{Ananthanarayan:2000cp,
Ananthanarayan:2001uy} and since then it has been intensively used
in various contexts. Here we use the general form of the theorem,
more detailed discussion of which will be presented elsewhere
\cite{Zdrahal:200x}.

For the scattering of pseudoscalar mesons $P_1P_2\rightarrow
P_3P_4 $, let us  denote the $s-$, $t-$ and $u-$channel amplitudes as $%
S(s,t;u)$, $T(s,t;u)$ and $U(s,t;u)$ and write their partial wave
expansion as
\begin{equation}
A(s,t;u)=32\pi \sum_{l=0}^\infty (2l+1)A_l(s)P_l(\cos \theta _A),
\label{partialwave}
\end{equation}
where $A=S$, $T$, $U$ and
\begin{equation}
\cos \theta _A=\frac{s(t-u)+\Delta _{A_i}\Delta _{A_f}}{\lambda
_{A_i}^{1/2}(s)\lambda _{A_f}^{1/2}(s)}.
\end{equation}
Here $A_l(s)$ are the partial waves,
\begin{equation}
\lambda
_{A_{i,f}}(s)=(s-(M_{P_j}+M_{P_k})^2)(s-(M_{P_j}-M_{P_k})^2)
\end{equation}
is the triangle function which corresponds to the initial/final
state $A_{i,f}$ (consisting of the pseudoscalars $P_jP_k$) of the
process in the channel $A$ and
\begin{equation}
\Delta _{A_{i,f}}=M_{P_j}^2-M_{P_k}^2.
\end{equation}
According to the theorem, we get the following representation for
the amplitude $S(s,t;u)$
\begin{equation}
S(s,t;u)=\mathcal{S}(s,t;u)+\mathcal{S}_{unit}(s,t;u)+O(p^8),
\label{Srec}
\end{equation}
where $\mathcal{S}(s,t;u)$ is a third order polynomial with the
same symmetries as the whole amplitude $S(s,t;u)$. The nontrivial
analytical properties are incorporated in the unitarity part
$\mathcal{S}_{unit}(s,t;u)$, which can be expressed as
\begin{eqnarray}
\mathcal{S}_{unit}(s,t;u) &=&\Phi ^S(s)+\Phi ^T(t)+\Phi ^U(u)  \nonumber \\
&&+[s(t-u)+\Delta _{12}\Delta _{34}]\Psi ^S(s) \nonumber \\
&&+[t(s-u)+\Delta _{13}\Delta _{24}]\Psi ^T(t) \nonumber \\
&&+[u(t-s)+\Delta _{14}\Delta _{23}]\Psi ^U(u).
\end{eqnarray}
In the last expression, $\Delta _{ij}=M_{P_i}^2-M_{P_j}^2$. The functions $%
\Phi ^A(s)$ and $\Psi ^A(s)$ with $A=S,T,U$ are analytic in the
cut complex plane with the right hand cut from $\tau
_A=\min_{i,j}(M_{P_i}+M_{P_j})^2$
(where $P_iP_j$ are the possible intermediate states in the given channel $A$%
) to infinity with discontinuities given by the formulae
\begin{eqnarray}
\mathrm{disc}\,\Phi ^A(s) &=&32\pi \theta (s-\tau
_A)\mathrm{disc}\,A_0(s)
\label{disc0} \\
\mathrm{disc}\,\Psi ^A(s) &=&96\pi \theta (s-\tau _A)\mathrm{disc}\frac{A_1(s)%
}{\lambda _{A_i}^{1/2}(s)\lambda _{A_f}^{1/2}(s)}.  \label{disc1}
\end{eqnarray}
Here  $A_0(s)$,  $A_1(s)$ are the corresponding $l=0,1$ partial
waves.

Consequently, once the right hand sides of (\ref{disc0},
\ref{disc1}) are known, the unitarity part
$\mathcal{S}_{unit}(s,t;u)$ of the amplitude can be uniquely
reconstructed to $O(p^6)$ up to the polynomial, which encompass
subtraction polynomials for the dispersion integrals.

Let us now assume the chiral expansion of the amplitudes in the
form
\begin{eqnarray}
A(s,t;u) &=&A^{(2)}(s,t;u)+A^{(4)}(s,t;u)+A\delta _A,
\label{Amp_expansion}
\\
A^{(n)}(s,t;u) &=&32\pi \sum_{l=0}^\infty
(2l+1)A_l^{(n)}(s)P_l(\cos \theta _s(t)). \label{Amp_l_expansion}
\end{eqnarray}
Starting from the $O(p^2)$ amplitudes, we can use the two particle
partial wave unitarity to get the discontinuity of the partial
waves $A_l^{(4)}(s)$
along the right hand cut\footnote{%
It can be shown that more than two particle intermediate states
yield contribution of the order $O(p^8)$ and higher.}
\begin{equation}
\mathrm{disc}A_l^{(4)}(s)=\sum_{ij}\frac
2{{z}_{ij}}\frac{\lambda
_{ij}^{1/2}(s)}sA_l^{(2)ij\rightarrow A_f}(s)A_l^{(2)ij\rightarrow
A_i}(s)^{*}+O(p^6)
\end{equation}
Here ${z}_{ij}=1,2$ is a  symmetry factor taking into
account the possibility of identical particles in the intermediate
state $ij$. Inserting this into the
dispersive integrals we easily\footnote{%
Note that $A^{(2)}(s,t;u)$ are real polynomials of the first order in $s$, $%
t$ and $u$.} get a minimal form for the $O(p^4)$ unitarity
corrections in terms of the functions $\Phi ^{(4)A}(s)$ and $\Psi
^{(4)A}(s)$ reconstructed from the $O(p^2)$ amplitudes
\begin{eqnarray}
\Phi ^{(4)A}(s) &=&(32\pi )^2\sum_{ij}\frac 1{{z}_{ij}}\overline{%
\overline{J}}_{ij}(s)A_0^{(2)A_i\rightarrow
ij}(s)A_0^{(2)ij\rightarrow
A_f}(s)^{*} \label{phi_4}\\
\Psi ^{(4)A}(s) &=&\frac{(96\pi )^2}3\sum_{ij}\frac 1{{z}_{ij}}%
\overline{\overline{J}}_{ij}(s)\frac{A_1^{(2)A_i\rightarrow
ij}(s)A_1^{(2)ij\rightarrow A_f}(s)^{*}}{\lambda
_{A_i}^{1/2}(s)\lambda _{A_f}^{1/2}(s)}.\label{psi_4}
\end{eqnarray}
$\overline{\overline{J}}_{ij}(s)=J_{ij}^r(s)-J_{ij}^r(0)-sJ_{ij}^{r^{\prime }}(s)$
corresponds to the twice subtracted scalar bubble with internal line masses $M_{P_i,P_j}$.
Provided $\Delta _{A_i}=0$ or $\Delta_{A_f}=0$, which will be our case, it can be shown that we
only need one subtraction, $\overline{J}_{ij}(s)=J_{ij}^r(s)-J_{ij}^r(0)$ 
instead of $\overline{\overline{J}}_{ij}(s) $. The explicit form of the function 
$\overline{J}_{ij}(s)$ is given in the Appendix \ref{appendix Jbar}. 

The above formulae can be used to write a dispersive
representation of the ``good'' observable
$G(s,t;u)=\prod_{i=1}^4F_{P_i}S(s,t;u)$ to the next-to-leading
order in the form
\begin{equation}
G(s,t;u)=\mathcal{G}(s,t;u)+\mathcal{G}_{unit}(s,t;u),
\label{dispG}
\end{equation}
where $\mathcal{G}(s,t;u)$ is the polynomial part and the unitarity corrections up to $O(p^6)$ are included in 
\begin{eqnarray}
\mathcal{G}_{unit}(s,t;u)
&=&\phi ^S(s)+\phi ^T(t)+\phi ^U(u)  \nonumber \\
&&+[s(t-u)+\Delta _{12}\Delta _{23}]\psi ^S(s)  \nonumber \\
&&+[t(s-u)+\Delta _{13}\Delta _{24}]\psi ^T(t)  \nonumber \\
&&+[u(t-s)+\Delta _{14}\Delta _{23}]\psi ^U(u).  \label{Gunit}
\end{eqnarray}
Our goal is to write down a representation of $\phi^{(4)A}$ and $\psi^{(4)A}$, which, notice, are distinct quantities from $\Phi^{(4)A}$ and $\Psi^{(4)A}$,  analogous to
(\ref{phi_4}, \ref{psi_4}). Note, however, that while the relation of $G(s,t;u)$ and $S(s,t;u)$
is unambiguously fixed to all orders by (\ref{G}), the amplitude can be defined order by order
in various ways.  For example, for the ``good'' observable $G$, the leading order
piece $G^{(2)}$ of its strict chiral expansion is fixed by the lowest order
Lagrangian $\mathcal{L}^{(2)}$, but the corresponding $O(p^2)$ piece of the amplitude $S$ 
can be related in various ways. Similarly, the same is true 
order by order, where the amplitude at the given order can be defined up to higher order corrections.

The most straightforward way is to write a safe expansion for $S(s,t;u)$ in the form
\begin{equation}
S(s,t;u)=\left( \prod_{i=1}^4F_{P_i}\right) ^{-1}\left(
G^{(2)}(s,t;u)+G^{(4)}(s,t;u)+G\delta _G\right) 
\end{equation}
with physical values of $F_{P_i}$, thus satisfying the relation (\ref{G}) order by order
\begin{equation}
S^{(n)}(s,t;u)=\left( \prod_{i=1}^4F_{P_i}\right)
^{-1}G^{(n)}(s,t;u). \label{Sgood}
\end{equation}
As we'll see, the minimal modification of the form derived from the generating functional 
is obtained by using an alternative, potentially ``dangerous'' expansion
\begin{eqnarray}  \label{G/FFFF_dangerous}
S(s,t;u) &=&\left( \prod_{i=1}^4F_{P_i}\right) ^{-1}G(s,t;u)  \nonumber \\
&=&\left( \prod_{i=1}^4F_0(1+\frac
12\frac{(F_{P_i}^2)^{(4)}}{F_0^2}+\ldots ) \right) ^{-1}\left(
G^{(2)}(s,t;u)+G^{(4)} (s,t;u)+\ldots\right)  \nonumber
\\
&=&F_0^{-4}G^{(2)}(s,t;u)- \frac
12F_0^{-6}G^{(2)}(s,t;u)\sum_{i=1}^4(F_{P_i}^2)^{(4)}+F_0^{-4}G^{(4)}
(s,t;u)+ \ldots,
\end{eqnarray}
which defines
\begin{eqnarray}
\widetilde{S}^{(2)}(s,t;u)&=&F_0^{-4}G^{(2)}(s,t;u)
\label{S2dangerous}\\
\widetilde{S}^{(4)}(s,t;u)&=&F_0^{-4}G^{(4)}(s,t;u) - 
\frac12F_0^{-6}G^{(2)}(s,t;u)\sum_{i=1}^4(F_{P_i}^2)^{(4)}.  
\label{S4dangerous}
\end{eqnarray}

The representation of $\phi^{(4)A}$ and $\psi^{(4)A}$ is therefore
not unique. According to our definitions of the amplitude we get either 
(we assume partial wave expansion of $G(s,t;u)$ analogous to (\ref{partialwave}))
\begin{eqnarray}
\phi ^{(4)A}(s) &=&(32\pi )^2\sum_{ij}\frac 1{{z}_{ij}}\frac{\overline{%
\overline{J}}_{ij}(s)}{F_{P_i}^2F_{P_j}^2}G_0^{(2)A_i\rightarrow
ij}(s)G_0^{(2)ij\rightarrow A_f}(s)^{*}  \label{phiA} \\
\psi ^{(4)A}(s) &=&\frac{(96\pi )^2}3\sum_{ij}\frac 1{{z}_{ij}}\frac{%
\overline{\overline{J}}_{ij}(s)}{F_{P_i}^2F_{P_j}^2}\frac{%
G_1^{(2)A_i\rightarrow ij}(s)G_1^{(2)ij\rightarrow
A_f}(s)^{*}}{\lambda _{A_i}^{1/2}(s)\lambda _{A_f}^{1/2}(s)},
\label{psiA}
\end{eqnarray}
corresponding to the definition (\ref{Sgood}) or
\begin{eqnarray}
\widetilde{\phi }^{(4)A}(s) &=&(32\pi )^2F_0^{-4}\sum_{ij}\frac 1
{{z}_{ij}}\overline{\overline{J}}_{ij}(s)G_0^{(2)A_i\rightarrow
ij}(s)G_0^{(2)ij\rightarrow A_f}(s)^{*}  \label{phiA1} \\
\widetilde{\psi }^{(4)A}(s) &=&\frac{(96\pi )^2}3F_0^{-4}\sum_{ij}\frac 1{%
{z}_{ij}}\overline{\overline{J}}_{ij}(s)\frac{G_1^{(2)A_i\rightarrow
ij}(s)G_1^{(2)ij\rightarrow A_f}(s)^{*}}{\lambda
_{A_i}^{1/2}(s)\lambda _{A_f}^{1/2}(s)},  \label{psiA1}
\end{eqnarray}
when reconstructing the bare expansion of $G$ from the
``dangerous'' expansion (\ref{G/FFFF_dangerous}) and using the
definitions (\ref {S2dangerous}, \ref{S4dangerous}) for the $O(p^2)$ and $O(p^4)$ amplitudes.

\subsection{Matching the strict chiral expansion to the dispersive
representation}

The dispersive representation (\ref{dispG}) can be now matched to
the formula (\ref{G_strict}). As we have mentioned above, the positions
of the cuts in the formulas (\ref{G_strict}) and (\ref{dispG}) are not
the same; in the former case they correspond to the $O(p^2)$ masses
(\ref{Op2M}), which ensures the renormalization scale
independence, while in the latter they are determined by the physical
ones, as required by the unitarity conditions. In order to
reconcile both these requirements, one can  proceed as follows
(c.f. also \cite{Descotes-Genon:2003cg}).

In (\ref{G_strict}), the nonanalytic terms are generally of the form $%
P(s)J_{ij}^r(s)$, where $J_{ij}^r(s)$ is the renormalized scalar
bubble defined in Appendix \ref{appendix Jbar} and $P(s)$ is some
second order polynomial. As the first step, one rewrites these
expressions in terms of $\overline{\overline{J}}_{ij}(s)$ writing $%
J_{ij}^r(s)=J_{ij}^r(0)+s\overline{J}_{ij}^{^{\prime }}(0)+\overline{%
\overline{J}}_{ij}(s)$. This adjustment allows us to split $G$
uniquely into a polynomial part $G_{pol}$ and a
nonanalytic part $G_{cut}$ which accumulates the unitarity cuts
\begin{equation}
G(s,t;u)=G_{pol}(s,t;u)+G_{cut}(s,t;u)+G\delta _G,
\end{equation}
where
\begin{equation}
G_{pol}(s,t;u)=\left( G(s,t;u)-G\delta _G\right) |_{\overline{J}_{ij}=%
\overline{\overline{J}}_{ij}=0}.
\end{equation}
Both parts are now renormalization scale independent.

As a second step, we replace the $G_{cut}(s,t;u)$ with $\mathcal{G}%
_{unit}(s,t;u)$ from (\ref{dispG}). This means we write
\begin{equation}
G(s,t;u)=G_{pol}(s,t;u)+\mathcal{G}_{unit}(s,t;u)+G\delta
_G^{^{^{\prime }}} \label{Gbare}
\end{equation}
where $\delta _G^{^{\prime }}$ is a new remainder defined by this
equation.
According to the naive chiral power counting, $G_{cut}(s,t;u)$ $-$ $\mathcal{%
G}_{unit}(s,t;u)=O(p^6)$.

The third step, not necessary from the point of view of
preserving unitarity and renormalization scale invariance,
consists of a further modification of $G_{pol}(s,t;u)$ by means
of replacement of the $O(p^2)$ masses $\stackrel{o}{M}_P^2$ in $%
J_{ij}^r(0)$ with the physical masses $M_P^2$. This replacement
does not spoil the renormalization scale independence of the
$G_{pol}(s,t;u)$ and corresponds to the convention introduced in
\cite{Descotes-Genon:2003cg,
Descotes-Genon:2007ta}. This again means a redefinition of the remainders $%
\delta _G^{^{\prime }}$, \emph{i.e}. re-shuffling of the terms of
the next-to-next-to-leading order.

Note that the origin of $J_{ij}^r(0)$'s in one loop generating functional (%
\ref{Zvaps4}) is twofold: they can stem either from the tadpole part $%
Z_{tad}^{(4)}$ or from the unitarity corrections $Z_{unit}^{(4)}$.
It was argued in \cite {Descotes-Genon:2007ta} that in the former
case the above mentioned replacement does not necessarily modify
the numerical value of the remainders much. The reason should be that
the chiral logs appear only in the combination $\mu _P\propto
\stackrel{o}{M}_P^2\ln (\stackrel{o}{M}_P^2/\mu ^2)$. The
replacement here means
\begin{equation}
\stackrel{o}{M}_P^2\ln (\stackrel{o}{M}_P^2/\mu ^2)\rightarrow \stackrel{o}{M%
}_P^2\ln (M_P^2/\mu ^2).
\end{equation}
Because $\stackrel{o}{M}_P^2\propto Y=X/Z$, the difference should
therefore either be small for $Y\sim 1$ (where $\stackrel{o}{M}_P^2$$\sim$$M_P^2$) or the contribution of $\mu _P$
itself is tiny for $Y\rightarrow 0$.

On the other hand, the logs from $Z_{unit}^{(4)}$ do not generally
come with such a prefactor.
Therefore, with  a replacement $\stackrel{o}{M}_P^2\rightarrow M_P^2$ inside $%
J_{ij}^r(0)$, one might create large differences between the
``old`` and ``new`` remainders due to the enhancement of the
contributions of chiral logs for small $Y$. However, without
the replacement inside the chiral logs of this type we could
expect an unphysical increase (and irregularities) of the observables for $%
Y\rightarrow 0$. Also, here the replacement is natural physically,
remember that the matching with the dispersive representation
consist essentially of an analogous replacement within the unitary
corrections. Let us also note that the splitting of the
generating functional into the tadpole and unitarity part is not
unique (it depends \emph{e.g. }on the parametrization of the
fluctuations around the classical solution of the $O(p^2)$ field
equations in the functional integral), though the sum must be
independent on this and therefore it is more consistent to use the same rule for both%
\footnote{%
Also notice that the offending $Y$ dependence of the chiral logs with $%
O(p^2)$ masses inside comes always in the combination $Y/\mu ^2$
where $\mu $ is the renormalization scale. Provided we were able
to reparametrize the bare expansion in such a way that all the
running $O(p^4)$ constants were completely expressed in terms of
the physical observables, the explicit independence on $\mu $
would at the same time guarantee an elimination of the irregularities for $Y\rightarrow
0$. Such a treatment has to include the reparametrization of
$L_1$-$L_3$, which is, however,  beyond the scope of our paper.}.
Nevertheless, it could be of some worth to test the differences
between various treatments of the chiral logs numerically (see
Subsection \ref{resummed_numerics}).

The resulting \emph{bare} expansion
(\ref{Gbare}) now not only meets the requirement of the exact
scale independence of the remainder $\delta _G^{^{\prime }}$, it
has also correct physical location of the unitarity cuts. Of
course, we could achieve the last property simply by inserting
physical masses into the functions $\overline{\overline{J}}_{ij}(s)$ in $%
G_{cut}(s,t;u)$. The replacement $G_{cut}(s,t;u)\rightarrow \mathcal{G}%
_{unit}(s,t;u)$ has however another advantage. Namely, using the
prescription (\ref{phiA}), (\ref{psiA}), the corresponding
amplitude, written in the form (without any expansion of the
denominator)
\begin{equation}
S(s,t;u)=\frac{G(s,t;u)}{\prod_{i=1}^4F_{P_i}}
\end{equation}
satisfies the relations of perturbative unitarity (with $S^{(2)}$ and $S^{(4)}$ given by (\ref{Sgood}))
\begin{equation}
\mathrm{disc}S_l^{(4)}(s)=\sum_{ij}\frac
2{{z}_{ij}}\frac{\lambda
_{ij}^{1/2}(s)}sS_l^{(2)ij\rightarrow A_f}(s)S_l^{(2)ij\rightarrow
A_i}(s)^{*}
\end{equation}
\emph{exactly }(\emph{i.e.} not only modulo the
next-to-next-to-leading correction), which can be sometimes
technically useful (\emph{e.g.} for the unitarization by means of
the inverse amplitude method, cf. \cite
{GomezNicola:2001as}). The same is true using the prescription (\ref{phiA1}%
), (\ref{psiA1}) with $\widetilde{S}^{(2)}$ and $\widetilde{S}^{(4)}$ given by
(\ref{S2dangerous}, \ref{S4dangerous}). As we shall see in what follows, the latter
prescription gives a minimal modification of the strict expansion
(\ref{G_strict}) compatible with exact perturbative unitarity.

\section{\boldmath General properties of $\pi \eta $ scattering amplitude}

\subsection{Basic notation}

Let us denote the $s-$and $u-$ channel amplitude in the isospin
conservation limit as
\begin{equation}
\langle \pi ^b(p^b)\eta (q)_{out}|\pi ^a(p^a)\eta (p)_{in}\rangle =\mathrm{i}%
(2\pi )^4\delta (P_f-P_i)\delta ^{ab}S(s,t;u)
\end{equation}
and the crossed amplitude in the $t-$ channel as
\begin{equation}
\langle \eta (p)\eta (q)_{out}|\pi ^a(p^a)\pi ^b(p^b)_{in}\rangle =\mathrm{i}%
(2\pi )^4\delta (P_f-P_i)\delta ^{ab}T(s,t;u).
\end{equation}
Crossing and Bose symmetries then yield
\begin{eqnarray}
T(s,t;u) &=&S(t,s;u) \nonumber \\
S(s,t;u) &=&S(u,t;s) \nonumber \\
T(s,t;u) &=&T(s,u;t).
\end{eqnarray}
Writing the partial wave expansion as
\begin{eqnarray}
S(s,t;u) &=&32\pi \sum_{l=0}^\infty (2l+1)P_l(\cos \theta _s)S_l(s), \nonumber \\
\cos \theta _s &=&\frac{(t-u)s+\Delta _{\eta \pi }^2}{\lambda _{\eta \pi }(s)%
},
\end{eqnarray}
the scattering lengths $a_l$ and phase shifts $\delta _l(s)$ are
given by the formulae
\begin{eqnarray}
\mathrm{Re}S_l(s) &=&\frac{\sqrt{s}}4P^{2l}(a_l+O(P^2))\,\,\,\,\mathrm{for}%
\,\,\,\,P\rightarrow 0\,,\,s\rightarrow (M_\eta +M_\pi )^2 \nonumber \\
\delta _l(s) &=&\arctan \left(
\frac{4P}{\sqrt{s}}\mathrm{Re}S_l(s)\right) ,
\end{eqnarray}
where $P=\lambda _{\eta \pi }^{1/2}(s)/2\sqrt{s}$ is the
CMS\thinspace momentum. I.e., in the units of (pion Compton
wavelength)$^{2l+1}$
\begin{equation}
a_l=M_\pi ^{2l+1}\lim_{P\rightarrow 0}\frac 4{\sqrt{s}P^{2l}}\mathrm{Re}%
A_l(s).
\end{equation}
Let us also define the subthreshold parameters $c_{ij}$ in terms
of the expansion of the amplitude in the point of analyticity
$t=0$, $s=u=\Sigma
_{\eta \pi }=M_\eta ^2+M_\pi ^2$%
\begin{equation}
S(s,t;u)=\sum_{i,j}c_{ij}t^i\nu ^{2j}
\end{equation}
where
\begin{equation}
\nu =\frac{s-u}{4M_\eta }=\frac{2s+t-2\Sigma _{\eta \pi }}{4M_\eta
}.
\end{equation}
The dimension $c_{ij}$ is $\dim [c_{ij}]=mass^{-2i-2j}$, in what
follows we will refer to the dimensionless numbers $c_{ij}M_\pi
^{2i+2j}$. Let us note that in the limit $m_u=m_d=0$ we have two
Adler zeros at $p^a=0$ and $p^b=0$, which implies the following
$SU(2)_L\times SU(2)_R$ theorem
\begin{equation}
\lim_{m_u=m_d\rightarrow 0}c_{00}=0.  \label{c00teorem}
\end{equation}
We can also quote the low-energy current algebra result
\cite{Osborn:1970sm}
\begin{equation}
S(s,t;u)=\frac{M_\pi ^2}{3F_\eta ^2},  \label{low-energy}
\end{equation}
which is in agreement with (\ref{c00teorem}).

\subsection{Dispersive representation}

As a result of the symmetry properties of the amplitudes, the
dispersive representation to the next-to-leading order
(\ref{Gunit}) for $G_{\pi \eta }(s,t;u)=F_\pi ^2F_\eta ^2S(s,t;u)$
simplifies, namely $\phi ^S=\phi ^U\equiv \phi $ and $\psi ^S=\psi
^U\equiv \psi $. The intermediate states
in (\ref{phiA}, \ref{psiA}) are\footnote{%
Here we assume isospin conservation.} $\pi \eta $ and
$\overline{K}K$ in the $s$ and $u$ channels and $\pi \pi $, $\eta
\eta $ and $\overline{K}K$ in the $t-$channel. This implies $\psi
(s)=O(p^6)$, because the $P-$waves in the $s-$channel start at
$O(p^4)$ due to the low-energy theorem (\ref {low-energy}) for
the $\pi \eta \rightarrow \pi \eta $ amplitude and as a result of
charge conjugation invariance of the $\pi \eta \rightarrow
\overline{K}K$ amplitude. Moreover, $\psi ^T=0$, because the
partial wave decomposition of the $t-$channel amplitude $T(s,t;u)$
contains only even partial waves due to Bose symmetry and
charge conjugation. We therefore get
\begin{equation}
G_{\pi \eta }(s,t;u)=G_{\pi \eta ,\,pol}(s,t;u)+\mathcal{G}_{\pi
\eta ,unit}(s,t;u)+O(p^6),
\end{equation}
where the polynomial part has the following general form
\begin{equation}
G_{\pi \eta ,\,pol}(s,t;u)=\alpha +\beta t+\gamma t^2+\omega
(s-u)^2. \label{Gpietapol}
\end{equation}
Note that the parameters $\alpha ,\ldots ,\omega $ are related to
the expansion of the Green function $G_{\pi \eta }(s,t;u)$ at the
point of analyticity $t=0$, $s=u=\Sigma _{\eta \pi }=M_\eta
^2+M_\pi ^2$ and therefore they represent ``good observables''
according to our classification.

The dispersive part is
\begin{equation}
\mathcal{G}_{\pi \eta ,unit}(s,t;u)=\phi ^T(t)+\phi (s)+\phi (u),
\end{equation}
where $\phi ^T(t)$ and $\phi (s)$ are given by the formula
(\ref{phiA}). A complete list of relevant leading order contributions
$G_{0,1}^{(2)12\rightarrow ij}$ and $G_{0,1}^{(2)ij\rightarrow
34}$ can be found in Appendix \ref {appendix dispersive}, here
we give the resulting expressions (transcription to the convention
(\ref{phiA1}, \ref{psiA1}) is straightforward)
\begin{eqnarray}
\phi (s) &=&F_0^4\left\{ \frac 19\stackrel{o}{M}_\pi ^4\frac{\overline{J}%
_{\pi \eta }(s)}{F_\pi ^2F_\eta ^2}\right.  \nonumber \\
&&+\frac 38[(s-\frac 13M_\eta ^2-\frac 13M_\pi ^2-\frac 23M_K^2)
\nonumber
\\
&&\left. -\frac 13(2\stackrel{o}{M}_K^2-\stackrel{o}{M}_\pi ^2-\stackrel{o}{M%
}_\eta ^2)]^2\frac{\overline{J}_{KK}(s)}{F_K^4}\right\} ,  \nonumber \\
\phi ^T(s) &=&F_0^4\left\{ \frac 13\stackrel{o}{M}_\pi ^2[(s-\frac
43M_\pi
^2)+\frac 56\stackrel{o}{M}_\pi ^2]\frac{\overline{J}_{\pi \pi }(s)}{F_\pi ^4%
}\right.  \nonumber \\
&&-\frac 1{18}\stackrel{o}{M}_\pi ^2\left( \stackrel{o}{M}_\pi ^2-4\stackrel{%
o}{M}_\eta ^2\right) \frac{\overline{J}_{\eta \eta }(s)}{F_\eta
^4}
\nonumber \\
&&+\frac 18[(s-\frac 23M_\pi ^2-\frac 23M_K^2)+\frac 23(\stackrel{o}{M}_K^2+%
\stackrel{o}{M}_\pi ^2)]  \nonumber \\
&&\left. \times [(3s-2M_K^2-2M_\eta ^2)+(2\stackrel{o}{M}_\eta ^2-\frac 23%
\stackrel{o}{M}_K^2)]\frac{\overline{J}_{KK}(s)}{F_K^4}\right\} .
\label{disp_res}
\end{eqnarray}
In terms of these functions, we have (notice that $\phi ^T(0)=0$)
\begin{eqnarray}
a_0 &=&\frac 1{8\pi F_\eta ^2F_\pi ^2}\frac{M_\pi }{(M_\pi +M_\eta
)}\left( \alpha +16\omega M_\eta ^2M_\pi ^2+\phi ((M_\pi +M_\eta
)^2)+\phi ((M_\eta
-M_\pi )^2)\right)  \nonumber \\
a_1 &=&\frac 1{12\pi F_\eta ^2F_\pi ^2}\frac{M_\pi ^3}{(M_\pi +M_\eta )}%
\left( \beta +8\omega M_\eta M_\pi +\phi ^{T^{\prime }}(0)-\phi
^{^{\prime }}((M_\eta -M_\pi )^2)\right)  \label{a0a1}
\end{eqnarray}
and
\begin{eqnarray}
c_{00} &=&\frac 1{F_\eta ^2F_\pi ^2}\left( \alpha +2\phi (\Sigma
_{\eta \pi
})\right)  \nonumber \\
c_{10} &=&\frac 1{F_\eta ^2F_\pi ^2}\left( \beta +\phi ^{T^{\prime
}}(0)-\phi ^{^{\prime }}(\Sigma _{\eta \pi })\right)  \nonumber \\
c_{20} &=&\frac 1{F_\eta ^2F_\pi ^2}\left( \gamma +\frac 12\phi
^{T^{\prime \prime }}(0)+\frac 14\phi ^{^{^{\prime \prime
}}}(\Sigma _{\eta \pi })\right)
\nonumber \\
c_{01} &=&\frac{16M_\eta ^2}{F_\eta ^2F_\pi ^2}\left( \omega
+\frac 14\phi ^{^{^{\prime \prime }}}(\Sigma _{\eta \pi })\right).
\label{cij}
\end{eqnarray}
While the scattering lengths, being related to the value of the
amplitude at the threshold, are not candidates for ``good
observables'', the situation is a
little bit more subtle in the case of the subthreshold parameters. Provided
the $\eta $ decay constant was known from experiments as accurately as $%
F_\pi $, then (similarly to $\alpha ,$ $\beta ,\ldots $) also the
$c_{ij}$ could be treated as ``good observables''. However, this
is not the case, and we should rather use a chiral expansion of
$F_\eta $ in the above formulae. Therefore, the subthreshold
parameters are typical examples of the dangerous ratios, which
should be treated with care.

\subsection{Bare expansion for $G(s,t;u)$}

For the strict expansion in terms of LEC's (i.e. without any
reparametrization
in terms of physical observables) derived from (\ref{Zvaps}, \ref{Zvaps4}%
), we have confirmed the results of the article \cite{Bernard:1991xb} by independent calculation. 
The $O(p^4)$ expansion can be written in the form
\begin{equation}
G_{\pi \eta
}=G^{(2)}+G_{ct}^{(4)}+G_{tad}^{(4)}+G_{unit}^{(4)}+G\delta _G,
\end{equation}
where
\begin{eqnarray}
G^{(2)}(s,t;u) &=&\frac{F_0^2}3\stackrel{o}{M}_\pi ^2  \nonumber \\
&&  \nonumber \\
G_{ct}^{(4)}(s,t;u) &=&8(L_1^r(\mu )+\frac 16L_3^r(\mu ))(t-2M_\pi
^2)(t-2M_\eta ^2)  \nonumber \\
&&+4(L_2^r(\mu )+\frac 13L_3^r(\mu ))[(s-M_\pi ^2-M_\eta
^2)^2+(u-M_\pi
^2-M_\eta ^2)^2]  \nonumber \\
&&+8L_4^r(\mu )[(t-2M_\pi ^2)\stackrel{o}{M}_\eta ^2+(t-2M_\eta ^2)\stackrel{%
o}{M}_\pi ^2]  \nonumber \\
&&-\frac 83L_5^r(\mu )(M_\pi ^2+M_\eta ^2)\stackrel{o}{M}_\pi ^2+8L_6^r(\mu )%
\stackrel{o}{M}_\pi ^2(\stackrel{o}{M}_\pi
^2+5\stackrel{o}{M}_\eta ^2)
\nonumber \\
&&+32L_7^r(\mu )(\stackrel{o}{M}_\pi ^2-\stackrel{o}{M}_\eta ^2)\stackrel{o}{%
M}_\pi ^2+\frac{64}3L_8^r(\mu )\stackrel{o}{M}_\pi ^4  \nonumber \\
&&  \nonumber \\
G_{tad}^{(4)}(s,t;u) &=&-\frac{F_0^2}3\stackrel{o}{M}_\pi ^2\left(
3\mu _\pi
+2\mu _K+\frac 13\mu _\eta \right)  \nonumber \\
&&  \nonumber \\
G_{unit}^{(4)}(s,t;u) &=&\frac 19\stackrel{o}{M}_\pi ^4[J_{\pi
\eta
}^r(s)+J_{\pi \eta }^r(u)]  \nonumber \\
&&+\frac 38[s-M_\pi ^2-M_\eta ^2+\frac 23\stackrel{o}{M}_\pi
^2]^2J_{KK}^r(s)+\frac 38[u-M_\pi ^2-M_\eta ^2+\frac
23\stackrel{o}{M}_\pi
^2]^2J_{KK}^r(u)  \nonumber \\
&&+\frac 13\stackrel{o}{M}_\pi ^2[t-2M_\pi ^2+\frac
32\stackrel{o}{M}_\pi
^2]J_{\pi \pi }^r(t)  \nonumber \\
&&+\frac 29\stackrel{o}{M}_\pi ^2(\stackrel{o}{M}_\eta ^2-\frac 14\stackrel{o%
}{M}_\pi ^2)J_{\eta \eta }^r(t)  \nonumber \\
&&+\frac 18[t-2M_\pi ^2+2\stackrel{o}{M}_\pi ^2][3t-6M_\eta ^2+4\stackrel{o}{%
M}_\eta ^2-\frac 23\stackrel{o}{M}_\pi ^2]J_{KK}^r(t)
\label{Gpietachiral}
\end{eqnarray}
are the $O(p^2)$, counterterm, tadpole and unitarity contributions
respectively. In the above formulae, the masses within the loop functions $%
J_{PQ}^r(t)$ are the $O(p^2)$ masses
\begin{equation}
\stackrel{o}{M}_\pi ^2 = 2B_0\widehat{m},\quad 
\stackrel{o}{M}_K^2 = B_0\widehat{m}\left( r+1\right),\quad
\stackrel{o}{M}_\eta ^2 = \frac 23B_0\widehat{m}\left(
2r+1\right).
\end{equation}
The chiral logs $\mu _P$ can be expressed using $J^r_{PP}(0)$
\begin{equation}
\mu _P=\frac{\stackrel{o}{M}_P^2}{32\pi ^2F_0^2}\ln \frac{\stackrel{o}{M}_P^2%
}{\mu
^2}=-\frac{\stackrel{o}{M}_P^2}{2F_0^2}\left(J^r_{PP}(0)+\frac{1}{16\pi^2}\right).
\end{equation}
Written in such a form, the sum
$G_{ct}^{(4)}+G_{tad}^{(4)}+G_{unit}^{(4)}$ is \emph{exactly}
renormalization scheme independent by construction. Let us now
proceed as described in the previous section and write the bare
expansion of $G(s,t;u)$ in the form
\begin{equation}
G_{\pi \eta }(s,t;u)=G_{\pi \eta ,\,pol}(s,t;u)+\mathcal{G}_{\pi
\eta ,unit}(s,t;u)+G\delta _G^{^{^{\prime }}}.
\end{equation}
Writing $J_{ij}^r(s)=J_{ij}^r(0)+\overline{J}_{ij}(s)$ in (\ref{Gpietachiral}),
we get  the renormalization scale independent polynomial part
\begin{eqnarray}
G_{\pi \eta ,\,pol}(s,t;u)
&=&G^{(2)}(s,t;u)+G_{ct}^{(4)}(s,t;u)+G_{tad}^{(4)}(s,t;u) \nonumber\\
&&+\frac 13\stackrel{o}{M}_\pi ^2[t-2M_\pi ^2+\frac
32\stackrel{o}{M}_\pi
^2]J_{\pi \pi }^r(0) \nonumber\\
&&+\frac 29\stackrel{o}{M}_\pi ^4J_{\pi \eta }^r(0)+\frac 29\stackrel{o}{M}%
_\pi ^2(\stackrel{o}{M}_\eta ^2-\frac 14\stackrel{o}{M}_\pi
^2)J_{\eta \eta
}^r(0) \nonumber\\
&&\frac 38\{[s-M_\pi ^2-M_\eta ^2+\frac 23\stackrel{o}{M}_\pi
^2]^2+[u-M_\pi
^2-M_\eta ^2+\frac 23\stackrel{o}{M}_\pi ^2]^2 \nonumber\\
&&+[t-2M_\pi ^2+2\stackrel{o}{M}_\pi ^2][t-2M_\eta ^2+\frac 43\stackrel{o}{M}%
_\eta ^2-\frac 29\stackrel{o}{M}_\pi ^2]\}J_{KK}^r(0).
\end{eqnarray}

Comparing this with the general form (\ref{Gpietapol}) of $G_{\pi
\eta ,\,pol}(s,t;u)$, we get for the bare expansions of the
parameters $\alpha -\omega $ the following manifestly
renormalization scale independent form
\begin{eqnarray}  \label{alpha_bare}
\alpha &=&\frac 13F_0^2\stackrel{o}{M}_\pi ^2+\frac 1{96\pi ^2}\stackrel{o}{M%
}_\pi ^2\left( \frac 72\stackrel{o}{M}_\pi
^2+\frac{11}6\stackrel{o}{M}_\eta
^2\right)  \nonumber \\
&&+4[8(L_1^r(\mu )+\frac 16L_3^r(\mu ))+\frac 38J_{KK}^r(0)]M_\pi
^2M_\eta ^2
\nonumber \\
&&-[16L_4^r(\mu )+J_{KK}^r(0)]M_\pi ^2\stackrel{o}{M}_\eta
^2-[16L_4^r(\mu )+\frac 83L_5^r(\mu )+\frac 32J_{KK}^r(0)]M_\eta
^2\stackrel{o}{M}_\pi ^2
\nonumber \\
&&-[\frac 83L_5^r(\mu )-\frac 16J_{KK}^r(0)+\frac 23J_{\pi \pi
}^r(0)]M_\pi
^2\stackrel{o}{M}_\pi ^2  \nonumber \\
&&+[40L_6^r(\mu )+\frac 5{18}J_{\eta \eta }^r(0)+\frac 54J_{KK}^r(0)]%
\stackrel{o}{M}_\pi ^2\stackrel{o}{M}_\eta ^2  \nonumber \\
&&+32L_7^r(\mu )\stackrel{o}{M}_\pi ^2(\stackrel{o}{M}_\pi ^2-\stackrel{o}{M}%
_\eta ^2)  \nonumber \\
&&+[8L_6^r(\mu )+\frac{64}3L_8^r(\mu )+J_{\pi \pi }^r(0)+\frac
29J_{\pi \eta
}^r(0)-\frac 1{18}J_{\eta \eta }^r(0)+\frac 14J_{KK}^r(0)]\stackrel{o}{M}%
_\pi ^4+\frac 13F_\pi ^2M_\pi ^2\delta _\alpha  \nonumber \\
\\
&&  \nonumber \\
\beta &=&-2\Sigma _{\eta \pi }[8(L_1^r(\mu )+\frac 16L_3^r(\mu
))+\frac
38J_{KK}^r(0)]  \nonumber \\
&&+[8L_4^r(\mu )(\stackrel{o}{M}_\eta ^2+\stackrel{o}{M}_\pi
^2)+\frac 12J_{KK}^r(0)\stackrel{o}{M}_\eta ^2+\frac 13(J_{\pi \pi
}^r(0)+\frac 12J_{KK}^r(0))\stackrel{o}{M}_\pi ^2]+\beta \delta
_\beta  \label{beta_bare}
\\
\gamma &=&[8(L_1^r(\mu )+\frac 16L_3^r(\mu ))+\frac
38J_{KK}^r(0)]+[2(L_2^r(\mu )+\frac 13L_3^r(\mu ))+\frac
3{16}J_{KK}^r(0)]+\gamma \delta _\gamma  \label{gama_bare} \\
\omega &=&[2(L_2^r(\mu )+\frac 13L_3^r(\mu ))+\frac
3{16}J_{KK}^r(0)]+\omega \delta _\omega .  \label{omega_bare}
\end{eqnarray}

\section{Reparametrization of the bare expansion}

Let us now discuss the various possibilities of the
reparametrization of the bare expansion.

\subsection{$\pi \eta $ scattering within the standard chiral
perturbation theory to $O(p^4)\label{standard_rules}\label{subsection_std}%
\label{standard}$}

The standard way of dealing with the chiral expansion consists of two ``dangerous'' steps. The first
one involves using the inverted expansions of the type
(\ref{inverted}) in order to express the amplitude in terms of the
masses and decay constants instead of
the parameters $B_0\widehat{m}$, $F_0$ and $r=m_s/\widehat{m}$ of the $%
O(p^2) $ chiral Lagrangian. Here one encounters an ambiguity
connected with
different possibilities how to choose the observable $G$ in (\ref{inverted}%
), the chiral expansion of which starts with the desired $O(p^2)$
parameter $G_0$. 

Let us fix this ambiguity by using the expansions of $F_\pi^2$, $M_\pi^2$ and $M_K^2$, inverting of which leads to\footnote{%
Instead of $M_K^2$ we could use the chiral expansion of $M_\eta^2$ to obtain
\begin{equation*}
r=\widetilde{r}_2=\frac 32\left( \frac{M_\eta ^2}{M_\pi ^2}-\frac
13\right)
\end{equation*}
or even $F_K^2 M_K^2$ to get
\begin{equation*}
r=r_2^{*}=2\frac{F_K^2M_K^2}{F_\pi ^2M_\pi ^2}-1.
\end{equation*}
The latter choice, formally as good as the previous
two, could also involve the redefinition of the loop masses to $\stackrel{o}{M}_P^2=F_P^2M_P^2/F_\pi^2$ instead of the simple $\stackrel{o}{M}_P^2=M_P^2$ as in the case of the other standard reparametrizations. Even then, however,
it suffers from numerically large $O(p^4)$ corrections
which could produce instabilities of the reparametrization based
on this observable.}
\begin{eqnarray}
		F_0^2 &=& F_{\pi}^2\left(1+4\mu_{\pi}+2\mu_K\right) 
			- 8M_{\pi}^2\left(L_4^r(\mu)(2+r)+L_5^r(\mu)\right)
\label{standard_F0}\\
		2B_0\widehat{m} &=& M_{\pi}^2\left[1-\mu_{\pi}+\frac{1}{3}\mu_{\eta}
			-\frac{8M_{\pi}^2}{F_{\pi}^2}\Big( 
			2L_8^r(\mu)+2(2+r)L_6^r(\mu)-L_5^r(\mu)-(2+r)L_4^r(\mu)\Big) \right]
\label{standard_B0}\\
		r &=& r_2\, =\, \frac{2M_K^2}{M_{\pi}^2}-1 + O(p^2).
\label{standard_r}
\end{eqnarray}
Inserting the inverted expansions (\ref{standard_F0}--\ref{standard_r}) into (\ref
{alpha_bare}) and (\ref{beta_bare}) and keeping terms up to the order $%
O(p^4)\,$we get
\begin{eqnarray}
\alpha &=&\frac 13F_\pi ^2M_\pi ^2+\frac{16}3M_\pi ^2M_\eta ^2L_3^r(\mu )-%
\frac{64}3L_7^r(\mu )M_\pi ^4(r_2-1)  \nonumber \\
&&+M_\pi ^2M_\eta ^2[32L_1^r(\mu )-16L_4^r(\mu )-\frac 83L_5^r(\mu
)]
\nonumber \\
&&+\frac 13(2r_2+1)M_\pi ^4[32L_6^r(\mu )-16L_4^r(\mu )+\frac
29J_{\eta \eta
}^r(0)]  \nonumber \\
&&+M_\pi ^4[-\frac 83L_5^r(\mu )+16L_8^r(\mu )-\frac 16J_{\pi \pi
}^r(0)+\frac 29J_{\pi \eta }^r(0)-\frac 1{18}J_{\eta \eta
}^r(0)+\frac
13J_{KK}^r(0)]+\alpha \delta _\alpha ^{st}\nonumber\\  \label{alpha_std} \\
&&  \nonumber \\
\beta &=&-2\Sigma _{\eta \pi }[8(L_1^r(\mu )+\frac 16L_3^r(\mu
))+\frac
38J_{KK}^r(0)]  \nonumber \\
&&+\frac 13M_\pi ^2[16L_4^r(\mu )(r_2+2)+J_{KK}^r(0)(r_2+1)+J_{\pi
\pi }^r(0)]+\beta \delta _\beta ^{st},  \label{beta_std}
\end{eqnarray}
with new remainders $\delta _\alpha ^{st}$ and $\delta _\beta
^{st}$, which might be, however, out of control as we have already
discussed. In fact, this first step involves three ``unsafe" manipulations from the point of view of resummed $\chi PT$: using ``dangerous" expansions for the masses as a starting point, the inversion and finally the negligence of all higher order terms generated by this procedure after the insertion. 

It's understood to use physical masses inside the chiral logarithms. Higher order LEC's are then fitted by using additional experimental input, no parameters are therefore left free. Also note that (\ref{standard_r}) effectively implements the classical Gell-Mann-Okubo formula

\begin{equation}
		3M_\eta^2-4M_K^2+M_\pi^2=0.
\end{equation}
This insures renormalization scale independence. We, however, leave $M_\eta$ at its physical value in cases when it was produced by on-shell mass on outer legs or inside chiral logarithms, which is compatible with the requirement of scale independence. 

The second step is connected to the fact that the amplitude is used in standard $\chi PT$ rather than $G(s,t;u)$. As was shown in Section 3.2, the expansion of the amplitude can be organized in various ways, of which only (\ref{Sgood}) is considered safe in the resummed approach. On the other hand, from the standard point of view it often seems more advantageous to use (\ref{S2dangerous}, \ref{S4dangerous}), as together with (\ref{standard_F0}) it leads to only the experimentally very well known pion decay constant being present in the formulae. This can be seen on the case of $F_\eta$, which is experimentally poorly known due to $\eta-\eta'$ mixing \cite{FEta} and thus if it's kept at its physical value as was done in \cite{Bernard:1991xb}, a significant uncertainty is introduced into the results. As the normalization (\ref{S2dangerous}, \ref{S4dangerous}) is used more often in NLO $S\chi PT$, we will adhere to this view and perform this second step by expanding the kaon and eta decay constants from the denominators and subsequently cutting off the higher orders.

Using therefore the prescription (\ref{phiA1}, \ref{psiA1}), the dispersive part of the $O(p^4)$ amplitude (\ref{disp_res}) simplifies using the reparametrization recipe
described above
\begin{eqnarray}
\phi (s) &=&\frac 19M_\pi ^4\overline{J}_{\pi \eta }(s)  \nonumber \\
&&+\frac 38[(s-\frac 13M_\eta ^2-\frac 13M_\pi ^2-\frac
23M_K^2)-\frac
19M_\pi ^2(r_2-1)]^2\overline{J}_{KK}(s),  \nonumber \\
\phi ^T(s) &=&\frac 13M_\pi ^2(s-\frac 12M_\pi
^2)\overline{J}_{\pi \pi }(s)+\frac 1{54}M_\pi ^4\left(
8r_2+1\right) \overline{J}_{\eta \eta }(s)
\nonumber \\
&&+\frac 18[(s-\frac 23M_\pi ^2-\frac 23M_K^2)+\frac 13M_\pi
^2(r_2+3)]
\nonumber \\
&&\times [(3s-2M_K^2-2M_\eta ^2)+\frac 13M_\pi ^2(3r_2+1)]\overline{J}%
_{KK}(s).  \label{disp_std}
\end{eqnarray}
The second step also propagates itself to the case of the subthreshold parameters $c_{ij}$ and the scattering lengths $a_i$,  where it consists of
the expansion of $F_\eta ^2$ in the denominator of the formulae (\ref{cij}) and (\ref{a0a1}). This step could in principle produce uncontrollable contribution to the remainders as well.

\subsection{Resummation of the vacuum fluctuation}

In order to preserve the global convergence, as was discussed, in the context of resummed $\chi PT$ we are not allowed to perform ``dangerous'' inverted
expansions and thus to express the $O(p^2)$ masses
$\stackrel{o}{M}_P$ and the decay constant $F_0$ in terms of the
physical ones in the way it is common within the standard $\chi PT$
calculations sketched above. Instead of this, the $O(p^2)$ LEC's are left free, or more precisely, rewritten using parameters directly related  to the
order parameters of the chiral symmetry breaking\footnote{%
Here we omit the explicit dependence of $X$, $Y$ and $Z$ on $N_f$
keeping in mind that $N_f=3$ in what follows.}
\begin{equation}
r = \frac{m_s}{\widehat{m}},\quad
X = \frac{\stackrel{o}{M}_\pi ^2F_0^2}{M_\pi ^2F_\pi ^2},\quad
Z = \frac{F_0^2}{F_\pi ^2}.
\end{equation}
The bare expansions for masses $F_P^2M_P^2$ and decay constants $F_P^2$ are used the reparametrize the NLO LEC's $L_4-L_8$. As the dependence is linear, it can be done in a purely non-perturbative algebraic way by introduction of an unknown higher order remainder to each observable used. The relevant formulae for $L_4-L_8$ can be found in Appendix \ref{appendix Li}.

As masses and decay constants do not depend on $L_1,\,L_2,\,L_3$, bare expansions of some additional, experimentally well known observables is needed for these LEC's. This is, however, even if highly desirable, out of the scope of our article and we make a shortcut and use the standard tabular values for these constants. We will make an analysis of the sensitivity of our results to a change in the value of $L_1-L_3$ in the next section devoted to numerical results.

For the resulting expression for the parameters $%
\alpha $ and $\beta $, we use the following
abbreviation for some repeatedly occurring combinations
\begin{eqnarray}
r_2^{*} &=&2\frac{F_K^2M_K^2}{F_\pi ^2M_\pi ^2}-1 \label{r_2*}\\
\varepsilon (r) &=&2\frac{r_2^{*}-r}{r^2-1} \label{epsilon}\\
\eta (r) &=&\frac 2{r-1}\left( \frac{F_K^2}{F_\pi ^2}-1\right) \label{eta}\\
\Delta _{GMO} &=&\frac{3F_\eta ^2M_\eta ^2+F_\pi ^2M_\pi ^2-4F_K^2M_K^2}{%
F_\pi ^2M_\pi ^2}, \label{DeltaGMO}
\end{eqnarray}
in terms of which we get
\begin{eqnarray}
\alpha &=&\frac 13XF_\pi ^2M_\pi ^2+\frac 13F_\pi ^2M_\pi ^2(1-X)\frac{5r+4}{%
r+2}+\frac 13F_\pi ^2M_\pi ^2\varepsilon (r)r\frac{2r+1}{r+2}-\frac 23\frac{%
F_\pi ^2M_\pi ^2}{r-1}\Delta _{GMO}  \nonumber \\
&&+2\frac{F_\pi ^2}{r+2}(Z-1)\left( \frac 13M_\pi ^2(2r+1)+M_\eta ^2\right) +%
\frac{F_\pi ^2}{r+2}\eta (r)\left( rM_\pi ^2-\frac 13(r-4)M_\eta
^2\right)
\nonumber \\
&&+\frac 1{96\pi ^2}\frac XZM_\pi ^4\left( 4r+5\right) +\frac
3{32\pi ^2}\frac XZM_\pi ^2M_\eta ^2-\frac 1{864\pi ^2}\left(
\frac XZ\right)
^2M_\pi ^4(44r+67)  \nonumber \\
&&-\frac{M_\pi ^4}{2(r+2)(r-1)}\frac XZ[J_{\eta \eta
}^r(0)(2r+1)+2J_{KK}^r(0)r-J_{\pi \pi }^r(0)(4r+1)]r  \nonumber \\
&&+\frac{M_\pi ^2M_\eta ^2}{6(r+2)(r-1)}\frac XZ[J_{\eta \eta
}^r(0)(2r+1)(r-4)+J_{\pi \pi }^r(0)(19r-4)-2J_{KK}^r(0)(r^2+6r-4)]
\nonumber
\\
&&+\frac{M_\pi ^4}{18(r+2)(r-1)}\left( \frac XZ\right) ^2[J_{\eta
\eta
}^r(0)(5r^2-10r-4)+6J_{KK}^r(0)(3r^2-2r-4)  \nonumber \\
&&+4J_{\pi \eta }^r(0)(r^2+r-2)-9J_{\pi \pi }^r(0)(3r^2-2r-4)]  \nonumber \\
&&+4[8(L_1^r(\mu )+\frac 16L_3^r(\mu ))+\frac 38J_{KK}^r(0)]M_\pi
^2M_\eta
^2+\frac 13F_\pi ^2M_\pi ^2\delta _\alpha ^{^{\prime }},  \nonumber \\
&&  \label{alpha_res} \\
\beta &=&\frac 23F_\pi ^2(1-Z-\eta (r))  \nonumber \\
&&+\frac 13\frac{M_\pi ^2}{r-1}\frac X Z[J_{\eta \eta
}^r(0)(2r+1)+J_{KK}^r(0)(r+1)-J_{\pi \pi }^r(0)(3r+2)-\frac
1{16\pi
^2}(r+2)(r-1)]  \nonumber \\
&&-2\Sigma _{\eta \pi }[8(L_1^r(\mu )+\frac 16L_3^r(\mu ))+\frac
38J_{KK}^r(0)]+\beta \delta _\beta ^{^{\prime }}.
\label{beta_res}
\end{eqnarray}
The new primed remainders are the following functions of the
original remainders entering the game
\begin{eqnarray}
\delta _\alpha ^{^{\prime }} &=&\delta _\alpha
-\frac{5r+4}{r+2}\delta _{F_\pi M_\pi }+\frac 2{r+2}\left(
(2r+1)+\frac{3M_\eta ^2}{M_\pi ^2}\right)
\delta _{F_\pi }  \nonumber \\
&&-\frac 2{(r+2)(r-1)}\left( 3r-(r-4)\frac{M_\eta ^2}{M_\pi ^2}\right) (%
\frac{F_K^2}{F_\pi ^2}\delta _{F_K}-\delta _{F_\pi })  \nonumber \\
&&-\frac{2r(2r+1)}{(r+2)(r^2-1)}\left( 2\frac{F_K^2M_K^2}{F_\pi ^2M_\pi ^2}%
\delta _{F_KM_K}-(r+1)\delta _{F_\pi M_\pi }\right)  \nonumber \\
&&+\frac 2{r-1}\left( 3\frac{F_\eta ^2M_\eta ^2}{F_\pi ^2M_\pi
^2}\delta
_{F_\eta M_\eta }+\delta _{F_\pi M_\pi }-4\frac{F_K^2M_K^2}{F_\pi ^2M_\pi ^2}%
\delta _{F_KM_K}\right) 
\label{dalpha_res}\\
\delta _\beta ^{^{\prime }} &=&\delta _\beta +\frac
23\frac{2F_K^2\delta _{F_K}-(r+1)F_\pi ^2\delta _{F_\pi }}{\beta
(r-1)}.  
\label{dbeta_res}
\end{eqnarray}
This is an alternative to the first step used in the standard approach to $\chi PT$. 
Because the value of $F_\eta$ is not very well known, we make in a sense a parallel
to the second step as well, i.e. using the chiral expansion for $F_\eta^2$ in the denominator of (\ref{disp_res}), (\ref{a0a1}) and (\ref{cij}). It involves the
reparametrization in terms of $X$, $Z$, $r$
(see Appendix \ref{appendix decay constants} for details), but,
contrary to the standard case, the denominator is not further
expanded and the result is given in a nonperturbative resummed
form of a ratio of two ``safe" expansions.

\subsection{$\pi \eta $ scattering within the generalized chiral
perturbation theory to $O(p^4)$ - the bare expansion of $G(s,t;u)$}

In analogy with (\ref{Gpietachiral}), the strict chiral expansion for $%
G(s,t;u)$ within the generalized $\chi PT$ can be
straightforwardly obtained by using the Lagrangian summarized in the
Appendix \ref{appendix Lagrangian}, where we use the traditional
notation for the LEC's. The result has the following structure
\begin{equation}
G_{\pi \eta
}=\widetilde{G}^{(2)}+\widetilde{G}^{(3)}+\widetilde{G}_{ct}^{(4)}+\widetilde{G}_{tad}^{(4)}+\widetilde{G}_{unit}^{(4)}+G\delta_G^{G\chi PT},
\end{equation}
where\footnote{%
All the LEC's in the following formulae are the renormalized LEC's at scale $%
\mu $. We have omitted explicit notation of this in order to
simplify the expressions.}
\begin{eqnarray}
\widetilde{G}^{(2)}(s,t;u) &=&\frac 13F_0^2[\widetilde{M}_\pi ^2+4\widehat{m}%
^2(3A_0-4(r-1)Z_0^P+2(2r+1)Z_0^S)] \nonumber\\
\widetilde{G}^{(3)}(s,t;u) &=&\frac 13F_0^2\left[ -2\,\widehat{m}\,\left(
6\,M_\eta
^2+M_\pi ^2\,\left( 2+4\,r\right) -2\,\left( 2+r\right) \,t\right) \,\tilde{%
\xi}-2\widehat{m}\,\Sigma _{\pi \eta }\,\xi \right. \nonumber\\
&&+\left. 81\,\widehat{m}^3\,\rho _1+\widehat{m}^3\,\rho _2+\left(
80-64\,\,r-16\,\,r^2\right) \,\widehat{m}^3\,\rho _3\right. \nonumber\\
&&+\left. \left( 100+64\,\,r+34\,\,r^2\right)
\,\widehat{m}^3\,\rho
_4+\left( 2\,+\,r^2\right) \widehat{m}^3\,\rho _5\right. \nonumber\\
&&+\left. \left( 96-96\,r\right) \,\widehat{m}^3\,\rho _6+\left(
144\,+288\,\,r+108\,\,r^2\right) \widehat{m}^3\,\rho _7\right] \nonumber\\
&& \nonumber\\
\widetilde{G}_{ct}^{(4)}(s,t;u) &=&8(L_1+\frac 16L_3)(t-2M_\pi ^2)(t-2M_\eta ^2) \nonumber\\
&&+4(L_2+\frac 13L_3)[(s-M_\pi ^2-M_\eta ^2)^2+(u-M_\pi ^2-M_\eta ^2)^2] \nonumber\\
&&+\frac 83\widehat{m}^2F_0^2\left\{ -(B_1-B_2)\Sigma _{\pi \eta
}+2D^PM_\pi
^2(r-1)-2C_1^PM_\eta ^2(r-1)\right. \nonumber\\
&&+\left. C_1^S(2r+1)t-D^S[\frac 12\Sigma _{\pi \eta
}(5r+4)-(2r+1)t]\right.
\nonumber\\
&&-\left. 2B_4[3M_\eta ^2+M_\pi ^2(2r^2+1)-(r^2+2)t]\right\} \nonumber\\
&&+\frac 13\widehat{m}^4F_0^2\left[
256E_1+16E_2+F_1^P(256-256r^2)+F_4^S(32+16r^2)\right. \nonumber\\
&&+\left. F_1^S(256+320r^2)+F_5^{SP}(192-320r+160r^2-32r^3)\right. \nonumber\\
&&+\left. F_2^P(240-216r-24r^3)+F_6^{SP}(32-32r+16r^2-16r^3)\right. \nonumber\\
&&+\left. F_3^P(16-8r-8r^3)+F_3^S(16+10r+10r^3)\right. \nonumber\\
&&+\left.
F_6^{SS}(32+40r+16r^2+20r^3)+F_7^{SP}(384-160r-256r^2+32r^3)\right.
\nonumber\\
&&+\left. F_2^S(400+234r+74r^3)+F_5^{SS}(576+720r+480r^2+168r^3)\right] \nonumber\\
&& \nonumber\\
\widetilde{G}_{tad}^{(4)}(s,t;u) &=&-\frac 19F_0^2\left[ 2\widehat{m}B_0(\mu
_\eta +6\mu _K+9\mu _\pi )+8A_0\widehat{m}^2(8\mu _\eta +3\mu
_K(r+8)+48\mu _\pi )\right.
\nonumber\\
&&+\left. 4Z_0^S\widehat{m}^2(\mu _\eta (16+41r)+\mu
_K(48+90r)+\mu _\pi
(96+45r))\right. \nonumber\\
&&-\left. 16Z_0^P\widehat{m}^2(2\mu _\eta (5r-2)+3\mu
_K(6r-4)+3\mu _\pi
(3r-8)\right] \nonumber\\
&& \nonumber\\
\widetilde{G}_{unit}^{(4)}(s,t;u) &=&\frac 19[\widetilde{M}_\pi ^2+4\widehat{m}%
^2(3A_0-4(r-1)Z_0^P+2(2r+1)Z_0^S)]^2[J_{\pi \eta }^r(s)+J_{\pi
\eta }^r(u)]
\nonumber\\
&&+\frac 38[s-M_\pi ^2-M_\eta ^2+\frac 23\widetilde{M}_\pi
^2-\frac
83(r-1)\widehat{m}^2(A_0+2Z_0^P)]^2J_{KK}^r(s) \nonumber\\
&&+\frac 38[u-M_\pi ^2-M_\eta ^2+\frac 23\widetilde{M}_\pi
^2-\frac
83(r-1)\widehat{m}^2(A_0+2Z_0^P)]^2J_{KK}^r(u) \nonumber\\
&&+\frac 13[\widetilde{M}_\pi ^2+4\widehat{m}%
^2(3A_0-4(r-1)Z_0^P+2(2r+1)Z_0^S)] \nonumber\\
&&\times [t-2M_\pi ^2+\frac 32\widetilde{M}_\pi ^2+10\widehat{m}%
^2(A_0+2Z_0^S)]J_{\pi \pi }^r(t) \nonumber\\
&&+\frac 29 [\widetilde{M}_\pi ^2+4\widehat{m}%
^2(3A_0-4(r-1)Z_0^P+2(2r+1)Z_0^S)] \nonumber\\
&&\times [\widetilde{M}_\eta ^2-\frac 14\widetilde{M}_\pi ^2+\widehat{m}%
^2((8r^2+1)A_0+8r(r-1)Z_0^P+2(2r+1)^2Z_0^S)]J_{\eta \eta }^r(t) \nonumber\\
&&+\frac 18[t-2M_\pi ^2+2\widetilde{M}_\pi ^2+8(r+1)\widehat{m}%
^2(A_0+2Z_0^S))] \nonumber\\
&&\times [3t-6M_\eta ^2+6\widetilde{M}_\eta ^2-\frac
83\widetilde{M}_K^2
\nonumber\\
&&+\frac
83(r+1)\widehat{m}^2(3rA_0+2(r-1)Z_0^P+2(2r+1)Z_0^S]J_{KK}^r(t).
\end{eqnarray}
In the above formulae, the generalized $O(p^2)$ masses
(also present implicitly  in the chiral logs $\mu _P$ and the loop
functions $J_{PQ}^r(s)$) are
\begin{eqnarray}
\widetilde{M}_\pi ^2 &=&2[B_0+2\widehat{m}(r+2)Z_0^S]\widehat{m}+4A_0%
\widehat{m}^2  \nonumber \\
\widetilde{M}_K^2
&=&[B_0+2\widehat{m}(r+2)Z_0^S]\widehat{m}\left(
r+1\right) +A_0\widehat{m}^2\left( r+1\right) ^2  \nonumber \\
\widetilde{M}_\eta ^2 &=&\frac 23[B_0+2\widehat{m}(r+2)Z_0^S]\widehat{m}%
\left( 2r+1\right) +\frac 43A_0\widehat{m}^2\left( 2r^2+1\right)
+\frac 83Z_0^P\widehat{m}^2(r-1)^2.  \label{Op2_mass_G}
\end{eqnarray}

The unitarity part can be further split into the polynomial and
dispersive part
\begin{equation}
\widetilde{G}_{unit}^{(4)}=\widetilde{G}_{unit,\,pol}^{(4)}+\widetilde{G}_{unit,\,disp}^{(4)}
		=\widetilde{G}_{unit}^{(4)}|_{J^r\rightarrow J(0)}+
		 \widetilde{G}_{unit}^{(4)}|_{J^r\rightarrow \bar{J}}.
\end{equation}
According to the general prescription, the dispersive part can be
replaced with  that of the dispersive representation which has the
general form
\begin{equation}
		\mathcal{\widetilde{G}}_{unit}^{(4)}=\phi^T(t)+\phi(s)+\phi(u),
\end{equation}
where now
\begin{eqnarray}
\phi(s) &=&F_0^4\left\{ \left( \frac 13\alpha _{\pi \eta }\widetilde{M}%
_\pi ^2\right) ^2\frac{\overline{J}_{\pi \eta }(s)}{F_\pi ^2F_\eta
^2}\right.
\nonumber \\
&&\left. +\frac 38[s-\frac 13M_\eta ^2-\frac 13M_\pi ^2-\frac
23M_K^2-\frac 13(2\widetilde{M}_K^2-\widetilde{M}_\pi
^2-\widetilde{M}_\eta
^2+\alpha _{\pi \eta K}\widetilde{M}_\pi ^2)]^2\frac{\overline{J}_{KK}(s)}{%
F_K^4}\right\}  \nonumber \\
&&  \label{disp phi gen} \\
\phi^T(s) &=&F_0^4\left\{ \frac 13\alpha _{\pi \eta
}\widetilde{M}_\pi
^2[s-\frac 43M_\pi ^2+\frac 56\alpha _{\pi \pi }\widetilde{M}_\pi ^2]\frac{%
\overline{J}_{\pi \pi }(s)}{F_\pi ^4}\right.  \nonumber \\
&&\left. -\frac 1{18}\alpha _{\eta \eta }\alpha _{\pi \eta }{\widetilde{M}_\pi ^2}\left( \widetilde{M}_\pi ^2-4\widetilde{M}_\eta ^2\right) \frac{%
\overline{J}_{\eta \eta }(s)}{F_\eta ^4}\right.  \nonumber \\
&&\left. +\frac 18[s-\frac 23M_\pi ^2-\frac 23M_K^2+\frac 23((\widetilde{M}%
_K-\widetilde{M}_\pi )^2+2\alpha _{\pi K}\widetilde{M}_K\widetilde{M}%
_\pi ]\right.  \nonumber \\
&&\left. \times [3s-2M_K^2-2M_\eta ^2+\alpha _{\eta
K}(2\widetilde{M}_\eta ^2-\frac
23\widetilde{M}_K^2)]\frac{\overline{J}_{KK}(s)}{F_K^4}\right\}
\nonumber \\
&&  \label{disp phi_T gen}
\end{eqnarray}
for (\ref{phiA}, \ref{psiA}) and analogously for (\ref{phiA1}, \ref{psiA1}).
The coefficients $\alpha _{\pi \eta }$ $\ldots $ parametrize the difference
between the standard and the generalized cases, within the standard $%
O(p^4)$ chiral expansion their values are either one or zero. The dependence 
of these constants on the LEC's are given in the Appendix \ref {appendix coefficients}.

\subsection{Observables of the $\pi \eta $ scattering within $G\chi PT$
to $O(p^4)$ - the reparametrization \label%
{subsection generalized reparametrization}}

As it can be easily seen from the above formulae (in fact it is a
consequence of the construction of $G\chi PT$), after identifying
the parameters of the Lagrangians,
\begin{eqnarray}
\frac{B_0\widehat{m}}{F_0^2}L_4^r(\mu ) &\rightarrow &\frac 18\widehat{m}%
\widetilde{\xi }  \nonumber \\
\frac{B_0\widehat{m}}{F_0^2}L_5^r(\mu ) &\rightarrow &\frac
18\widehat{m}\xi
^r  \nonumber \\
\frac{B_0^2\widehat{m}^2}{F_0^2}L_6^r(\mu ) &\rightarrow &\frac 1{16}%
\widehat{m}^2Z_0^S  \nonumber \\
\frac{B_0^2\widehat{m}^2}{F_0^2}L_7^r(\mu ) &\rightarrow &\frac 1{16}%
\widehat{m}^2Z_0^P  \nonumber \\
\frac{B_0^2\widehat{m}^2}{F_0^2}L_8^r(\mu ) &\rightarrow &\frac 1{16}%
\widehat{m}^2A_0  \label{identification}
\end{eqnarray}
and defining the remainders using the \emph{physical} masses
inside the chiral logarithms and the loop functions $J_{PQ}^r$,
the generalized {\em{bare}} chiral expansions contain all the terms of
the standard one. More precisely, the generalized $O(p^4)$ bare
expansions include extra $O(p^4)$ terms, which are counted as
$O(p^6)$ and $O(p^8)$ within the standard chiral power counting
scheme. As a consequence, after writing the generalized bare chiral
expansion of a generic ``good'' observable $g$ in the form
\begin{equation}
g=g^{(2),\,G\chi PT}+g^{(3),\,G\chi PT}+g^{(4),\,G\chi PT}
+g\delta_g^{G\chi PT}
\end{equation}
and then collecting the ``standard'' terms together, this expansion can
be formally rewritten as 
\begin{equation}
		g=g^{(2),\,std}+g^{(4),\,std}+g\delta_g,\quad
		g\delta_g=g\delta_g^{(G)}+g\delta_g^{G\chi PT},
\label{generalized_standard_bare}
\end{equation}
where the identification (\ref{identification}) is assumed. The
extra $O(p^4)$ terms mentioned above are now accumulated in
$g\delta_g^{(G)}$. In the case of the polynomial parameters $\alpha\dots\,\omega$ (\ref{alpha_bare}--\ref{omega_bare}), the two versions of the chiral expansion coincide for $\gamma$ and $\omega$\footnote{The reason is that they stem from the terms quadratic in the Mandelstam variables.} 
\begin{equation}
		\delta _\gamma =\delta _\gamma^{G\chi PT}\qquad \delta _\omega =\delta _\omega ^{G\chi PT},
\end{equation}		
while the ``standard'' remainders $\delta _\alpha $,  $\delta
_\beta $ can be split into an explicitly known part, which includes the
extra ``nonstandard'' terms, and the unknown remainders inherent to $G\chi PT$
\begin{eqnarray}
\delta _\alpha &=&\delta _\alpha {}^{loops}(\mu )+3\frac{\widehat{m}^2F_0^2}{%
F_\pi ^2M_\pi ^2}\delta _\alpha ^{CT}(\mu )+\delta _\alpha ^{G\chi PT}
\label{GChPT remainders alpha} \\
\beta \delta _\beta &=&\beta \delta _\beta {}^{loops}(\mu )+\widehat{m}%
^2F_0^2\delta _\beta ^{CT}(\mu )+\beta \delta _\beta ^{G\chi PT}.
\label{GChPT remainders beta}
\end{eqnarray}
Here the first terms correspond to the new loops and the second to
the new counterterm contributions. The explicit expressions for
them can be easily extracted from the formulae of the previous
subsection, the results are however rather lengthy and we postpone
them to Appendix \ref{appendix alpha}.

Let us note that both $\delta _{\alpha ,\beta }^{loop}(\mu )$ and
$\delta _{\alpha ,\beta }^{CT}(\mu )$ are generally
renormalization scale dependent.
However, due to the running of the $G\chi PT$ LEC's $A_0$, $Z_0^S$, $Z_0^P$, $%
\xi $ and $\widetilde{\xi }$ ,which, after identification (\ref
{identification}), is the same as in the standard case, the
``standard'' remainders $\delta _\alpha $ and $\delta _\beta $
given by (\ref{GChPT remainders alpha}, \ref{GChPT remainders beta}) are $\mu $-independent. Of
course, the ``true $G\chi PT$'' remainders $\delta _\alpha ^{G\chi
PT},\ldots ,\delta _\omega ^{G\chi PT}$ are scale independent by
construction. That means the sum of the loop and counterterm
contributions to the ``standard'' remainders is $\mu $-independent
too.

The usual way how to handle the reparametrization of the $G\chi PT$ bare expansions is
quite similar to the standard one. The difference is that as there are three additional $O(p^2)$ LEC's, not all of them can be reparametrized using inverted mass and decay constant expansions. The solution is to leave two of them free (\emph{e.g.} $r$ and $\zeta=Z_0^S/A_0$). Consequently, the expansion is performed according to the generalized power counting scheme and the terms of order higher than $O(p^4)$ are discarded.

We will, however, not use this approach, but rather exploit the relation (\ref{generalized_standard_bare}), \emph{i.e.} sew the standard and generalized bare expansions together. The reparametrization is then an extension of the resummed one (Appendix \ref{appendix Li}), where all the remainders of mass and decay constant bare expansions are split according to (\ref{generalized_standard_bare}). We can use all the resummed formulae as they are exact algebraic identities, valid independently on the version of $\chi PT$. The generalized contributions to the remainders can be found in Appendices \ref{appendix decay constants} and \ref{appendix generalized Li}. The outcome for the parameters $\alpha$ and $\beta$ is then obtained by simply inserting all the generalized results for the remainders (Appendices \ref{appendix decay constants}, \ref{appendix alpha}, \ref{appendix generalized Li}) into the expression for $\delta _\alpha ^{^{\prime }}$ and $\delta _\beta ^{^{\prime }}$ (\ref{dalpha_res}, \ref{dbeta_res}).

After this procedure, the generalized LEC's are present only in the formulae for the standard remainders. Also note that $\delta _\alpha {}^{loops}(\mu )$, $\delta
_\beta{}^{loops}(\mu )$ as well as the generalized loop contributions to mass and decay constant remainders depend explicitly on the $O(p^2)$ LEC's $B_0=XM_\pi ^2/2\widehat{m}$,
$F_0^2=ZF_\pi ^2$, $A_0$, $Z_0^S$ and $Z_0^P$.\footnote{
More precisely, the loops depend on the ``true $O(p^2)$ LEC's''$\overline{A}%
_0 $, $\overline{Z}_0^{S,P}$(cf. Appendix \ref{appendix
Lagrangian}), the difference is however of the higher order in the
generalized power counting.} So as the last step of the reparametrization, the
remaining dependence of the generic ``loop'' remainders $\delta
_\alpha {}^{loops}(\mu ),\ldots ,etc.$ on the $O(p^2)$ LEC's $F_0$, $A_0$,
$Z_0^S$ and $Z_0^P $ can be removed up to the order $O(p^4)\,$using the leading order expressions
\begin{eqnarray}
F_0^2=F_K^2=F_\eta^2&\rightarrow &F_\pi^2 \nonumber \\
\widehat{m}^2F_0^2Z_0^S &\rightarrow &\frac 14\frac{F_\pi ^2M_\pi ^2}{r+2}%
(1-X-\varepsilon (r))  \nonumber \\
\widehat{m}^2F_0^2Z_0^P &\rightarrow &-\frac 18F_\pi ^2M_\pi
^2\left(\varepsilon (r)-\frac{\Delta _{GMO}}{(r-1)^2}\right)  \nonumber \\
\widehat{m}^2F_0^2A_0 &\rightarrow &\frac 14F_\pi ^2M_\pi
^2\varepsilon (r). \label{L_i_leading}
\end{eqnarray}

As a summery, our handling of the generalized bare expansion can be viewed in two ways - either as a partial estimate of the standard remainders present in the resummed approach or as a special treatment within the generalized framework, where the $O(p^2)$ (and partly $O(p^3)$) LEC's are reparametrized algebraically at the leading order, while treated perturbatively at the $O(p^4)$ one. The numerical results including a simple estimate of the remaining NLO and NNLO LEC's are presented in Subsection 6.6, also  Appendix \ref{appendix decay constants} contains an illustrative example of applying this procedure on $F_\eta$.

\section{Numerical results}

In this section we shall present the numerical analysis of
the observables connected to the $\pi
\eta -$ scattering amplitude and the results which
illustrate the subtleties of the various versions of the chiral
expansions described above. In the numerical estimates we use $M_\pi =135%
\mathrm{MeV}$, $M_\eta =548\mathrm{MeV}$, $M_K=496\mathrm{MeV}$,
$\mu =M_\rho =770\mathrm{MeV}$, $F_\pi =92.4\mathrm{MeV}$ and
$F_K=113\mathrm{MeV} $. For the calculation within the standard
$\chi PT$, the $O(p^4)$ LEC's are taken from ref.
\cite{Bijnens:1994qh} and \cite{Amoros:2000mc}. In the alternative
reparametrization schemes, where only $L_1$, $L_2$ and $L_3$
remain among the  free parameters and the other $O(p^4)$ LEC's are
expressed in terms of physical masses, decay constants and the indirect remainders,
we again keep (though rather non-systematically) the values of
$L_1$, $L_2$ and $L_3$ from the same references. The sensitivity
on this LEC's might be then estimated by means of the variation
around these central values. In this chapter we insert the
physical masses into the functions $J^{r}_{PQ}(0)$ unless stated
otherwise.

\subsection{The standard chiral perturbation theory \label{standard numerics}}

\begin{table}[t] \hspace{4cm}
\begin{tabular}{|c|c|c|c|c|} \hline {$L_i$} &
${{\alpha}/{\alpha^{CA}}}$ & $10^3\beta/M_\eta^2 $ & $10^3\gamma $
& $10^4\omega $  \\ \hline \cite{Bijnens:1994qh}
& $\text{1.68}$ & 0$\text{.90}$ & $-$1.52 & 2.24  \\
\cite{Amoros:2000mc} & $\text{1.91}$ & $-\text{0.68}$ & $-$0.23 & $-$5.03 \\
$\Delta $ & $\text{2.48}$ & $\text{7.49}$ & 3.31 & 9.48  \\ \hline
\end{tabular}
\caption{Standard $O(p^4)$ values of the polynomial parameters
for the two sets of LEC's taken form references
\protect\cite{Bijnens:1994qh} and \protect\cite{Amoros:2000mc}. In
the last row, the sensitivity on the LEC's is (over)estimated by
adding the uncertainties associated with the LEC's
\protect\cite{Bijnens:1994qh} in quadrature (This is of course
only a rough estimate, because in fact not all the uncertainties
of  the $L_i$'s are independent).}\label{table_std1}
\end{table}

This subsection discusses the predictions of the
standard chiral expansion to the order $O(p^4)$, which are
summarized in Table \ref {table_std1} and \ref{table_std2}. Let us
start with the parameter $\alpha $ of the polynomial part of the
amplitude. The relevant formulae from the Subsection \ref
{standard} and the LEC's taken from \cite{Bijnens:1994qh}\footnote{%
This set of LEC's is used in numerical estimates
unless stated otherwise.} result in the following value
\begin{equation}
\alpha =\frac 13\left( 1+0.683+\delta _\alpha ^{st}\,\,\right)
F_\pi ^2M_\pi ^2  \label{alpha_std_num}
\end{equation}
In this expression, the first term corresponds to the current
algebra result
$\alpha ^{CA}=$ $F_\pi ^2M_\pi ^2/3$, while the second one represents the $%
O(p^4)$ correction. The third term is the standard remainder,
which might be out of control when $X,Z\ll 1$ and $r$ far from
$r_2$, even if the bare expansion of $\alpha $ were globally
convergent (let us remind that $\alpha $ is a ``good`` observable)
as we have discussed in Section \ref{motivation and notation}. Let
us also notice the unusually large next-to-leading correction,
which could also indirectly indicate the numerical importance of
the remainder in this scheme.

The actual numerical value of the NLO correction is very
sensitive to a shift in the $O(p^4)$ LEC's. The corresponding
variation $\Delta \alpha $ is numerically
\begin{equation}
\frac{\Delta \alpha }{\alpha ^{CA}}=(3.38\Delta L_1+0.56\Delta
L_3-3.50\Delta L_4-0.30\Delta L_5+3.62\Delta L_6-3.42\Delta
L_7+0.10\Delta L_8)\times 10^3.  \label{dalpha_Li}
\end{equation}
For example, using the $O(p^6)$ analysis based LEC's from \cite{Amoros:2000mc} instead of those from \cite{Bijnens:1994qh}, we get (cf. Table \ref
{table_std1})\footnote{%
Because the values of the LEC's $L_i$ based on the $O(p^6)$ fit
include implicitly parts of the $O(p^6)$ corrections, the large
variation can be interpreted as a signal of the importance of the
NNLO contributions to the parameter $\alpha $. The same is true
for other observables from the Table \ref{table_std1}.}
\begin{equation}
\frac{\Delta \alpha }{\alpha ^{CA}}=0.23.
\end{equation}
Note that the large coefficients in front of the $L_4$ and $L_6$ contributions indicate
sensitivity of this observable to the vacuum fluctuations of
$\overline{s}s$ pairs as mentioned in the Introduction.

\begin{table}[t] \hspace{3cm}
\begin{tabular}{|c|c|c|c|c|c|c|}
\hline {$L_i$} & $c_{00}/c_{00}^{CA}$ & $10^3c_{10}$ & $10^3c_{20}$ & $%
10^3c_{01}$ & $a_0/a_0^{CA}$ & $10^3a_1$ \\ \hline
\cite{Bijnens:1994qh} &  $%
\text{1.06}$ & 0.$\text{91}$ & $-\text{1.23}$ & $\text{8.27}$ &
$\text{1.96}$
& $\text{0.59}$ \\
\cite{Amoros:2000mc} &  $%
\text{1.51}$ & $-\text{0.67}$ & $\text{0.07}$ & $-\text{3.36}$ &
$\text{1.18}
$ & $-\text{0.60}$ \\
$\Delta $ & $\text{2.49}$ & $%
\text{7.49}$ & $\text{3.31}$ & $\text{15.16}$ & $\text{3.21}$ &
$\text{2.80} $ \\ \hline
\end{tabular}
\caption{Standard $O(p^4)$ values of the subthreshold and
threshold parameters as in Tab.\ref{table_std1}. The $c_{ij}$
parameters are given in their natural units described in the main
text. Analogously to Tab.\ref{table_std1}, $\Delta$ is the sensitivity on the LEC's (over)estimated by
adding the uncertainties associated with the LEC's
\protect\cite{Bijnens:1994qh} in quadrature}\label{table_std2}
\end{table}

Let us compare this case with the related ``dangerous``
observable, namely the subthreshold parameter $c_{00}$. From
(\ref{cij}) we get
\begin{equation}
c_{00}=\frac 13\left( 1+0.683-0.625+0.006\right) \frac{M_\pi ^2}{F_\pi ^2}%
=1.064c_{00}^{CA},
\end{equation}
where the individual terms are the leading order contribution $%
c_{00}^{CA}=M_\pi ^2/3F_\pi ^2$, the next-to-leading correction to
the parameter $\alpha $, the next-to-leading correction to
$F_\eta ^2$ induced by the expansion of the denominator and the
contribution stemming from the unitarity correction $\phi (s)$
respectively. The first two large corrections accidentally cancel
here, this, however, does not automatically imply similar
cancellation of the potentially large remainders (we have not
written them down explicitly here). Also, the strong sensitivity of
$\alpha $ to the variation of the LEC's propagates here giving
\begin{equation}
\frac{\Delta c_{00}}{c_{00}^{CA}}=\frac{\Delta \alpha }{\alpha ^{CA}}%
-0.28\Delta L_5\times 10^3
\end{equation}
and furthermore increases the uncertainty of the $O(p^4)$
correction. This strong sensitivity supports the possibility
that the standard remainders  for $c_{00}$ might be numerically
larger than the next-to-leading correction. Namely using the LEC's
from the $O(p^6)$ fit \cite{Amoros:2000mc}, which generates part
of the $O(p^6)$ corrections to the reparametrized expansion of
$c_{00}$, we get
\begin{equation}
\frac{\Delta c_{00}}{c_{00}^{CA}}=0.45.
\end{equation}

We can also check the sensitivity of the
next-to-leading order contributions to the way we rewrite them in
terms of the physical masses and decay constants (\emph{i.e.}
how we use the $O(p^2)$ relations generating here a difference
of the order $O(p^6)$). Provided we insert the alternative $O(p^2)$ expressions
for $r$ into the chiral expansions of $\alpha$ and $c_{00}$
\begin{eqnarray}
\widetilde{r}_2 &=&\frac 12\left( \frac{3M_\eta ^2}{M_\pi
^2}-1\right) =24.2
\\
r_2^{*} &=&2\frac{F_K^2M_K^2}{F_\pi ^2M_\pi ^2}-1=39.4,
\end{eqnarray}
instead of the standard $O(p^2)\,$value $r=r_2=2M_K^2/M_\pi
^2-1=25.9$, we get as a result
\begin{eqnarray}
		\widetilde{\alpha} &=& \frac 13(1+0.601)\,F_\pi ^2M_\pi ^2 \\
		\widetilde{c}_{00} &=& \frac 13(1+0.031)\frac{M_\pi ^2}{F_\pi ^2}
\end{eqnarray}
and
\begin{eqnarray}
		\alpha^{*} &=& \frac 13(1+1.297)\,F_\pi ^2M_\pi ^2 \\
		c_{00}^{*} &=& \frac 13(1+0.325)\frac{M_\pi ^2}{F_\pi ^2}.
\end{eqnarray}

The remaining parameters of the polynomial part start at $O(p^4)$
and we get them from (\ref{beta_std}), (\ref{gama_bare}) and
(\ref{omega_bare}). Their numerical values and the
related subthreshold parameters $c_{ij}$ in natural units
(chosen in such a way to make the comparison with the polynomial
parameters easy, \emph{i.e.} we take $c_{10}$ and $c_{01}$ in
units of $M_\eta ^2/F_\pi ^2$ and $c_{20}$ in units of $F_\pi ^{-4}$, cf. (%
\ref{cij})) are shown in Tables \ref{table_std1} and
\ref{table_std2} for the two sets of $O(p^4)$ LEC's.

All the considered parameters are strongly sensitive to the
variations of the LEC's. For instance the parameter $\beta $ varies
with the $L_i$'s as
\begin{equation}
\Delta \beta =(17.0\Delta L_1-2.8\Delta L_3+9.1\Delta L_4)M_\eta
^2. \label{dbeta_Li}
\end{equation}
For the LEC's \cite{Bijnens:1994qh} we get $\beta =0.90\times
10^{-3}M_\eta ^2 $. Using the set \cite{Amoros:2000mc} we get a
drastic change
\begin{equation}
\Delta \beta =-1.58\times 10^{-3}M_\eta ^2.
\end{equation}
Let us turn to the ``doubly dangerous'' observables
represented by the scattering lengths now. For the s-wave we obtain
from (\ref{a0a1}) and the LEC's
\cite{Bijnens:1994qh} 
\begin{eqnarray}\label{a0std}
a_0 &=&\frac 1{24\pi F_\pi ^2}\frac{M_\pi ^3}{M_\eta +M_\pi }%
(1+0.683+0.378-0.625+0.527) \nonumber\\
&=&\frac 1{24\pi F_\pi ^2}\frac{M_\pi ^3}{M_\eta +M_\pi }\left(
1+0.963\right) =11.0\times 10^{-3}.
\end{eqnarray}
Here the individual terms in the first line represent the
current algebra result, the correction stemming from the $O(p^4)$
contributions to the parameters $\alpha $ and $\omega $, the next-to-leading correction to
$F_\eta ^2$ induced by the expansion of the denominator and the
correction induced by the dispersive part of the amplitude $\phi
(s)$, in this order. This result confirms the expectations about a
bad convergence of the chiral expansion for the observables which
are connected to the threshold values of the amplitude - even if the polynomial NLO corrections were small, which are not, the dispersive part would be still as large as 50\% of the leading order term. 

The sensitivity to the $O(p^4)$ LEC's is illustrated in Table
\ref{table_std2}. The p-wave scattering length then starts at $O(p^4)$, 
we get the values in the last column of the table from (\ref{a0a1}).

When comparing our standard $\chi PT$  results for the scattering lengths (first row of Table \ref{table_std2})
\begin{equation}
		a_0=11.0\times 10^{-3}\qquad a_1=5.9\times10^{-4}
\end{equation}
with those of \cite{Bernard:1991xb}, quoted in the Introduction 
\begin{equation}\label{BKM}
		a_0^{BKM}=7.2\times 10^{-3}\qquad a_1^{BKM}=-5.2\times10^{-4},
\end{equation}
we can see a seemingly large discrepancy. The difference is produced by a different set of $O(p^4)$ LEC's, the alternative treatment of the $F_\eta$ in the denominator and by another form of
the unitarity corrections - the authors do not use a matching
with a dispersive representation. Taken these distinctions into
account, we get more consistent numbers (with our inputs for the masses and decay constants): 
\begin{equation}
		a_0=7.0\times 10^{-3}\qquad a_1=-5.0\times10^{-4}.
\end{equation}
As we can see, a slightly different treatment of the standard chiral expansion may lead to a significant shift in the results. This does not necessarily mean that the standard counting is not consistent, though. As follows from from Table \ref{table_std2}, the nominal uncertainty associated with $O(p^4)$ LEC error bars encompasses the difference
\begin{equation}
		\Delta a_0=18.0\times 10^{-3}\qquad \Delta a_1=28.0\times10^{-4}.
\end{equation}
What can be concluded is that the standard approach has a large theoretical uncertainty attached, which is hard to estimate. The sensitivity to the $L_i^r$ values also leads to a considerable difference when one uses the $O(p^6)$ fit (second row in Table \ref{table_std2}). As the two fits effectively differ only in a rearrangement of the expansion, both cannot have small higher order corrections at the same time.

\subsection{Resummation of vacuum fluctuations - basic properties\label{resummed_numerics}}

In the resummed case, the free parameters are $X$, $Z$, and $r$
together with the remaining LEC's $L_1$, $L_2$ and $L_3$ and the
direct and indirect remainders $\delta _\alpha $\dots$\delta _\omega$ and $\delta _{F_P}$, $\delta _{F_P M_P}$. Because $F_\eta $ is experimentally not
known with enough accuracy, we also have to fix how
to treat the observable $\Delta _{GMO}$ which was introduced to
eliminate the LEC $L_7$ using the bare expansion for $F_\eta^2M_\eta^2$. Let us remind that our definition of $\Delta _{GMO}$ follows the ref. \cite
{Descotes-Genon:2003cg}, where it is based on the good observables $F_P^2M_P^2$ instead of $M_P^2$ and differs from that originally defined in \cite{Gasser:1984gg}. One possibility is to treat $\Delta _{GMO}$ as an additional independent parameter. The another, similarly to the treatment of $F_\eta$ in the denominators of (\ref{disp_res}), (\ref{a0a1}) and (\ref{cij}) discussed earlier, is to use a (resummed) chiral
expansion of $F_\eta^2$ inserted to $\Delta _{GMO}$ for the numerical estimates, \emph{i.e.} to insert the following exact algebraic identity into (\ref{alpha_res})  
(cf. Appendix \ref{appendix decay constants} for details)
\begin{eqnarray}
F_\pi ^2M_\pi ^2\Delta _{GMO} &=&F_\pi ^2M_\pi
^2-4M_K^2F_K^2+M_\eta ^2(1-\delta _{F_\eta })^{-1}\bigg(
4F_K^2(1-\delta _{F_K})-F_\pi ^2(1-\delta
_{F_\pi }) \bigg. \nonumber\\
 && \left. -M_\pi ^2\left( \frac XZ\right) \left( J_{\pi \pi
}^r(0)-2(r+1)J_{KK}^r(0)+(2r+1)J_{\eta \eta }^r(0)\right) \right).
\end{eqnarray}
This generates the indirect remainder $\delta _{F_\eta }$ in a nonlinear way.

\begin{figure}[t] \hspace{4cm}
\scalebox{0.7}{\epsfig{figure=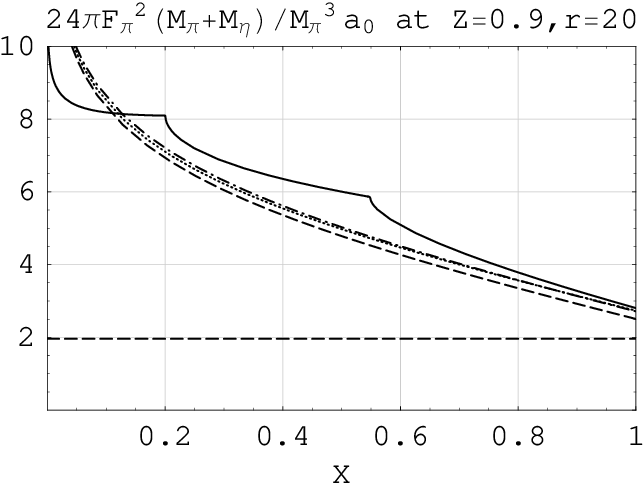}}
\caption{\emph{Comparison of the numerical impact of the various
forms of the dispersive part on the scattering length $a_0$. The
full line represents the strict chiral expansion, dotted, dashed
and dash-dotted lines the ``minimal'' modification,
(\ref{phiA}, \ref{psiA}) and (\ref{phiA1}, \ref{psiA1})
respectively. The horizontal line shows our standard NLO prediction.}} \label{disp}
\end{figure}

Before doing a more detailed analysis, let us first
illustrate the numerical sensitivity connected with the subtleties of the definition of the bare expansion. As we have discussed in Section 3, there is still some freedom
how to define the amplitudes entering the dispersive part of the $%
G_{\pi \eta }$ (cf. (\ref{phiA}, \ref{psiA}) and (\ref{phiA1},
\ref{psiA1}) and also how to treat the masses inside the chiral
logs.  Based on general considerations it was argued
\cite{Descotes-Genon:2007ta}, that in the latter case the different
prescriptions should not make much difference. Nevertheless, it might be
interesting to test this assumption numerically in our concrete
case and also to check what is the numerical influence of the varying amplitude definition.

In Figure \ref{disp} we plot the comparison of various definitions
of the dispersive part of the amplitude using the scattering
length $a_0$ as an example, {\em{i.e}} we illustrate its
sensitivity  on the various versions of the unitarity corrections.
The cusps on the full line, which uses the {\em{strict}} chiral expansion
with the unphysical choice of the $O(p^2)$ masses in all $J^r_{ij}$, originate in the conflict of the physical masses used for the on-shell outer legs and the unphysical location of the thresholds. This illustrates the fact that the original strict chiral expansion is unsuitable for realistic physical predictions and its redefinition into a bare one is necessary. The dotted line shows the ``minimal'' physical
modification of the strict expansion by means of insertion
physical masses into all $J^r_{ij}$. While the ``minimal''
version and the unitary choice (\ref{phiA1}, \ref{psiA1}) give numerically
almost the same result, the difference between the these two and
the third possibility (\ref{phiA}, \ref{psiA}) is up to $\sim 0.3\;a_0^{CA}$.

\begin{figure}[h] \hspace{0.75cm}
\scalebox{0.6}{\epsfig{figure=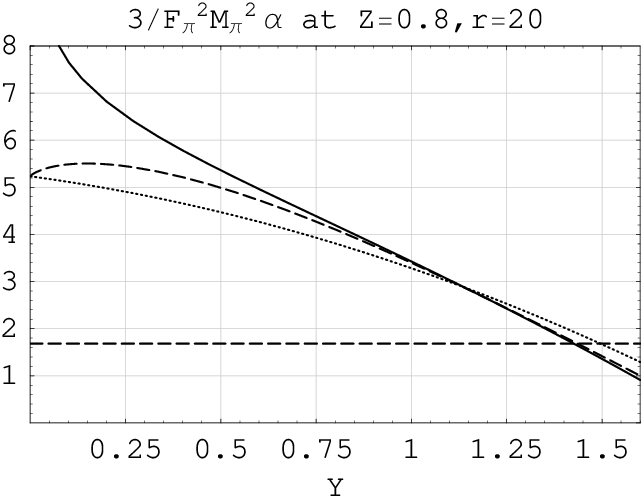}}
\hspace{1cm}
\scalebox{0.6}{\epsfig{figure=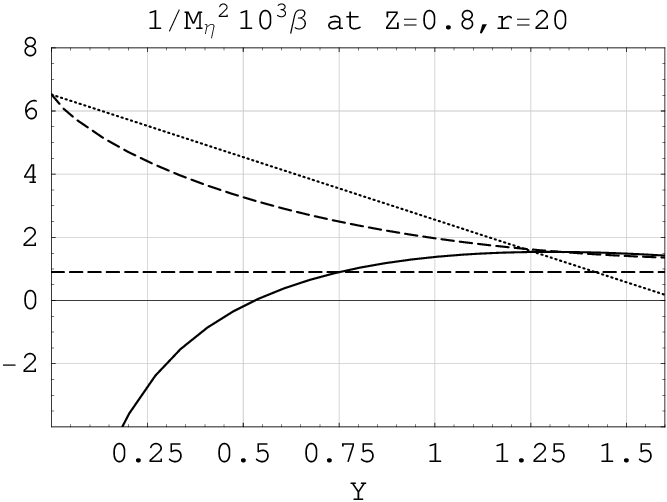}} \caption{\emph{In this
figure we illustrate the sensitivity of the ``good'' variables
$\alpha$ and $\beta$ to the treating of the chiral logs. The full
line corresponds to the $O(p^2)$ masses in all $J^r_{ij}(0)$'s, while
doted and dashed lines to the physical masses either in
all $J^r_{ij}(0)$'s or only in $J^r_{ij}(0)$'s originating from
the unitarity corrections. Horizontal lines are standard NLO predictions.}} \label{chilogs}
\end{figure}

Figure \ref{chilogs} shows the dependence of the polynomial parameters $%
\alpha $ and $\beta $ on $Y=X/Z$ for $Z=0.8$ and $r=20$, using the the various possible treatments of chiral logarithms in the bare expansion. The results demonstrate that the difference might be numerically important in some range of $Y$. For $\alpha $ the
various possibilities do not differ drastically in comparison with
the value of $\alpha$ itself, on the other hand the differences
become  comparable with $\alpha^{CA}$ at $Y\sim 0.5$. As we have discussed in Section 2, in the region of small $Y$ the case with $O(p^2)$ masses in the tadpoles only should not differ
drastically from the case when all the masses are physical.
However, the convergence to the common value at $Y=0$ is rather
slow and in the intermediate region of $Y$ the difference of this
two cases for  $\alpha$ is $\sim 0.5\;\alpha^{CA}$ in a relatively
wide interval. Keeping $O(p^2)$ masses also in the unitarity
corrections produce instabilities for $Y\rightarrow 0$ as
expected. The parameter $\beta $ (which starts at $O(p^4)$) is even much more
sensitive.

In the following numerical analysis we take a pragmatic point of
view and fix the bare expansion in such a way that the
comparison of the resummed and standard reparametrizations remains as
simple as possible, \emph{i.e. }we  insert physical masses into
$J_{ij}^r(0)$ and define the amplitude according to
(\ref{S2dangerous}, \ref{S4dangerous}) and (\ref{phiA1}, \ref{psiA1}).

\subsection{Numerical comparison of the resummed and standard
reparametrization\label{resummed vs standard}}

Within the standard $\chi PT$ we have an $O(p^4)$ prediction for $X$, $Z$, $%
r $ and $\Delta _{GMO}$ based on the standard formal $O(p^4)$
chiral expansion (\ref{XstdZstdYstdrstd}) and
(\ref{Delta_GMO_std}), cf. Appendix \ref{appendix_standard}. Using
the LEC's from \cite{Bijnens:1994qh} and \cite{Amoros:2000mc}, we get numerically the following central values, which should confirm the self-consistency of the standard chiral
expansion scheme
\begin{equation*}
\begin{tabular}{|c|c|c|c|c|c|}
\hline {$L_i$ set} & $X^{std}$ & $Z^{std}$ & $r^{std}$ &
$r^{*std}$ & $\Delta _{GMO}^{std}$ \\ \hline
\cite{Bijnens:1994qh} & 0.902 & 0.865 & 25.2 & 26.7 & 6.41 \\
\cite{Amoros:2000mc} & 0.726 & 0.734 & 25.9 & 31.7 & 3.31 \\
\hline
\end{tabular}
\label{standard_Table}
\end{equation*}
As we can see, while expectations are fulfilled in the first case, there is a considerable shift when using the $O(p^6)$ fitted constants. These numbers, moreover, should be
taken with some caution, because they originate in the expansions
of the ``dangerous`` observables and can be therefore plagued with
large $O(p^6)$ remainders as well as with strong
sensitivity to the $O(p^4)$ LEC's\footnote{%
As it was analyzed in detail  in \cite{Descotes-Genon:2003cg}, the
actual values of $X^{std}$ and $Z^{std}$ are strongly sensitive to
the values of
the LEC's $L_6$ and $L_4$ connected with the vacuum fluctuation of the $%
\overline{s}s$ pairs, the same is true for the sensitivity of $r^{std}$ and $%
\Delta _{GMO}^{std}$ to $L_8$ and $L_7$. This causes large error
bars to be attached to these values. Nevertheless, in the
following we take these central values as a reference point for an
illustrative numerical comparison of the two versions of the
chiral expansion.}. In the above table $r^{std}$ stems from the
chiral expansion of $r_2$ while $r^{*std}$ uses expansion of $r_2^{*}$.
\footnote{%
Though the difference between the values of $r^{std}$ and
$r^{*std} $ is within the standardly expected accuracy of the
$O(p^4)$ approximation, note, however, that for $r^{*std}$ the
$O(p^4)$ correction is much larger than in the first alternative
($r_2=25.9$ while $r_2^{*}=39.4$).}

\begin{table}[h] \hspace{1.25cm}
\begin{tabular}{|c|c|c|c|c|c|c|c|c|c|} \hline
$X$ & $Z$ & $r$ & $\Delta _{GMO}$ & $\alpha /\alpha ^{CA}$ & $10^3\beta $ & $%
c_{00}/c_{00}^{CA}$ & $10^3c_{10}$ & $a_0/a_0^{CA}$ & $10^3a_1$ \\
\hline $X^{std}$ & $Z^{std}$ & $r^{std}$ & $\Delta _{GMO}^{std}$ &
1.88 & 0.69 &
1.11 & 0.41 & 1.64 & 0.30 \\
$X^{std}$ & $Z^{std}$ & $r^{*std}$ & $\Delta _{GMO}^{std}$ & 1.61 & 0.55 & $%
$0.95 & 0.33 & 1.47 & 0.26 \\
$X^{std}$ & $Z^{std}$ & $r_2$ & $\Delta _{GMO}^{std}$ & 1.74 &
0.62 &
1.03 & 0.37 & 1.55 & 0.28 \\
$Z^{std}$ & $Z^{std}$ & $r^{*std}$ & $\Delta _{GMO}^{std}$ & 1.76
& 0.75 &
1.04 & 0.45 & 1.57 & 0.31 \\
$Z^{std}$ & $Z^{std}$ & $r^{std}$ & $\Delta _{GMO}^{std}$ & 2.02 &
0.89 &
1.20 & 0.53 & 1.72 & 0.35 \\
$X^{std}$ & $Z^{std}$ & $r^{std}$ & $\Delta _{GMO}$ & 2.07 & 0.69
& 1.22 & 0.41 & 1.75 & 0.30 \\
$X^{std}$ & $Z^{std}$ & $r^{*std}$ & $\Delta _{GMO}$ & 1.78 & 0.55
& 1.05 & 0.33 & 1.57 & 0.26 \\ \hline
\end{tabular}
\caption{\emph{The values of the the polynomial parameters $\alpha$ and $%
\beta$ and the related subthreshold and threshold parameters near
the standard reference point. For $\Delta_{GMO}$ we take either
the standard value ($\Delta _{GMO}^{std}$) or the resummed
prediction described ($\Delta _{GMO}$) in the main text.}}
\label{table1}
\end{table}

Let us now illustrate the relationship of the resummed and
standard approach using the observables from Subsection
\ref{standard numerics}.

\begin{figure}[!t] \mbox{} \hspace{1cm}
\scalebox{0.56}{\epsfig{figure=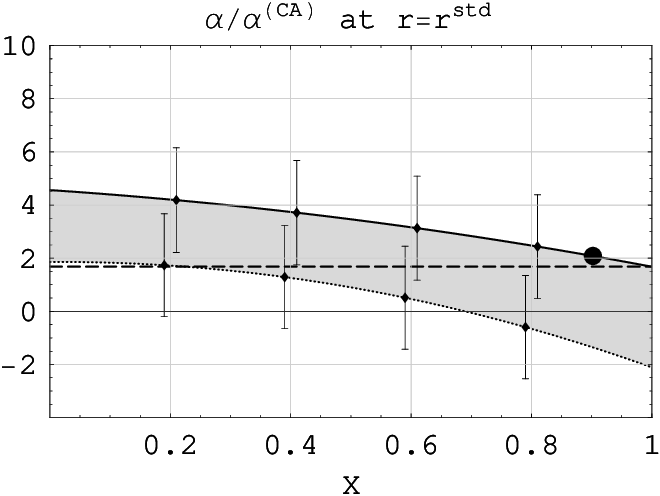}} \hspace{1cm}
\scalebox{0.56}{\epsfig{figure=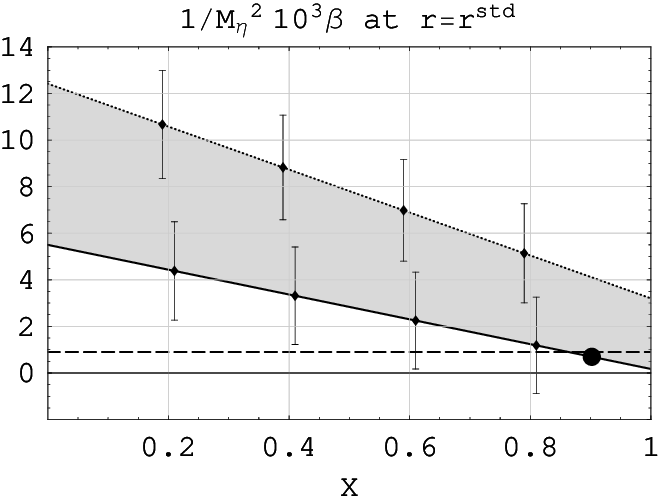}}\\ \mbox{} \hspace{1cm}
\scalebox{0.56}{\epsfig{figure=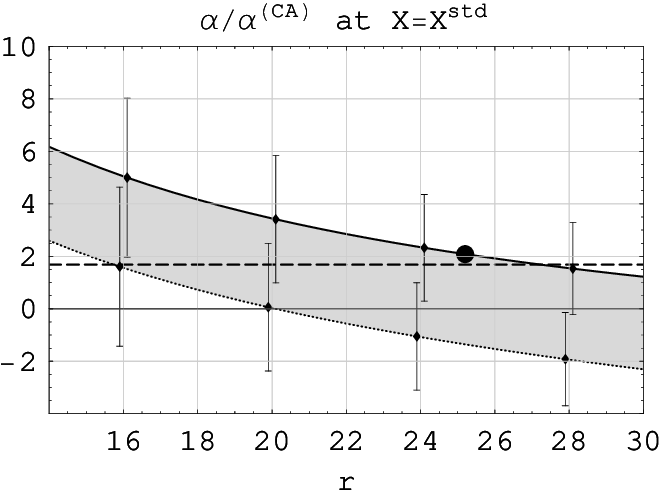}}  \hspace{1cm}
\scalebox{0.56}{\epsfig{figure=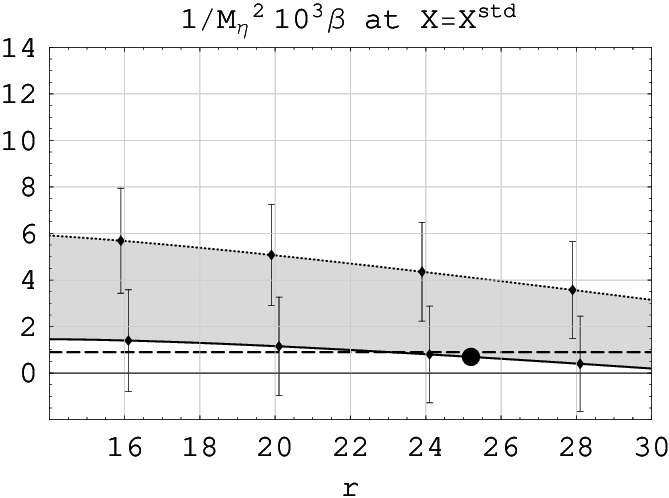}}\\ \mbox{} \hspace{1cm}
\scalebox{0.56}{\epsfig{figure=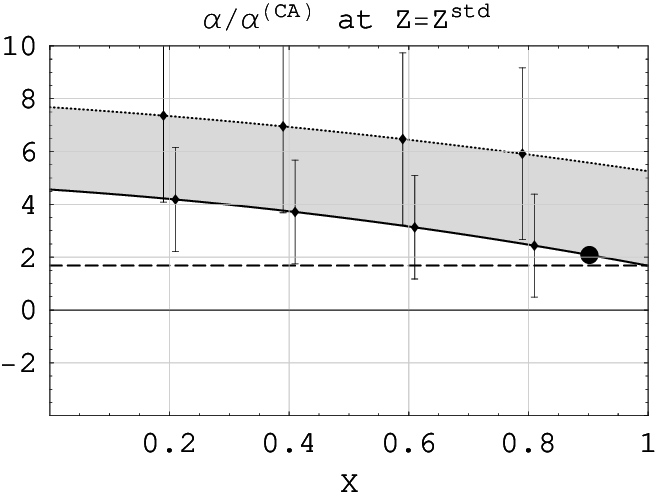}} \hspace{1cm}
\scalebox{0.56}{\epsfig{figure=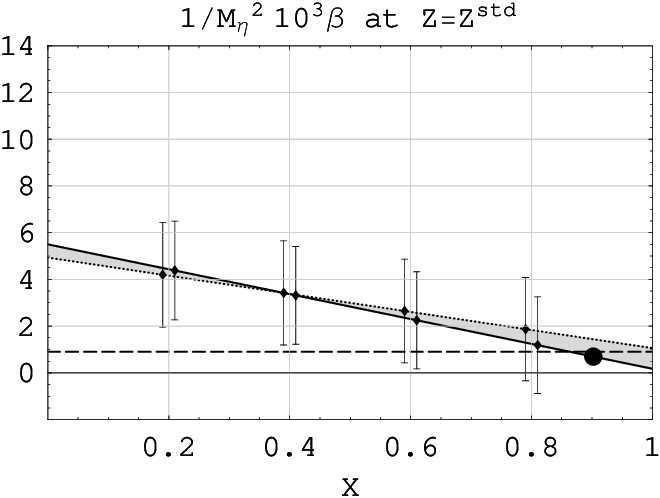}}
\caption{\emph{The dependence of the
parameters $\alpha$ and $\beta$ on $r$, $X$ and $Z$ is plotted, one of the parameters being fixed at its standard reference value in each figure. The dashed horizontal line shows the standard values from the
first row of the Table \ref{table_std1}, the full circle depicts the corresponding resummed value at the
standard reference point $[r^{std}, X^{std},Z^{std}]$. The error bars represent the $10\%$
uncertainties from the direct and indirect remainders added in
quadrature. In the first row, $r$ is fixed at $r^{std}$, the filled are highlight the dependence on $Z$ between $Z=Z^{std}$ (solid line) and $Z=0.5$ (dotted one). Similarly, in the second row $X=X^{std}$, the filled area shows the dependence on $Z$ again. $Z$ is fixed at $Z^{std}$ in the last row, the solid line shows the case with $r=r^{std}$ while the dotted the one with a lower value $r=15$.}} \label{figure1}
\end{figure}

\begin{figure}[h] \mbox{} \hspace{1cm}
\scalebox{0.56}{\epsfig{figure=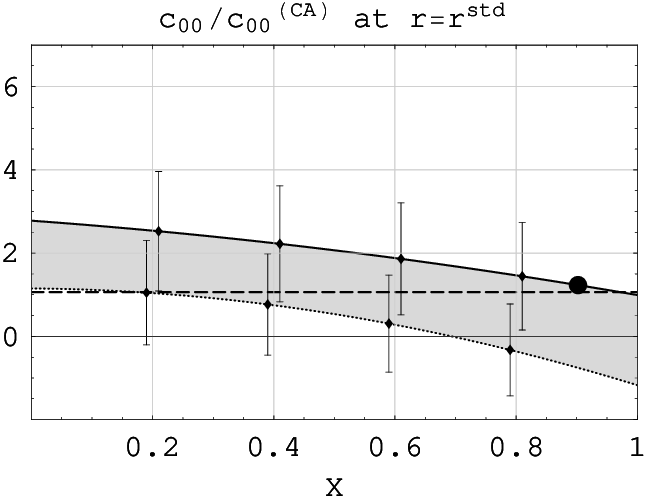}} \hspace{1cm}
\scalebox{0.56}{\epsfig{figure=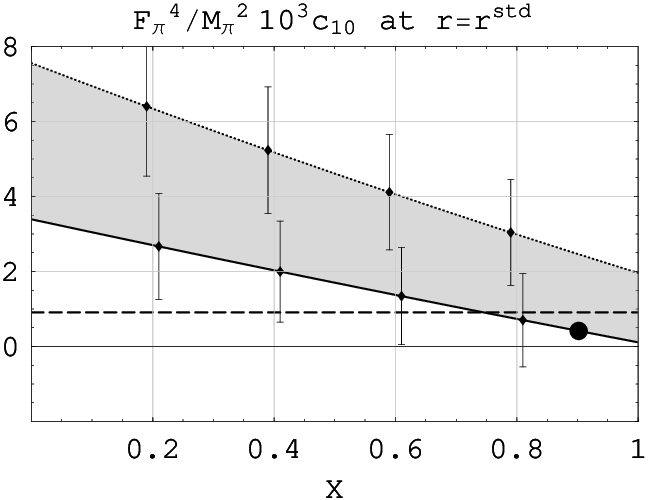}}\\ \mbox{} \hspace{1cm}
\scalebox{0.56}{\epsfig{figure=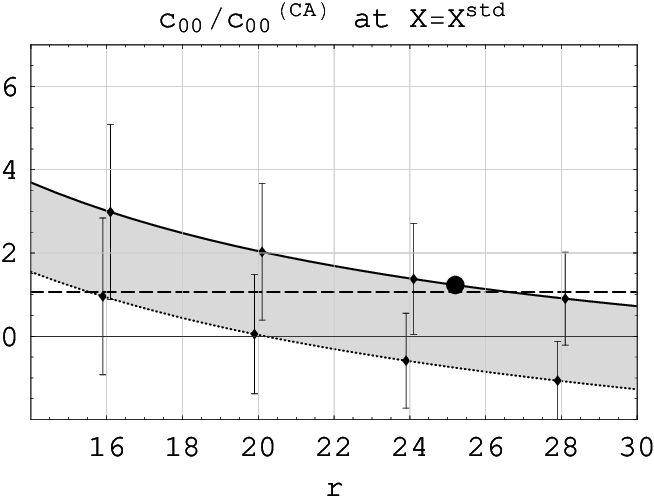}} \hspace{1cm}
\scalebox{0.56}{\epsfig{figure=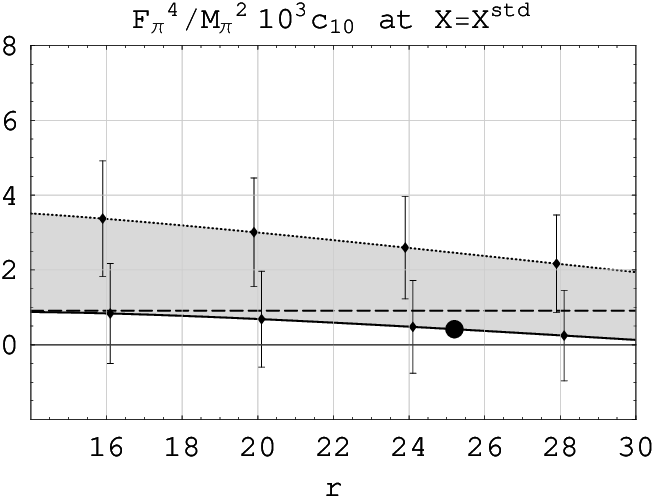}}\\ \mbox{} \hspace{1cm} 
\scalebox{0.56}{\epsfig{figure=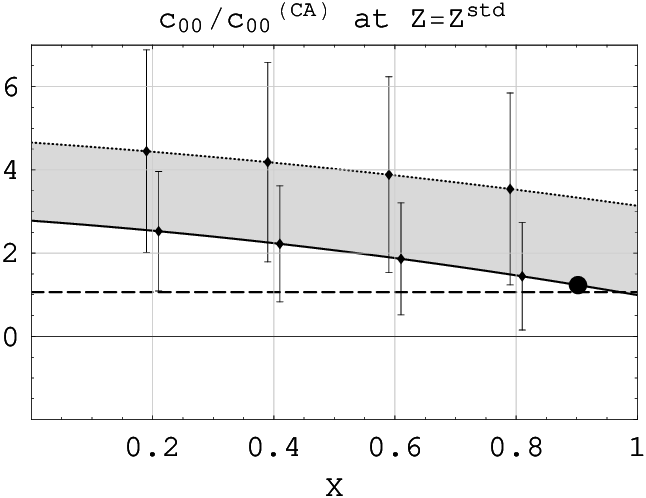}} \hspace{1cm}
\scalebox{0.56}{\epsfig{figure=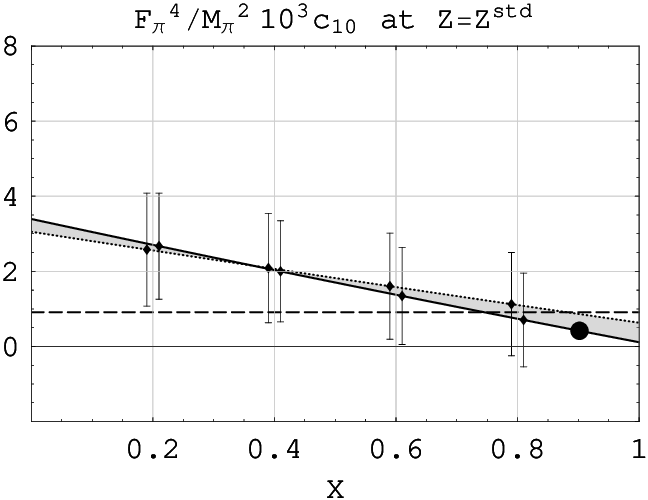}}
\caption{\emph{Dependence of the subthreshold parameters $c_{00}$ and $%
c_{10} $ related to the polynomial parameters $\alpha$ and
$\beta$. The figures are in one-to one correspondence to those in
Figure \ref{figure1}.}} \label{figure2}
\end{figure}

\begin{figure}[h] \mbox{} \hspace{1cm}
\scalebox{0.56}{\epsfig{figure=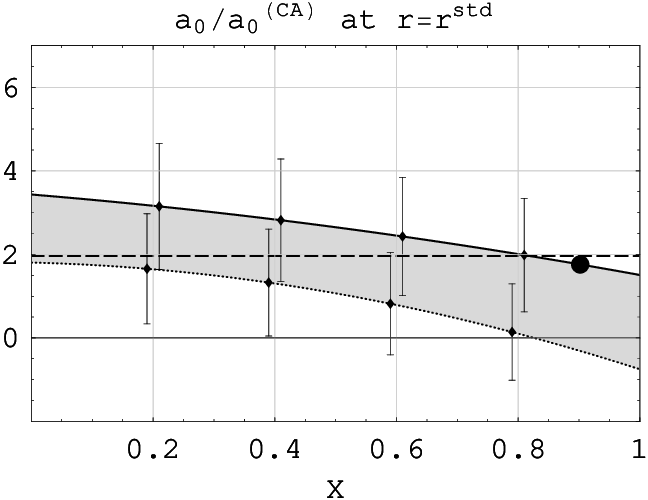}} \hspace{1cm}
\scalebox{0.56}{\epsfig{figure=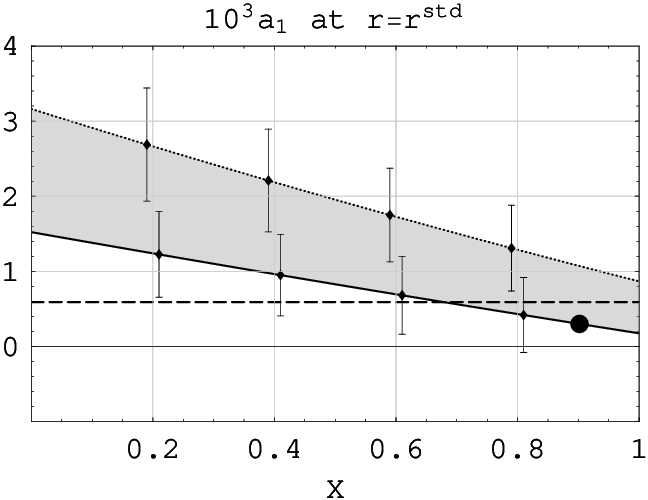}}\\ \mbox{} \hspace{1cm} 
\scalebox{0.56}{\epsfig{figure=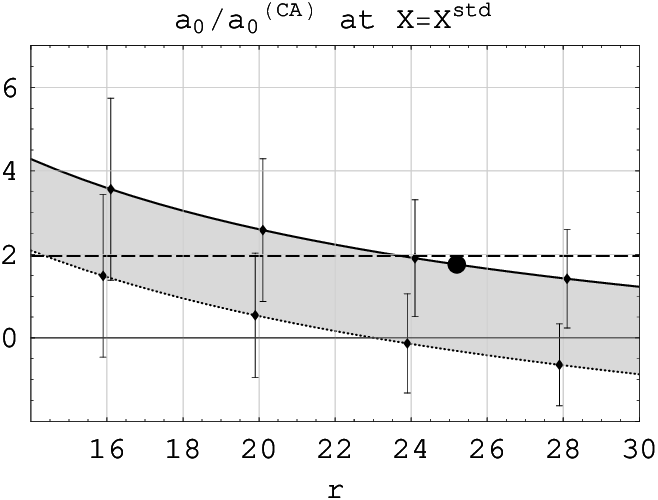}} \hspace{1cm}
\scalebox{0.56}{\epsfig{figure=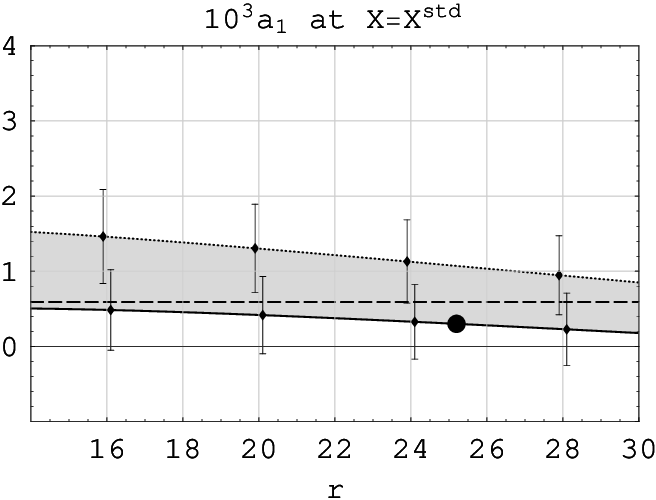}}\\ \mbox{} \hspace{1cm} 
\scalebox{0.56}{\epsfig{figure=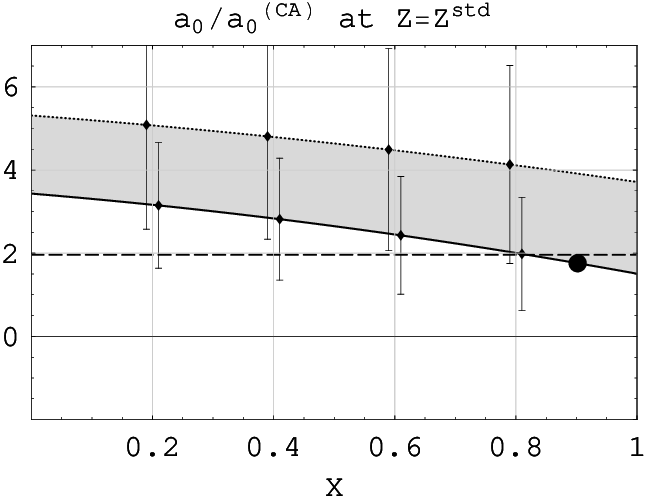}} \hspace{1cm}
\scalebox{0.56}{\epsfig{figure=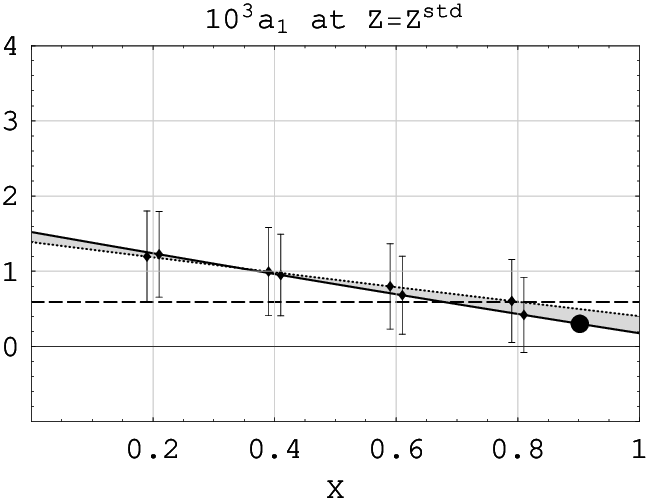}}
\caption{\emph{Dependence of the scattering lengths $a_{0}$ and
$a_{1} $ related to the polynomial parameters $\alpha$ and
$\beta$. The figures are in one-to one correspondence to those in
Figure \ref{figure1}.}} \label{figure3}
\end{figure}

For the ``good'' observables $\alpha$ and $\beta$ we can expect
that the numerical values of $X^{std}$, $Z^{std}$, $r^{std}$ and
$\Delta _{GMO}^{std}$, with $L_1$, $L_2$ and $L_3$ taken from
\cite{Bijnens:1994qh} for definiteness, should produce numbers consistent with
the first row of Table \ref{table_std1} when inserted into
(\ref{alpha_res}, \ref{beta_res}). The results for the various
possibilities how to approach the standard predictions for $\alpha
$ and $\beta $ (which is independent on $\Delta _{GMO}$) within
the resummed version of $\chi PT$ are summarized\footnote{In this
and the following tables in this subsection we ignore the
uncertainty stemming from the remainders and $L_i$, $i=1,2,3$ and
give only the central values (assuming the central values of the
remainders to be zero).} in the Table \ref{table1}. The last row
corresponds to the resummed treating of $\Delta _{GMO}$
explained above. The dependence of the central values of $\alpha $ and $%
\beta $ on the parameters $r$, $X$ and $Z$ in the broader vicinity
of their standard values is illustrated in  Fig.\ref{figure1}.
These results can be interpreted as a consistency of both variants
of reparametrization for
good observables near the standard reference point $X^{std}$, $Z^{std}$ and $%
r^{std}$, where the predictions of the resummed version almost
coincide
with the standard results\footnote{%
As a rule, the point $X^{std}$, $Z^{std}$ and $r^{std}$ cannot
give the best coincidence with the standard values in all cases, the reason can
be understood
\emph{e.g.} by having a closer look on the resummed reparametrization of $%
\beta $ (cf. (\ref{beta_res})). In order to reproduce the
dependence of $\beta $ on $L_4$ satisfactorily, we need
$Z=Z^{std}$ and $r=r^{std}$, on
the other hand to reproduce the chiral logs we need rather $X/Z=1$ and $%
r=r_2 $. This can explain why $\beta $ approaches the standard
value best for $X=Z=Z^{std}$.}. This coincidence together with the
working hypothesis about the controllable remainders of good
observables within the resummed reparametrization scheme confirms
again a self-consistency of the standard expansion based on the
assumption $X\sim 1$, $Z\sim 1$ and $r\sim r_2$. Away from the
standard reference point, however, the standard reparametrization
might be dangerous in the sense that the difference between the
standard and
the resummed prediction diverges rapidly and the importance of the standard $%
O(p^6)$ remainders might therefore increase considerably.

\begin{table}[h] \hspace{5cm}
\begin{tabular}{|c|c|c|c|c|}
\hline $X$ & $Z$ & $r$ & $10^4c_{20}$ & $10^3c_{01}$ \\ \hline
$X^{std}$ & $Z^{std}$ & $r^{std}$ & $-$7.10 & 4.86 \\
$X^{std}$ & $Z^{std}$ & $r^{*std}$ & $-$6.97 & 4.79 \\
$X^{std}$ & $Z^{std}$ & $r_2$ & $-$7.04 & 4.83 \\ \hline
\end{tabular}
\caption{\emph{The values of the subthreshold parameters $c_{20}$ and $%
c_{01} $ related to the polynomial parameters $\gamma$ and
$\omega$ at the standard reference point.}} \label{table2}
\end{table}

For the ``dangerous`` observables like $c_{ij}$ we cannot \emph{a
priori}
expect coincidence of both expansion even near the standard values of $X$, $%
Z $, and $r$ due to the different treatment of the denominators,
which contain large $O(p^4)$ corrections and are not expanded in
the resummed case. Comparison of both approaches is illustrated
in  Table \ref{table1} (with the same treatments of $\Delta
_{GMO}$ as above), Table \ref{table2} and in  Fig. \ref{figure2}.
For the dispersive part we use the prescription (\ref{phiA1},
\ref{psiA1}), which differs from the corresponding standard
contributions of the unitarity corrections to $c_{ij}$ for $X=Z=1$
and $r=r_2$ by a factor $F_\pi ^2/F_\eta ^2\approx 0.6$. This
is reflected by the values of those $c_{ij}$ that start at
$O(p^4)$ (cf. Table \ref{table1}, \ref{table2}). Namely in this
case the contribution of the polynomial part is reduced near the
reference point roughly by the same factor with respect to the
standard value (which includes only the first term of the
expansion of the denominator). On the other hand, $c_{00}$ is
compatible with the standard value, because of the large $O(p^2)$
contribution, tiny dispersive contribution and the fact that
within the standard reparametrization of the bare expansion also
the second term from the expansion of the denominator is taken
into account.

Let us now proceed to the ``doubly dangerous`` observables $a_0$
and $a_1$. These are related to the values of the amplitude at the
threshold and receive therefore large contribution from the
dispersive part of the
amplitude. While $a_0$ is reproduced well at the standard reference point, $%
a_1$ (which starts at the NLO)  is off the standard value roughly
by a factor $0.6$ from the same reasons as for the $c_{ij}$
parameters. The dependence of this observables on $X$, $Z$ and $r$
is depicted in Fig. \ref{figure3}.

\subsection{The role of the remainders}

Up to now we have not discussed the uncertainties of the
observables calculated within the resummed scheme. They are
connected with the direct and indirect remainders as well as with
the LEC's $L_i$, $i=1,2,3$. As the first illustration, we have
added the error bars stemming from the remainders to the central
values of the various observables depicted in the figures
\ref{figure1}-\ref{figure3}. These illustrate the rough
estimate of the remainders  $\delta \sim (30\%)^2\sim 0.1$ as
suggested in \cite {Descotes-Genon:2003cg} and adding the
uncertainties in quadrature.

In more detail, at the standard reference point $X=X^{std}$, $%
Z=Z^{std}$ and $r=r^{std}$,  using (\ref{dalpha_res}, \ref{dbeta_res}) and
(\ref{dalpha_Li}, \ref {dbeta_Li}), we numerically get   for the
corresponding variations (to the first order in the remainders)
\begin{eqnarray}
\frac{\Delta \alpha }{\alpha ^{CA}} &=&(\delta _\alpha +6.92\delta
_{F_\eta M_\eta }-12.71\delta _{F_KM_K}-0.76\delta _{F_\pi M_\pi
}+9.37\delta
_{F_K}+5.22\delta _{F_\pi }-6.92\delta _{F_\eta }) \nonumber\\
&&+(3.38\Delta L_1+0.56\Delta L_3)\times 10^3 \\
\Delta \beta &=&[(0.69\delta _\beta +2.34\delta _{F_K}-20.52\delta
_{F_\pi })+(17.0\Delta L_1-2.8\Delta L_3)\times 10^3]\times
10^{-3}M_\eta ^2.
\end{eqnarray}
This reveals strong sensitivity on both $\delta $s and LEC's.
Assuming again the typical size of the remainders to be $\delta
\sim 0.1$ and adding all the uncertainties in quadrature (for
$\Delta L_i$ we take the error bars form \cite{Bijnens:1994qh}) we
obtain rough (over)estimates
\begin{eqnarray}
\left| \frac{\overline{\Delta \alpha }}{\alpha ^{CA}}\right| &=&\sqrt{%
1.93^2+1.19^2}=2.27 \\
\left| \overline{\Delta \beta }\right|
&=&\sqrt{2.06^2+5.96^2}\times 10^{-3}M_\eta ^2=6.31\times
10^{-3}M_\eta ^2,
\end{eqnarray}
where the first number under the square root represents the
contribution
of the remainders while the second accumulates the uncertainty from $L_{1,3}$%
. Though these numbers are a little bit more optimistic than those
in the last row of Table \ref{table_std1} (note that the latter
originated purely from the uncertainties of $\Delta L_i$ and did
not include any estimates of the higher order corrections to
$\alpha $ and $\beta $), it is clear that,
without more restrictive information on the remainders (and $L_i$, $i=1,2,3$)%
\footnote{%
As already discussed, the explicit dependence on these constants could be
eliminated by means of reparametrization similar to those for
$L_i$, $i=4,\ldots ,6$ using further experimental input
\emph{e.g.} form $K_{e4}$ decay. The price to pay is to introduce
additional remainders connected with observables used for such a
reparametrization.}, the predictive power of $\chi PT$ is reduced
considerably in the case of $\pi \eta $ scattering even for
``good'' observables. In other words, small remainders are not a guarantee of an equivalently small final uncertainty. In what follows, we therefore try to gain
some additional information outside the (resummed) $\chi PT$
expansion to get further estimates of the size of the remainders.

The sources of the remainders are twofold: on one hand there are
the unknown terms of the pure derivative expansion, on the other
hand the contributions coming from the expansion in the quark
masses. We try to get estimates for both of them from different
sources, namely using the resonance estimate for the first type as
well as independent information from generalized $\chi PT$ for the
second.

\subsection{Resonance estimate of the direct remainders}

In order to partially estimate the derivative part of the higher
order corrections to the chiral expansion, we use the assumption,
that the process under consideration is saturated by the exchange
of the lowest laying resonances, the interactions of which can be
described by the Lagrangian of
the resonance chiral theory ($R\chi T$). The leading order Lagrangian of $%
R\chi T$ was originally formulated in the seminal paper
\cite{Ecker:1988te} and applied to $\pi \eta $ scattering in
\cite{Bernard:1991xb}. To this process, only scalar resonances as
well as $\eta _8-\eta _0$ mixing contribute. Our result for the amplitude agree with \cite{Bernard:1991xb} (cf. Appendix
\ref{Appendix resonances}), which we can rewrite in terms of the resonance contribution
$G_{\pi \eta }^R$ to $G_{\pi \eta }$ in the form
\begin{equation}
G_{\pi \eta }^R(s,t;u)=\alpha _R^{(4)}+\beta _R^{(4)}t+\gamma
_R^{(4)}t^2+\omega _R^{(4)}(s-u)^2+\Delta G_{\pi \eta }^R(s,t;u).
\label{remaider_subtracted}
\end{equation}
The polynomial part with the coefficients (in what follows, $M_S$ and $%
M_{S_1}$ are the octet and singlet scalar mass respectively, $c_d$, $c_m$, $%
\widetilde{c}_d$, $\widetilde{c}_m$and $\widetilde{d}_m$ are the
couplings defined in \cite{Ecker:1988te})

\begin{eqnarray}
\alpha _R^{(4)} &=&8M_\pi ^2M_\eta ^2\left( -\frac{c_d^2}{3M_S^2}+\frac{2%
\widetilde{c}_d^2}{M_{S_1}^2}\right) +16M_\pi
^2\stackrel{o}{M}_\eta
^2\left( \frac{c_dc_m}{3M_S^2}-\frac{\widetilde{c}_d\widetilde{c}_m}{%
M_{S_1}^2}\right) \nonumber\\
&&+8M_\eta ^2\stackrel{o}{M}_\pi ^2\left( \frac{c_dc_m}{3M_S^2}-\frac{2%
\widetilde{c}_d\widetilde{c}_m}{M_{S_1}^2}\right) -16\frac{\widetilde{d}_m^2%
}{M_{\eta _1}^2}\stackrel{o}{M}_\pi ^2\left( \stackrel{o}{M}_\pi ^2-%
\stackrel{o}{M}_\eta ^2\right) \nonumber\\
&&+20\stackrel{o}{M}_\pi ^2\stackrel{o}{M}_\eta ^2\left( \frac{\widetilde{c}%
_m^2}{M_{S_1}^2}-\frac{c_m^2}{3M_S^2}\right) -M_\pi ^2\stackrel{o}{M}_\pi ^2%
\frac{8c_dc_m}{3M_S^2}+4\stackrel{o}{M}_\pi ^4\left( \frac{7c_m^2}{M_S^2}+%
\frac{\widetilde{c}_m^2}{M_{S_1}^2}\right) \\
\beta _R^{(4)}
&=&-\frac{8\widetilde{c}_d\widetilde{c}_d}{M_{S_1}^2}\Sigma
_{\eta \pi }+\frac 43\frac{c_d^2}{M_S^2}\Sigma _{\eta \pi }+8\left( \frac{%
\widetilde{c}_d\widetilde{c}_m}{M_{S_1}^2}-\frac{c_dc_m}{3M_S^2}\right)
\left( \stackrel{o}{M}_\pi ^2+\stackrel{o}{M}_\eta ^2\right) \\
\gamma _R^{(4)}
&=&\frac{4\widetilde{c}_d^2}{M_{S_1}^2}-\frac{c_d^2}{3M_S^2}
\\
\omega _R^{(4)} &=&\frac{c_d^2}{3M_S^2}
\end{eqnarray}
gathers the complete $O(p^4)$ resonance contribution (here we
can recognize the resonance saturation of the LEC's in
(\ref{alpha_bare})-(\ref{omega_bare})). This part of the amplitude
is already included in our resummed version of $\chi PT$, either
explicitly through the LEC's $L_1$\dots$L_3$ or implicitly using the
reparametrization in terms of the masses, decay constants and parameters $r$%
, $X$ and $Z$. On the other hand, $\Delta G_{\pi \eta ,R}(s,t;u)$
can be formally understood as an infinite sum of the higher order
corrections in the (purely) derivative expansion, summed up to
\begin{eqnarray}
\Delta G_{\pi \eta ,R}(s,t;u)
&=&\frac{4t}{M_{S_1}^2(M_{S_1}^2-t)}\left(
\widetilde{c}_d(t-2M_\pi ^2)+2\widetilde{c}_m\stackrel{o}{M}_\pi
^2\right) \left( \widetilde{c}_d(t-2M_\eta
^2)+2\widetilde{c}_m\stackrel{o}{M}_\eta
^2\right)  \nonumber \\
&&+\frac 23\frac s{M_S^2(M_S^2-s)}\left( c_d(s-M_\pi ^2-M_\eta ^2)+2c_m%
\stackrel{o}{M}_\pi ^2\right) ^2  \nonumber \\
&&+\frac 23\frac u{M_S^2(M_S^2-u)}\left( c_d(u-M_\pi ^2-M_\eta ^2)+2c_m%
\stackrel{o}{M}_\pi ^2\right) ^2  \nonumber \\
&&-\frac 23\frac t{M_S^2(M_S^2-t)}\left( c_d(t-2M_\pi ^2)+2c_m\stackrel{o}{M}%
_\pi ^2\right)  \nonumber \\
&&\times \left( c_d(t-2M_\eta ^2)+2c_m(2\stackrel{o}{M}_\eta ^2-\stackrel{o}{%
M}_\pi ^2)\right)  \nonumber \\
&&+16\frac{\widetilde{d}_m^2M_\eta ^2}{M_{\eta _1}^2(M_{\eta _1}^2-M_\eta ^2)%
}\stackrel{o}{M}_\pi ^2\left( \stackrel{o}{M}_\eta
^2-\stackrel{o}{M}_\pi ^2\right) .  \label{dG_res}
\end{eqnarray}
Of course, this does not exhaust all of the possible higher order
corrections (note \emph{e.g.} that the resonance Lagrangian we use
contains only the leading order interaction terms with one
resonance field and chiral building blocks of the order $O(p^2)$),
nevertheless we can use it at least as a rough estimate of the
effect of higher orders of the derivative expansion. This is in
some sense a procedure opposite to the usual resonance saturation;
instead of LEC's we ``saturate`` the remainders by means of sewing
together the resummed chiral expansion $G_{\pi \eta }^{\chi PT}$
(without remainders) with the resonance chiral theory writing the full $%
R\chi T$ amplitude as
\begin{equation}
G_{\pi \eta }^{R\chi T}(s,t;u)=G_{\pi \eta }^{\chi
PT}(s,t;u)+\Delta G_{\pi \eta }^R(s,t;u).
\end{equation}
and identifying $G_{\pi \eta }^{R\chi T}$ with the full $\chi PT$
amplitude, the remainder being $\Delta G_{\pi \eta }^R$. Under this
assumption, we can derive the following higher order contributions to the direct remainders from $\Delta G_{\pi \eta }^R$

\begin{eqnarray}
\Delta \alpha _R&=&\frac 13F_\pi ^2M_\pi ^2\delta _\alpha ^R =\frac{16}3c_m^2%
\stackrel{o}{M}_\pi ^4\frac{\Sigma _{\eta \pi
}}{M_S^2(M_S^2-\Sigma _{\eta \pi
})}+16\frac{\widetilde{d}_m^2M_\eta ^2}{M_{\eta _1}^2(M_{\eta
_1}^2-M_\eta ^2)}\stackrel{o}{M}_\pi ^2\left( \stackrel{o}{M}_\eta ^2-%
\stackrel{o}{M}_\pi ^2\right)  \nonumber \\
&&  \label{dalpha_Res} \\
\Delta \beta _R&=&\beta \delta _\beta ^R
=\frac{16}{M_{S_1}^4}\left( \widetilde{c}_dM_\pi
^2-\widetilde{c}_m\stackrel{o}{M}_\pi ^2\right) \left(
\widetilde{c}_dM_\eta ^2-\widetilde{c}_m\stackrel{o}{M}_\eta
^2\right)
\nonumber \\
&&+\frac 83\frac 1{M_S^4}\left( c_dM_\pi ^2-c_m\stackrel{o}{M}_\pi
^2\right) \left( c_m(2\stackrel{o}{M}_\eta ^2-\stackrel{o}{M}_\pi
^2)-c_dM_\eta
^2\right)  \nonumber \\
&&-\frac 83\frac{c_m^2\stackrel{o}{M}_\pi ^4}{M_S^2(M_S^2-\Sigma
_{\eta \pi
})}-\frac 83\frac{c_dc_m\stackrel{o}{M}_\pi ^2\Sigma _{\eta \pi }}{%
M_S^2(M_S^2-\Sigma _{\eta \pi })}-\frac
83\frac{c_m^2\stackrel{o}{M}_\pi ^4\Sigma _{\eta \pi
}}{M_S^2(M_S^2-\Sigma _{\eta \pi })^2}  \label{dbeta_Res}
\end{eqnarray}
and similarly for $\delta _\gamma ^R$ and $\delta _\omega ^R$ (see
Appendix \ref{Appendix resonances}). Note that the dependence on
$X$ and $Z$ is exclusively through the ratio $Y=X/Z$ here.

One can notice that there are two distinct features of this procedure as compared to the usual LEC saturation. First, there is no need to fix a saturation scale, which is the result of ``saturating'' the renormalization scale independent remainder instead of the scale dependent LEC's. And second, as the resonance contribution are resummed to all chiral orders, the resonance poles are explicitly present in our result, as can be seen in (\ref{dG_res}), (\ref{dalpha_Res}) and (\ref{dbeta_Res}) as well as the formulae for $\delta _\gamma ^R$ and $\delta _\omega ^R$ in the Appendix \ref{Appendix resonances}.

\begin{figure}[t] \hspace{0.75cm}
\scalebox{0.56}{\epsfig{figure=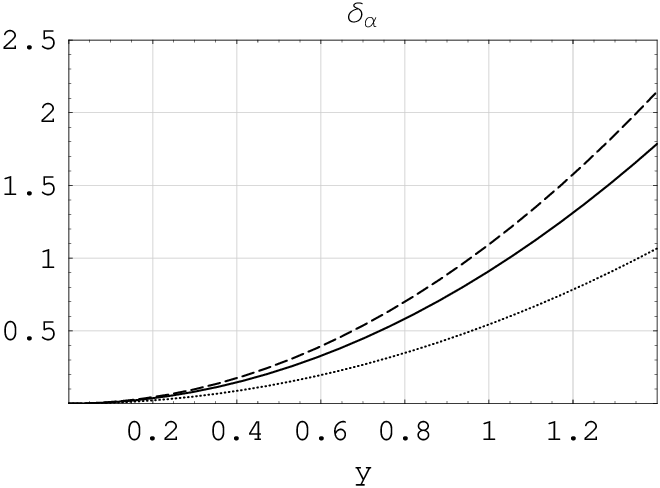}} \hspace{1cm}
\scalebox{0.56}{ \epsfig{figure=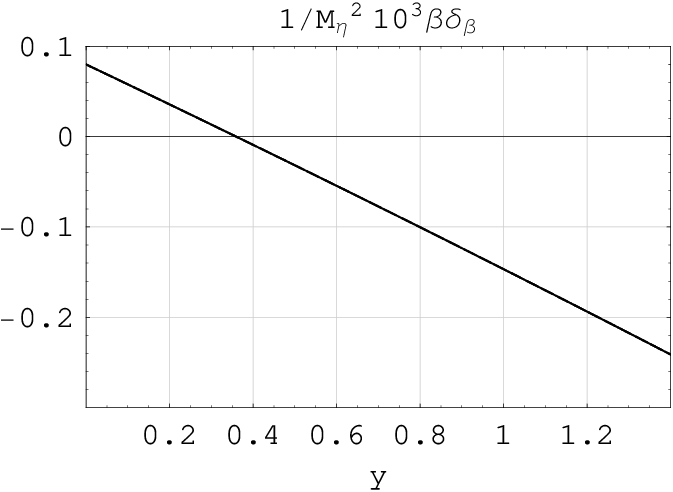}}
\caption{\emph{Dependence of the resonance estimates of the direct
remainders on $Y=X/Z$ for $r=15$ (dots), $r_2$ (solid) and $30$
(dashed).
Note that for $M_{S_1}=M_S$ and $\widetilde{c}_{m,d}=c_{m,d}/\protect\sqrt{3%
}$, the remainder $\delta _\beta ^R$ is exactly independent on
$r$.}} \label{figure4}
\end{figure}

\begin{figure}[h] \hspace{0.75cm}
\scalebox{0.56}{\epsfig{figure=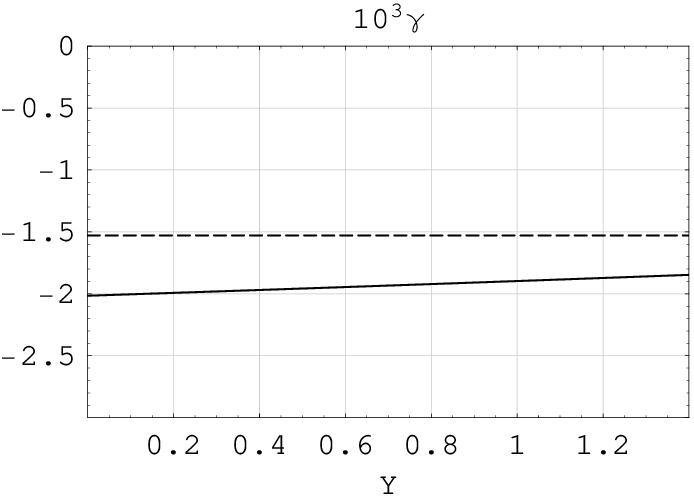}} \hspace{1cm}
\scalebox{0.56}{ \epsfig{figure=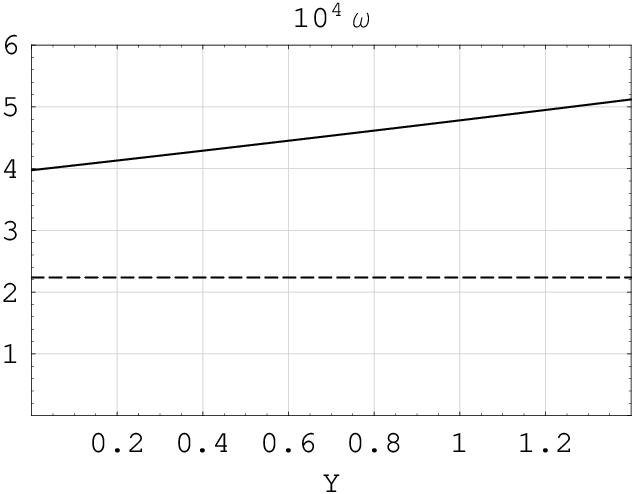}}
\caption{\emph{Polynomial parameters $\gamma$ and $\omega$
depending on $Y$. Horizontal dashed line: standard $O(p^4)$
and central R$\chi$PT value. The result with resonance remainder
estimates is shown by the solid line.}} \label{Rem2}
\end{figure}

For the rough numerical estimates we use $M_S$, $M_{S_1}$, $M_{\eta _1}$ and
the couplings $c_d$, $c_m$, $\widetilde{c}_d$, $\widetilde{c}_m$, $\widetilde{d}_m$ from \cite{Ecker:1988te}. This gives at the standard reference point $X=X^{std}$, $Z=Z^{std}$ and $r=r^{std}$
\begin{eqnarray}
\delta _\alpha ^R &=&1.00 \\
\beta \delta _\beta ^R &=&-0.15\times 10^{-3}M_\eta ^2,
\end{eqnarray}
which represents roughly $55\%$ and $20\%$ correction to values in
the first row of  the Table \ref{table1} respectively. The
dependence of $\delta _\alpha ^R$ and $\delta _\beta ^R$ on
$Y=X/Z$ and $r$ is depicted in  Figure \ref{figure4}. The effect
of the resonance remainder estimate on the parameters $\alpha$ and $\beta$ in a wider
range of the $X$, $r$ and $Z$ is illustrated in the first column of Figure \ref{Rem1}, analogous plots for $\gamma$ and $\omega$ are in Figure \ref{Rem2}. As can be seen, these results suggests the conclusion that the derivative part of the expansion could in some cases produce higher order remainders with much bigger value than 10\%.

\subsection{Generalized $\chi PT$}

\begin{figure}[p] \mbox{} \hspace{1cm}
\scalebox{0.56}{\epsfig{figure=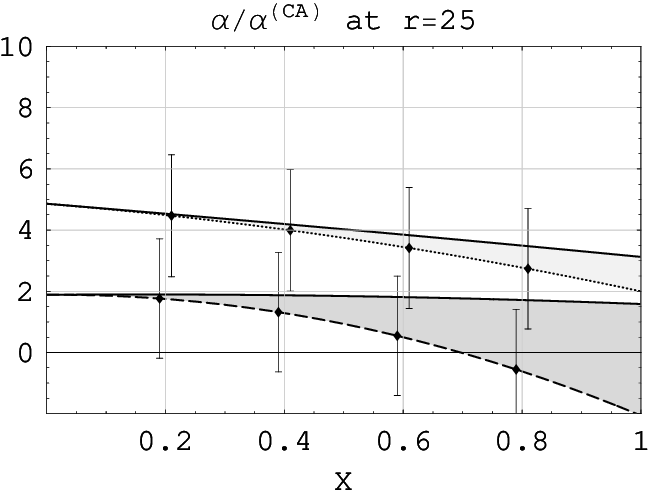}} \hspace{1cm}
\scalebox{0.56}{\epsfig{figure=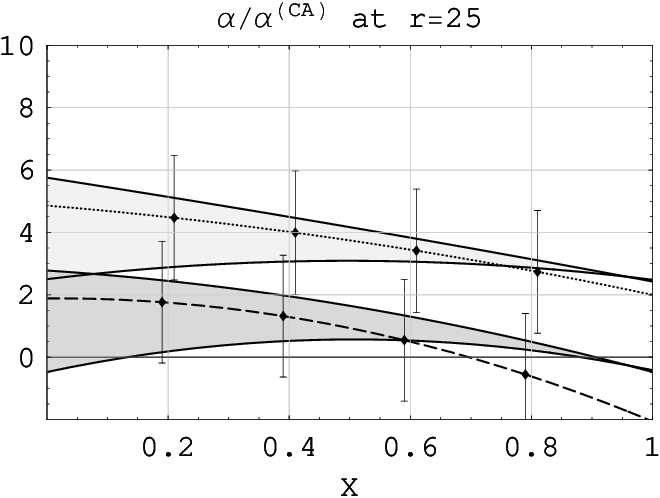}}\\ \mbox{} \hspace{1cm}
\scalebox{0.56}{\epsfig{figure=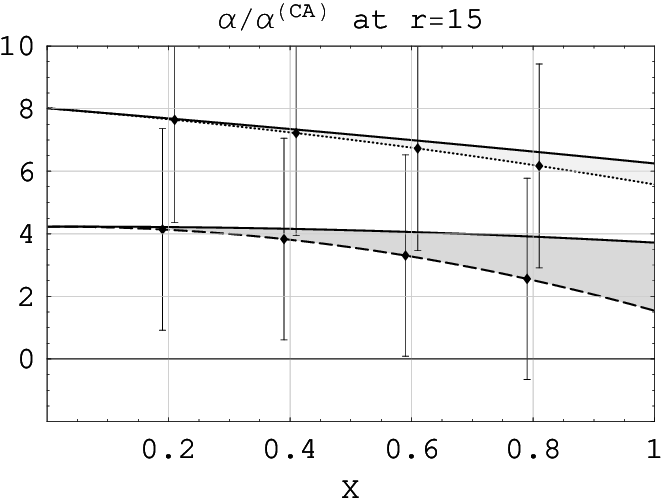}} \hspace{1cm}
\scalebox{0.56}{\epsfig{figure=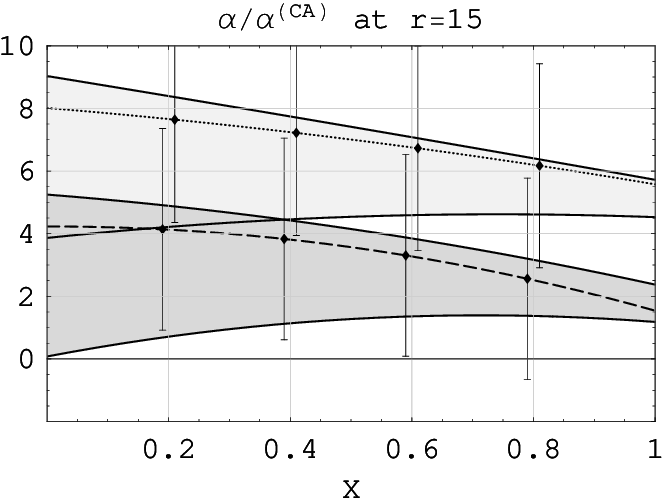}}\\ \mbox{} \hspace{1cm}
\scalebox{0.56}{\epsfig{figure=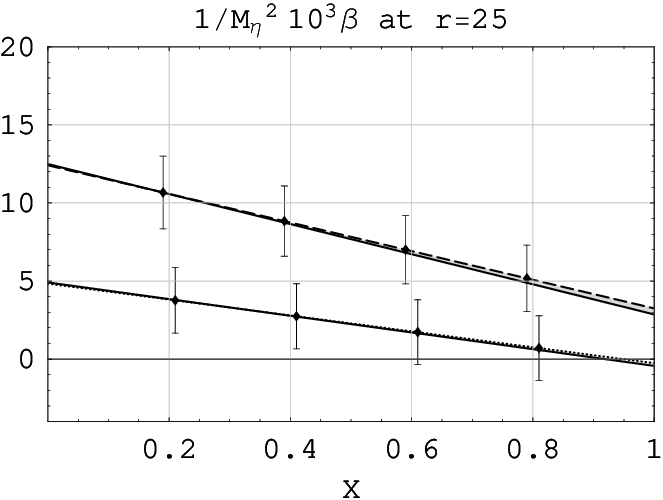}} \hspace{1cm}
\scalebox{0.56}{\epsfig{figure=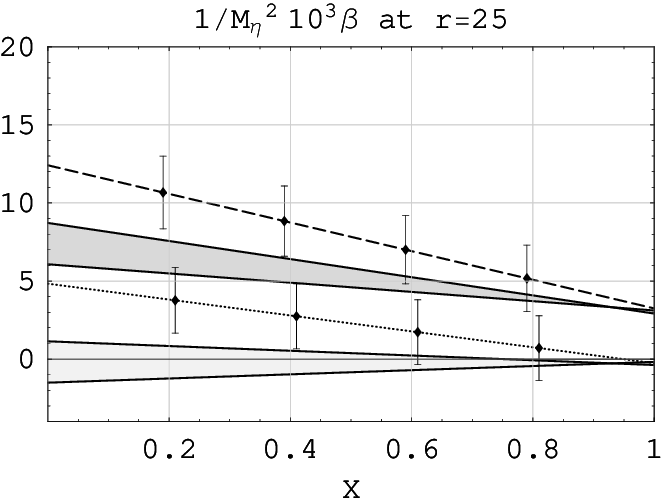}}\\ \mbox{} \hspace{1cm}
\scalebox{0.56}{\epsfig{figure=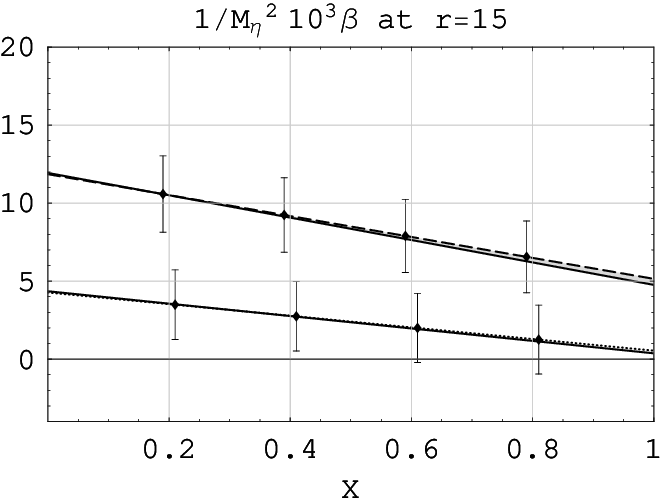}} \hspace{1cm}
\scalebox{0.56}{\epsfig{figure=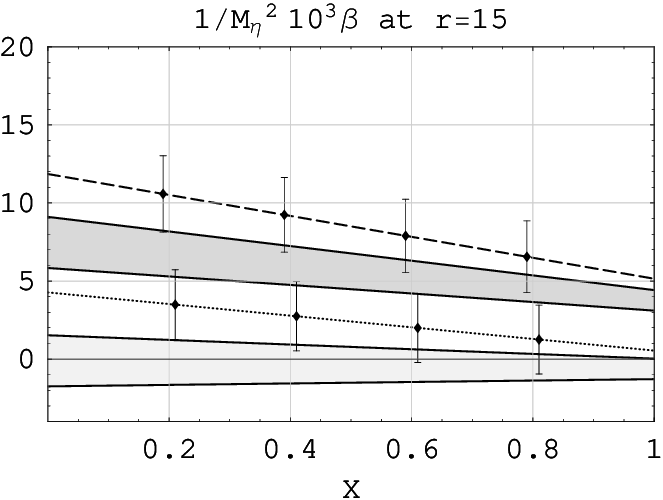}}
\caption{\emph{Polynomial parameters $\alpha$ and $\beta$
depending on $X$ and $Z$ for traditional and low value of $r$.
The dotted line shows the central value for $Z=0.9$, the dashed one is the
same for $Z=0.5$. The error bars correspond to the $10\%$
estimates of the remainders. Left column: resonance estimate,
filled areas highlight the $O(p^6)$ and higher corrections to the
amplitude generated by resonances (lighter for $Z=0.9$, darker for
$Z=0.5$). Right column: results with combined resonance and
G$\chi$PT estimate of remainders. Filled areas show the scale
dependence ($\mu\sim M_{\eta}-M_{\rho}$), lighter for $Z=0.9$ and
darker for $Z=0.5$.}} \label{Rem1}
\end{figure}
In the previous subsection we have tried to estimate the
contributions to the remainders generated by the derivative
expansion. The resulting expressions (\ref{dalpha_Res}) and
(\ref{dbeta_Res}) could, however, gather
only terms of at most the second order\footnote{%
Note that the physical masses in (\ref{dalpha_Res}),
(\ref{dbeta_Res}) originate in the derivative expansion.} in the
quark mass expansion due to the lowest order resonance Lagrangian
used. Also, the indirect remainders have not
been included in this way as there is no contribution to them in this simplest approach. 
For the appraisal of the importance of the missing terms we therefore need additional information. One possibility might be to use a resonance Lagrangian with
additional terms  of higher chiral order suited for saturation
of the $O(p^6)$ LEC's \cite{Cirigliano:2006hb} and/or to go to the
next-to-next-to leading order in the chiral expansion, this is,
however, beyond the scope of our paper.

Instead we try to get some flavour of the size of the effect by
means of comparison of our previous results with generalized $\chi
PT$, which was originally designed to handle  the badly convergent
quark mass expansion in the case $X\ll 1$ and therefore also
includes terms which correspond to higher orders in the standard chiral power counting.

In Subsection \ref{subsection generalized reparametrization}, we
have already rewritten the generalized expansion of the parameters $\alpha $ and $%
\beta $ (as well as that of the masses and decay constants in
the Appendices \ref{appendix decay constants}, \ref {appendix generalized Li} ) in the ``resummed''
form (\ref{alpha_res}--\ref{dbeta_res}) by means of
splitting  the ``standard`` remainders into the ``nonstandard'' extra terms $\delta ^{loops}(\mu )$ and $\delta ^{CT}(\mu )$
originating in $G\chi PT$ and the unknown part $\delta ^{G\chi PT}$. Therefore, neglecting the
latter, the sum $\delta ^{loops}(\mu )+\delta ^{CT}(\mu )$ could
be in a sense interpreted as the rough estimate of the
contribution to the standard remainders stemming from the higher
orders of the quark mass expansion.

While $\delta ^{loops}(\mu )$ are known, $\delta ^{CT}(\mu )$
depend on the unknown LEC's of the $G\chi PT$ Lagrangian (cf.
Appendix \ref{appendix Lagrangian}). We therefore set $\delta
^{CT}(\mu )=0$ at a fixed scale $\mu $ and by varying this scale in
$\delta ^{loops}(\mu )$ from $\mu=M_\eta$ to $\mu=M_\rho$ we can
get some information on the contribution of the unknown LEC's (note
that $\delta ^{loops}(\mu )+\delta ^{CT}(\mu )$ is renormalization
scale independent). We apply this procedure both to the direct and
indirect remainders.

The usual way of handling the generalized $\chi PT$ expansion is
to neglect the unknown remainders $\delta ^{G\chi PT}$. We can
repeat the considerations from the previous subsection and
partially appreciate them using the resonance estimate. In order
to avoid double counting, we have to further modify the resonance
contribution to the remainder (\ref{dG_res}) subtracting terms of
the order $O(p^4)$ within the generalized power counting in
the same way as it was done in the previous subsection (c.f.
(\ref{remaider_subtracted}))
\begin{eqnarray}
\Delta G_{\pi \eta ,R}^{G\chi PT}(s,t;u) &=&\Delta G_{\pi \eta ,R}(s,t;u)-%
\frac{16t}{M_{S_1}^4}\widetilde{c}_m^2\stackrel{o}{M}_\pi ^2\stackrel{o}{M}%
_\eta ^2-\frac{16}{M_S^4}c_m^2\stackrel{o}{M}_\pi
^2(\stackrel{o}{M}_\pi
^2(M_\eta ^2+M_\pi ^2)-\stackrel{o}{M}_\eta ^2t)  \nonumber \\
&&-\frac{16}{M_{\eta _1}^4}\widetilde{d}_m^2\stackrel{o}{M}_\pi
^2\left( \stackrel{o}{M}_\eta ^2-\stackrel{o}{M}_\pi ^2\right)
M_\eta ^2.
\end{eqnarray}
This combined $G\chi PT$ and resonance estimate of the
remainders is illustrated in Figure \ref{Rem1}, the
right column shows the result in the case of the
polynomial parameters $\alpha$ and $\beta$. The effect of the
unknown G$\chi$PT LEC's is estimated by their scale dependence.
The lines closer to the central $R\chi$PT results with neglected
remainders are the ones at the scale $\mu=M_{\rho}$ , {\em{i.e.}}
the constants are set to zero at the usually chosen scale. The
filled grey areas then show the change when the LEC's are set to
the difference when moving from the scale $\mu=M_\rho$ to
$M_\eta$. Admittedly, this assigns quite arbitrary numbers to the
LEC's, so the uncertainty should be viewed as a rough estimate
which can go both ways. The result can be interpreted as being
quite consistent with the 10\% estimate of the remainders, though
clearly exceeding it for some range of the free parameters $X,Z$
and $r$.

As for the parameters $\gamma$ and $\omega$, because their contribution in the polynomial expansion is quadratic in the Mandelstam variables, the G$\chi$PT estimate does not contribute here.

\section{Summary and conclusions\label{summary}}

In this paper we have studied the properties of various variants
of the chiral expansion, namely the recently introduced resummed $\chi PT$ 
as compared to the standard and partly generalized versions, 
on the concrete example of $\pi \eta $ scattering. Our calculations payed special 
attention to the possible reparametrization in terms of the physical observables. We have
tried to illustrate several issues in detail, specifically:

\begin{itemize}
\item  \emph{\ }the necessity of carefully choosing  a class of
``good`` observables for which the condition of global convergence
is believed to be satisfied in the sense that the $O(p^6)\,$and
higher \thinspace remainders are small and under control.

\item  the necessity to carefully define  the bare expansion of
``good`` observables. Here we have concentrated on the
requirements dictated by the exact renormalization scale
independence as well as the exact perturbative unitarity. As we
have shown, both these requirements can be met by means of sewing
together the strict chiral expansion in terms of the LEC's with the
dispersive representation for the corresponding Green function.
Nevertheless, the resulting bare expansion is not yet defined
uniquely; one has to fix the way how to treat the chiral logs and
also the $O(p^2)$ amplitudes entering the dispersive integrals.
Though the difference is formally of the same order as the
remainder itself, we have found that, it might be numerically
significant in some region of the free parameters.

\item  the\emph{\ }properties of the standard chiral expansion,
based on the potentially ``dangerous'' reparametrization of the
bare expansion implicitly assuming $X,Z\sim 1$ and $r\sim r_2$. In
this case we have established
strong sensitivity of the observables for $\pi \eta $ scattering on the $%
O(p^4)$ LEC's; this plagues the standard prediction with large
uncertainty. In the case of $L_4$ and $L_6$ this also means
strong sensitivity to the vacuum fluctuations of the
$\overline{s}s$ pairs and therefore to the deviation from the
standard scenario with $X,\,Z\sim 1$. The unusually large absolute
values of the NLO corrections as well as large variations achieved
for most of the observables (including the ``good`` ones) when
moving from the $O(p^4)$ fit of the LEC's $L_i$
\cite{Bijnens:1994qh} to the $O(p^6)$ based fit
\cite{Amoros:2000mc} might be interpreted as a signal of the
importance of the NNLO corrections within the standard chiral
expansion. This seems to be also supported by the sensitivity of
the NLO contributions to their form when expressed in terms of the
physical masses and decay constants (\emph{i.e.} how the $O(p^2)$ relations like\emph{\ e.g. }%
the Gell-Mann-Okubo formula are used).

\item  properties of the ``safe'' reparametrization and
resummation of the vacuum fluctuations. We have confirmed that,
for the ``good`` observables, the resummed and standard values
coincide near the standard reference point
$[X^{std},Z^{std},r^{std}]$. Under our working hypothesis, which
assumes the ``good`` observables to be accompanied with small and
controllable remainders, this can be interpreted as a consistence
of the standard $O(p^4)$ chiral expansion of ``good`` observables
in the sense that the potentially large higher order remainders
are in fact small. On the other hand, in most cases of the
``dangerous`` observables the standard and resummed values do not
meet at $[X^{std},Z^{std},r^{std}]$ (typically for those of them
that start at $O(p^4)$). Though this might indicate that the
standard expansion is convergent less satisfactorily in this case
and the higher order remainders might be important here, the
difference between the standard and resummed values lies within the estimated uncertainty of the resummed
prediction. Away from the standard reference point, however, we
have established that the central values diverge substantially
from those of the standard approach even for the ``good``
observables. This is a signal that, unless $X,\,Z\sim 1 $, the
higher order remainders of the standard chiral expansion might be
huge in comparison with LO+NLO value. Though this feature does not
exclude the possibility that in this case the standard remainders
might be saturated by the NNLO corrections, it could be
nevertheless interpreted as an indication of the instability of
the standard chiral expansion.

\item  the role of the remainders within the resummed approach.
We have found a strong sensitivity of the observables connected to $\pi \eta
$ scattering to the higher order remainders. This might reduce the
predictive power of this approach, unless additional information
on the actual size of the remainders is available. We have tried
to make an independent estimate of the remainders using the
simplest version of the resonance chiral Lagrangian as well as
making a comparison with $G\chi PT$. Both these  estimates seem to
be in accord with the rough expectation $\delta \sim $ $10\%$ for
the remainders only in some range of the parameters. For some observables and some corners of the parameter space, they can be substantially larger. Of course, the
convergence properties of the bare expansion deserves further
investigation by means of going to the NNLO, which is, however,
beyond the scope of our article.
\end{itemize}
Let us add some final remarks concerning the interpretation of the
above results from a practical point of view. The resummed version of the $\chi PT$ 
expansion not only seems to be a suitable framework for taking the effect of large 
$s\overline{s}$-pairs vacuum fluctuations into account, but by keeping the remainders 
as explicit parameters it effectively includes all orders of the chiral
expansion and thus it opens a space for incorporating further
improvements of the predictions using additional information from various
sources. As our analysis shows, $\pi \eta $ scattering allows to test the
plausibility of the standard assumption $X,Z\,$$\sim\,$1, $r\,$$\sim\,$25 due to
the sensitivity of the corresponding observables to the deviation of $X,Z$
and $r$ from these values. Provided experimental data were available, 
this could be done purely in the resummed framework using 
statistical methods similar to the ones used in the cases of $\pi\pi$ and $\pi K$ scattering \cite{Descotes-Genon:2003cg,Descotes-Genon:2007ta}.

On the other hand, to resolve a direct disagreement between the standard and resummed predictions 
is more delicate. At first sight, even though the $S\chi PT$ corrections at the NNLO are
still not available, the possible experimental data which were in conflict
with the standard $O(p^4)$ prediction but still compatible with that of
resumed $\chi PT$ might indicate problems with the standard chiral
expansion based on the assumption $X,Z\,$$\sim\,$1, $r\,$$\sim\,$25. This might show
itself either as unusually large $O(p^6)$ corrections or as $O(p^6)$
corrections too small to saturate the standard remainders. However, as we have illustrated in Subsection 6.1, the central values of the standard $O(p^4)$ predictions are plagued with large uncertainties even for the ``good'' observables. This feature together with the lack of information concerning the size of the standard $O(p^6)$ corrections would most likely prevents us from making a decisive conclusion concerning the possible deviations of the resummed $\chi PT$ from the standard chiral expansion. In the light of our results, this is expected - bad convergence of the standard chiral expansion does not necessarily manifest itself as a direct conflict with experimental data at NLO, but rather in large uncontrollable uncertainties attached to its predictions. 

Because of the current lack of low energy $\pi\eta$
scattering data, the comparison with experiment can only be done
indirectly. As we have mentioned in the Introduction, the promising process
here is the rare decay $\eta \rightarrow \pi^0\pi^0\gamma\gamma$, where
the off shell $\pi \eta \pi \eta ^{*}$ vertex enters the non-resonant part of
the amplitude. As the preliminary studies \cite{Kolesar:2003zh,Kolesar:2004ra} 
using $G\chi PT$ show, the effect of the $s\overline{s}$-pairs vacuum fluctuations parametrized 
by $X,\,Z\,\,$away form their standard values might give large deviations from
the prediction of $S\chi PT$ \cite{Drechsel,Bellucci,Ametller}, resulting in the increase
of the $\eta -$tail of the diphoton spectrum which can be in principle
observed. Based on the above results, the more careful analysis using
resummed version of $\chi PT$ expansion is expected to yield qualitatively
the same effect \cite{Kolesar:200x}.\\ \\
\textbf{Acknowledgment:} This work was supported in part by the Center for Particle Physics
(project no. LC 527) and by the EU Contract No. MRTN-CT-2006-035482, \lq\lq FLAVIAnet''.

\appendix

\section{\boldmath Standard chiral expansion of parameters $X$, $Z$ and
$r$\label{appendix_standard}}

Here we summarize the formulae leading to the standard values of $X$, $Z$, $%
Y $ and $r$ used in Subsection \ref{resummed vs standard}. Using
the standard reparametrization rules explained in Subsection
\ref{standard_rules}, we get up to the NLO order in terms of the $O(p^4) $ LEC's
\begin{eqnarray}  \label{XstdZstdYstdrstd}
X^{std} &=&1-\frac{M_\pi ^2}{2F_\pi ^2}\left(
32(L_6^r(r_2+2)+L_8^r)\right.
\nonumber \\
&&\left. +3J_{\pi \pi }^r(0)+(r_2+1)J_{KK}^r(0)+\frac
19(2r_2+1)J_{\eta \eta
}^r(0)+\frac{11r_2+37}{144\pi ^2}\right)  \nonumber \\
Z^{std} &=&1-\frac{M_\pi ^2}{2F_\pi ^2}\left( 16\left(
L_4^r(r_2+2)+L_5^r\right) +(r_2+1)J_{KK}^r(0)+4J_{\pi \pi }^r(0)+\frac{r_2+5%
}{16\pi ^2}\right)  \nonumber \\
r^{std} &=&r_2-\frac{M_\pi ^2(r_2+1)}{2F_\pi ^2}\left(
8(2L_8^r-L_5)(r_2-1)-\frac 13(2r_2+1)J_{\eta \eta }^r(0)+J_{\pi \pi }^r(0)-%
\frac{(r_2-1)}{24\pi ^2}\right).  \nonumber \\
\end{eqnarray}
For $r^{std}$ we can also use an alternative expression based on the
chiral expansion of $r_2^{*}$
\begin{eqnarray}
r^{*std} &=&r_2^{*}-\frac{M_\pi ^2(r_2+1)}{2F_\pi ^2}\left(
16L_8^r(r_2-1)\right.  \nonumber \\
&&\left. +\frac 16(2r_2+1)J_{\eta \eta }^r(0)+\frac
12(r_2+1)J_{KK}^r(0)-\frac 32J_{\pi \pi }^r(0)+\frac{5(r_2-1)}{96\pi ^2}%
\right).  \label{rstarstd}
\end{eqnarray}
$\Delta _{GMO}$ has the following standard chiral expansion
\begin{eqnarray}
\Delta _{GMO}^{std} &=&\frac{M_\pi ^2(r_2-1)}{2F_\pi ^2}\left(
32(2L_7^r+L_8^r)(r_2-1)\right.  \nonumber \\
&&\left. +\frac 13(2r_2+1)J_{\eta \eta
}^r(0)+(r_2+1)J_{KK}^r(0)-3J_{\pi \pi }^r(0)+\frac{5(r_2-1)}{48\pi
^2}\right) .  \label{Delta_GMO_std}
\end{eqnarray}

\section{\boldmath Chiral expansion of the $\eta $ decay constant \label%
{appendix decay constants}}

For the bare expansion of the ``good'' observables $F_P^2$ we
rewrite the standard formulae in the form
\begin{eqnarray}
F_\pi ^2 &=&F_0^2\left[ 1+\frac{B_0\widehat{m}}{F_0^2}\left(
16L_4^r(\mu )(r+2)+16L_5^r(\mu )+(r+1)J_{KK}^r(0)+4J_{\pi \pi
}^r(0)+\frac 1{16\pi
^2}(r+5)\right) \right]  \nonumber \\
&&+F_\pi ^2\delta _{F_\pi },  \label{F_pi_bare} \\
F_K^2 &=&F_0^2\left[ 1+\frac{B_0\widehat{m}}{F_0^2}\left(
16L_4^r(\mu )(r+2)+8L_5^r(\mu )(r+1)+\frac 32(r+1)\left(
J_{KK}^r(0)+\frac 1{16\pi
^2}\right) \right. \right.  \nonumber \\
&&\left. \left. +\frac 32\left( J_{\pi \pi }^r(0)+\frac 1{16\pi
^2}\right) +\frac 12(2r+1)\left( J_{\eta \eta }^r(0)+\frac 1{16\pi
^2}\right) \right)
\right]  \nonumber \\
&&+F_K^2\delta _{F_K}  \label{F_k_bare} \\
F_\eta ^2 &=&F_0^2\left[ 1+\frac{B_0\widehat{m}}{F_0^2}\left(
16L_4^r(\mu )(r+2)+\frac{16}3L_5^r(\mu )(2r+1)+3(r+1)\left(
J_{KK}^r(0)+\frac 1{16\pi
^2}\right) \right) \right]  \nonumber \\
&&+F_\eta ^2\delta _{F_\eta }.  \label{F_eta_bare}
\end{eqnarray}
Within the standard $\chi PT$, the $O(p^2)$ parameters $B_0$,
$F_0$ and $r$
are expressed using inverted expansions of the observables $%
F_P^2$, $M_P^2$,
as explained in subsection \ref{subsection_std}. This yields the
standard formula for $F_\eta ^2$
\begin{eqnarray}
F_\eta ^2 &=&F_\pi ^2\left[ 1+\frac{M_\pi ^2}{F_\pi ^2}\left( \frac{16}%
3L_5^r(\mu )(r_2-1)+(r_2+1)J_{KK}^r(0)-2J_{\pi \pi }^r(0)+\frac
1{16\pi
^2}(r_2-1)\right) \right]  \nonumber \\
&&+F_\eta ^2\delta _{F_\eta }^{st}.  \label{F_eta_std}
\end{eqnarray}
with a potentially large remainder $\delta _{F_\eta }^{st}$.
Numerically, with $L_5^r(M_\rho )$ taken from
\cite{Bijnens:1994qh} we get
\begin{equation}
F_\eta ^2=1.625F_\pi ^2.
\end{equation}
On the other hand, the ``safe'' reparametrization in terms of $r$, $X$
and $Z$ gives
\begin{eqnarray}
F_\eta ^2 &=&F_\pi ^2\left[ 1+\frac 23(r-1)\eta (r)-\frac 13\frac{M_\pi ^2}{%
F_\pi ^2}\left( \frac XZ\right) \left( J_{\pi \pi
}^r(0)-2(r+1)J_{KK}^r(0)+(2r+1)J_{\eta \eta }^r(0)\right) \right]
\nonumber
\\
&&+\frac 13\left( 3F_\eta ^2\delta _{F_\eta }+F_\pi ^2\delta
_{F_\pi }-4F_K^2\delta _{F_K}\right),  \label{F_eta_resumed}
\end{eqnarray}
which is valid as an exact algebraic identity\footnote{%
This identity can be also rewritten as
\begin{equation*}
4F_K^2(1-\delta _{F_K})-F_\pi ^2(1-\delta _{F_\pi })-3F_\eta
^2(1-\delta _{F_\eta })=\left( \frac XZ\right) M_\pi ^2\left(
J_{\pi \pi }^r(0)-2(r+1)J_{KK}^r(0)+(2r+1)J_{\eta \eta
}^r(0)\right) .
\end{equation*}
\par
Within the standard approach, the parameters on the r.h.s. of this
identity can be expressed to the order $O(p^4)$  in terms of the
physical observables and it is interpreted as a $O(p^4)$ sum rule
\begin{equation*}
4F_K^2-F_\pi ^2-3F_\eta ^2=M_\pi ^2\left( J_{\pi \pi
}^r(0)-2(r_2+1)J_{KK}^r(0)+(2r_2+1)J_{\eta \eta }^r(0)\right) .
\end{equation*}
This gives
\begin{equation*}
F_\eta ^2=1.697F_\pi ^2.
\end{equation*}
}.

Following the $G\chi PT$ procedure outlined in Section 5.4, after identifying the corresponding LEC's
in both approaches
\begin{eqnarray}
\frac{B_0\widehat{m}}{F_0^2}L_4^r(\mu ) &\rightarrow &\frac 18\widehat{m}%
\widetilde{\xi }  \nonumber \\
\frac{B_0\widehat{m}}{F_0^2}L_5^r(\mu ) &\rightarrow &\frac
18\widehat{m}\xi ^r,  \label{L_to_xi}
\end{eqnarray}
and defining the remainders using the physical masses inside the
chiral logs, we can use the exact formula (\ref{F_eta_bare}) and
write the remainder $\delta _{F_\eta }$ within $G\chi PT$ as
\begin{equation}
F_\eta ^2\delta _{F_\eta }=F_\eta ^2\delta _{F_\eta }^{loop}(\mu
)+F_0^2\delta _{F_\eta }^{(4)\,\,CT}(\mu )+F_\eta ^2\delta
_{F_\eta }^{G\chi PT}.  \nonumber
\end{equation}
Here
\begin{equation}
F_\eta ^2\delta _{F_\eta }^{loop}(\mu )=3\widehat{m}^2\left(
r+1\right) (A_0\left( r+1\right) +2(r+2)Z_0^S)\left(
J_{KK}^r(0)+\frac 1{16\pi ^2}\right)
\end{equation}
is the extra loop contribution and $\delta _{F_\eta }^{(4)\,CT}
(\mu )$ the contribution of the counterterms from
the $O(p^4)$ $G\chi PT$ Lagrangian renormalized at scale $\mu $
\begin{eqnarray}
\delta _{F_\eta }^{(4)\,CT}(\mu ) &=&\frac 23\widehat{m}^2\left[
\,\frac
1{\,2}(2A_1+A_2+4A_3+2B_1-2B_2)(1+2r^2)\right.  \nonumber \\
&&\left. +3(A_4\,+2B_4)(1+\frac
12r^2)\,-4\,C_1^P(r-1)^2+2D^S\left( r+2\right) \left( 2r+1\right)
\right].  \label{Fe_G_rem}
\end{eqnarray}
$\delta _{F_\eta }^{G\chi PT}$ is a new remainder, which is
exactly independent on the renormalization scale.

Analogously, for the $G\chi PT$ formula for $F_\pi ^2$ we have,
besides the substitution (\ref{L_to_xi}) to (\ref{F_pi_bare}), to
insert
\begin{equation}
F_\pi ^2\delta _{F_\pi }=F_\pi ^2\delta _{F_\pi }^{loop}(\mu
)+F_0^2\delta _{F_\pi }^{(4)\,\,CT}(\mu )+F_\pi ^2\delta _{F_\pi
}^{G\chi PT}, \label{Fpi_G_rem}
\end{equation}
where the loop and counterterm contribution are now
\begin{eqnarray}
F_\pi ^2\delta _{F_\pi }^{loop}(\mu ) &=&8\widehat{m}^2(A_0+(r+2)Z_0^S))%
\left( J_{\pi \pi }^r(0)+\frac 1{16\pi ^2}\right)  \nonumber \\
&&+\widehat{m}^2\left( r+1\right) (A_0\left( r+1\right)
+2(r+2)Z_0^S)\left(
J_{KK}^r(0)+\frac 1{16\pi ^2}\right)  \nonumber \\
\delta _{F_\pi }^{(4)\,\,CT}(\mu ) &=&2\widehat{m}^2\left[
A_1+\frac
12A_2+2A_3+(A_4\,+2B_4)(1+\frac 12r^2)\,\right.  \nonumber \\
&&\left. +B_1-B_2+2D^S\left( r+2\right) \right] .
\label{Fpi_G_rem_explicit}
\end{eqnarray}
Finally, we have the expression for $F_K^2$,
where the remainder is replaced with
\begin{equation}
F_K^2\delta _{F_K}=F_K^2\delta _{F_K}^{loop}(\mu )+F_0^2\delta
_{F_K}^{(4)\,\,CT}(\mu )+F_K^2\delta _{F_K}^{G\chi PT}
\label{Fk_G_rem}
\end{equation}
and the loops and counterterms contribute as
\begin{eqnarray}
F_K^2\delta _{F_K}^{loop}(\mu )
&=&3\widehat{m}^2(A_0+(r+2)Z_0^S))\left(
J_{\pi \pi }^r(0)+\frac 1{16\pi ^2}\right)  \nonumber \\
&&+\frac 32\widehat{m}^2\left( r+1\right) (A_0\left( r+1\right)
+2(r+2)Z_0^S)\left( J_{KK}^r(0)+\frac 1{16\pi ^2}\right)  \nonumber \\
&&+\widehat{m}^2(A_0\left( 2r^2+1\right)
+2(r-1)^2Z_0^P+(r+2)(2r+1)Z_0^S)\left( J_{\eta \eta }^r(0)+\frac
1{16\pi
^2}\right)  \nonumber \\
\delta _{F_K}^{(4)\,\,CT}(\mu ) &=&\widehat{m}^2\left[
(A_1+B_1)(r^2+1)+(A_2-2B_2)r+2(A_4\,+2B_4)(1+\frac 12r^2)\,\right.
\nonumber
\\
&&\left. +2D^S\left( r+2\right) (r+1)\right] .
\label{Fk_G_rem_explicit}
\end{eqnarray}
In order to reparametrize the $G\chi PT$ bare expansion in terms
of the masses
and decay constants, we can proceed as follows. Because the exact identity (%
\ref{F_eta_resumed}) is valid independently of the version of
$\chi PT$, we can also use it in the generalized case, provided we
rewrite the remainders according to (\ref{Fe_G_rem},
\ref{Fpi_G_rem}) and (\ref{Fk_G_rem}). This step eliminates the
LEC's $\xi $ and $\widetilde{\xi }$. Collecting the chiral logs
together we have
\begin{eqnarray}
F_\eta ^2 &=&F_\pi ^2\left[ 1+\frac 23(r-1)\eta (r)-\frac 1{3F_\pi
^2}\left( \widetilde{M}_\pi ^2\left( J_{\pi \pi }^r(0)+\frac
1{16\pi ^2}\right)
\right. \right. \nonumber\\
&&\left. \left. -4\widetilde{M}_K^2\left( J_{KK}^r(0)+\frac
1{16\pi ^2}\right) +3\widetilde{M}_\eta ^2\left( J_{\eta \eta
}^r(0)+\frac 1{16\pi ^2}\right) \right) \right] +F_\eta ^2\Delta
_{F_\eta }^{G\chi PT}(\mu )
\end{eqnarray}
where the $O(p^2)$ masses are given by (\ref{Op2_mass_G}) and
\begin{equation}
F_\eta ^2\Delta _{F_\eta }^{G\chi PT}(\mu )=\frac 13F_0^2\left(
3\delta _{F_\eta }^{(4)\,\,CT}+\delta _{F_\pi
}^{(4)\,\,CT}-4\delta _{F_K}^{(4)\,\,CT}\right) +\frac 13\left(
3F_\eta ^2\delta _{F_\eta }^{G\chi PT}+F_\pi ^2\delta _{F_\pi
}^{G\chi PT}-4F_K^2\delta _{F_K}^{G\chi PT}\right).
\end{equation}
The last step consists of replacing the LEC's $F_0$, $A_0$, $Z_0^S$
and $Z_0^P$ with the first term of their expansion in terms of the
masses and decay constants as described in Subsection \ref{subsection generalized reparametrization}.
This corresponds to a further redefinitions of the generalized remainders.

\section{\boldmath Dispersion representation of the $\pi \eta $ amplitude \label%
{appendix dispersive}}

For the dispersive representation of the amplitude we need the
$S-$ and $T-$channel discontinuities at $O(p^4)$. In the following
subsections we give a list of the relevant $O(p^2)$ amplitudes
$G^{(2)A_i \rightarrow ij }$ and $G^{(2)ij \rightarrow A_f }$ and
$O(p^4)$ discontinuities $\mathrm{disc}G_0^{ij }$ corresponding to
the different intermediate states $ij$.

\subsection{$S-$channel discontinuities at $O(p^4)$}
\begin{itemize}
\item  $\pi \eta $ intermediate state
\begin{eqnarray}
G^{(2)\pi \eta \rightarrow \pi \eta }&=&\frac 13F_0^2\stackrel{o}{M}%
_\pi ^2
\nonumber\\
\mathrm{disc}G_0^{\pi \eta }(s)&=&2\frac{\lambda ^{1/2}(s,M_\pi ^2,M_\eta ^2)}%
s\left( \frac 1{32\pi }\frac 13\stackrel{o}{M}_\pi ^2\right) ^2\frac{F_0^4}{%
F_\pi ^2F_\eta ^2}
\end{eqnarray}

\item  $\overline{K}K$ intermediate state
\begin{eqnarray}
G^{(2)\pi \eta \rightarrow \overline{K^0}K^0(K^{+}K^{-})} &=&-\frac{%
\sqrt{3}}4F_0^2(s-\frac 13M_\eta ^2-\frac 13M_\pi ^2-\frac 23M_K^2) \nonumber\\
&&+\frac 1{4\sqrt{3}}F_0^2(2\stackrel{o}{M}_K^2-\stackrel{o}{M}_\pi ^2-%
\stackrel{o}{M}_\eta ^2)
\nonumber\\
G^{(2)\overline{K^0}K^0(K^{+}K^{-})\rightarrow \pi \eta } &=&-\frac{%
\sqrt{3}}4F_0^2(s-\frac 13M_\eta ^2-\frac 13M_\pi ^2-\frac 23M_K^2) \nonumber\\
&&+\frac 1{4\sqrt{3}}F_0^2(2\stackrel{o}{M}_K^2-\stackrel{o}{M}_\pi ^2-%
\stackrel{o}{M}_\eta ^2)
\nonumber\\
\mathrm{disc}G_0^{\overline{K^0}K^0(K^{+}K^{-})}(s) &=&2\sqrt{1-\frac{4M_K^2}%
s}\left( \frac 1{32\pi }\right) ^2\frac 3{16}[(s-\frac 13M_\eta
^2-\frac
13M_\pi ^2-\frac 23M_K^2) \nonumber\\
&&-\frac 13(2\stackrel{o}{M}_K^2-\stackrel{o}{M}_\pi
^2-\stackrel{o}{M}_\eta ^2)]^2\frac{F_0^4}{F_K^4}
\end{eqnarray}
\end{itemize}

\subsection{$T-$channel discontinuities at $O(p^4)$}
\begin{itemize}
\item  $\pi \pi $ intermediate state
\begin{eqnarray}
G^{(2)\pi \pi \rightarrow \pi \pi ,\,I=0}&=&F_0^2[(s-\frac 43M_\pi
^2)+\frac 56\stackrel{o}{M}_\pi ^2]
\nonumber\\
\mathrm{disc}G_0^{\pi \pi ,\,I=0}(s)&=&2\sigma (s)\left( \frac
1{32\pi }\right)
^2\frac 13\stackrel{o}{M}_\pi ^2[(s-\frac 43M_\pi ^2)+\frac 56\stackrel{o}{M}%
_\pi ^2]\frac{F_0^4}{F_\pi ^4}
\end{eqnarray}

\item  $\eta \eta $ intermediate state
\begin{eqnarray}
G^{(2)\eta \eta \rightarrow \eta \eta }&=&-\frac 13F_0^2\left(
\stackrel{o}{M}_\pi ^2-4\stackrel{o}{M}_\eta ^2\right)
\nonumber\\
\mathrm{disc}G_0^{\eta \eta }(s)&=&-2\frac 12\sqrt{1-\frac{4M_\eta
^2}s}\left(
\frac 1{32\pi }\right) ^2\frac 19\stackrel{o}{M}_\pi ^2\left( \stackrel{o}{M}%
_\pi ^2-4\stackrel{o}{M}_\eta ^2\right) \frac{F_0^4}{F_\eta ^4}
\end{eqnarray}

\item  $\overline{K}K$ intermediate state\footnote{Let us note
\begin{equation*}
G^{I=0}(s,t;u)=-\frac 1{\sqrt{3}}\delta
^{ab}G^{ab}(s,t;u)=-\sqrt{3}G(s,t;u)
\end{equation*}}
\begin{eqnarray}
G^{(2)\pi \pi \rightarrow \overline{K^0}K^0(K^{+}K^{-}),\,I=0}
&=&\mp
\frac{\sqrt{3}}4F_0^2[(s-\frac 23M_\pi ^2-\frac 23M_K^2) \nonumber\\
&&+\frac 23(\stackrel{o}{M}_K^2+\stackrel{o}{M}_\pi ^2)]
\nonumber\\
G^{(2)\overline{K^0}K^0(K^{+}K^{-})\rightarrow \eta \eta ,\,I=0}
&=&\pm \frac 14F_0^2[(3s-2M_K^2-2M_\eta ^2) \nonumber\\
&&+(2\stackrel{o}{M}_\eta ^2-\frac 23\stackrel{o}{M}_K^2)]
\nonumber\\
\mathrm{disc}G_0^{\overline{K^0}K^0(K^{+}K^{-}),\,I=0}(s)
&=&\sqrt{1-\frac{4M_K^2}s}\left( \frac 1{32\pi }\right) ^2\frac
1{16}[(s-\frac 23M_\pi ^2-\frac 23M_K^2) \nonumber\\
&&+\frac 23(\stackrel{o}{M}_K^2+\stackrel{o}{M}_\pi ^2)] \nonumber\\
&&\times [(3s-2M_K^2-2M_\eta ^2)+(2\stackrel{o}{M}_\eta ^2-\frac 23\stackrel{%
o}{M}_K^2)]]\frac{F_0^4}{F_K^4}
\end{eqnarray}
\end{itemize}

\section{The scalar bubble \label{appendix Jbar}}

In this appendix we summarize the formulae for the scalar bubble,
defined as
\begin{eqnarray}
J_{PQ}(q^2) &=&-\mathrm{i}\int \frac{d^dk}{(2\pi )^d}\frac 1{(k^2-M_P^2+%
\mathrm{i}0)((k-q)^2-M_Q^2+\mathrm{i}0)} \nonumber\\
&=&-2\lambda _\infty +J_{PQ}^r(q^2).
\end{eqnarray}
Here, as usual
\begin{eqnarray}
\lambda _\infty &=&\frac{\mu ^{d-4}}{16\pi ^2}\left( \frac
1{d-4}-\frac
12(\ln 4\pi +\Gamma ^{^{\prime }}(1)+1)\right) \nonumber\\
&&
\end{eqnarray}
and $J_{PQ}^r(s)=J_{PQ}^r(0)+\overline{J}_{PQ}(s)$, where
\begin{equation}
J_{PQ}^r(0)=-\frac 1{16\pi ^2}\frac{M_P^2\ln (M_P^2/\mu
^2)-M_Q^2\ln (M_Q^2/\mu ^2)}{M_P^2-M_Q^2}
\end{equation}
and $\overline{J}_{PQ}(s)$, sometimes called Chew-Mandelstam
function, can be expressed by means of once subtracted dispersion
relation as

\begin{equation}
\overline{J}_{PQ}(s)=\frac s{16\pi ^2}\int_{(M_P+M_Q)^2}^\infty \frac{%
\mathrm{d}x}x\frac{\lambda ^{1/2}(x,M_P^2,M_Q^2)}x\frac 1{x-s}.
\end{equation}
The explicit form of $\overline{J}_{PQ}(s)$ reads
\begin{equation}
\overline{J}_{PQ}(s)=\frac 1{32\pi ^2}\left( 2+\frac{\Delta _{PQ}}s\ln \frac{%
M_Q^2}{M_P^2}-\frac{\Sigma _{PQ}}{\Delta _{PQ}}\ln \frac{M_Q^2}{M_P^2}+2%
\frac{(s-(M_P-M_Q)^2)}s\sigma _{PQ}(s)\ln \frac{\sigma
_{PQ}(s)-1}{\sigma _{PQ}(s)+1}\right),
\end{equation}
where
\begin{eqnarray}
\Delta _{PQ} &=&M_P^2-M_Q^2 \nonumber\\
\Sigma _{PQ} &=&M_P^2+M_Q^2 \nonumber\\
\sigma _{PQ}(t) &=&\sqrt{\frac{s-(M_P+M_Q)^2}{s-(M_P-M_Q)^2}}=\sqrt{1-\frac{%
4M_PM_Q}{s-(M_P-M_Q)^2}}\,.
\end{eqnarray}
In the limit $M_P\rightarrow M_Q$ we get
\begin{eqnarray}
J_{PP}^r(0) &=&-\frac 1{16\pi ^2}\left( \ln \frac{M_P^2}{\mu ^2}+1\right) \nonumber\\
\overline{J}_{PP}(s) &=&\frac 1{16\pi ^2}\left( 2+\sigma _{PP}(s)\ln \frac{%
\sigma _{PP}(s)-1}{\sigma _{PP}(s)+1}\right) .
\end{eqnarray}

\section{\boldmath $O(p^4)$ constants $L_4$-$L_8$ in terms of masses and decay
constants \label{appendix Li}}

In this Appendix we summarize the formulae used in the text for
the reparametrization of bare expansions of ``good'' observables.
We use the abbreviated notation (\ref{r_2*}--\ref{DeltaGMO}).
From the bare expansion of ``good'' variables $F_\pi ^2$ and $F_K^2$ we obtain
\begin{eqnarray}
4\stackrel{o}{M}_\pi ^2L_4^r(\mu ) &=&\frac 12(1-Z-\eta (r))\frac{F_\pi ^2}{%
r+2}  \nonumber \\
&&-\frac{M_\pi ^2}{4(r+2)(r-1)}\frac XZ[(4r+1)J_{\pi \pi
}^r(0)+(r-2)(r+1)J_{KK}^r(0)  \nonumber \\
&&-(2r+1)J_{\eta \eta }^r(0)+\frac{(r+2)(r-1)}{16\pi ^2}]  \nonumber \\
&&+\frac{2F_K^2\delta _{F_K}-(r+1)F_\pi ^2\delta _{F_\pi
}}{2(r+2)(r-1)},
\\
4\stackrel{o}{M}_\pi ^2L_5^r(\mu ) &=&\frac 12F_\pi ^2\eta (r)  \nonumber \\
&&+\frac{M_\pi ^2}{4(r-1)}\frac XZ[5J_{\pi \pi
}^r(0)-(r+1)J_{KK}^r(0)-(2r+1)J_{\eta \eta
}^r(0)-\frac{3(r-1)}{16\pi ^2}]
\nonumber \\
&&-\frac{F_K^2\delta _{F_K}-F_\pi ^2\delta _{F_\pi }}{(r-1)}.
\label{L_5_L_6}
\end{eqnarray}
In the same way, from the expansion of $F_P^2M_P^2$ we get

\begin{eqnarray}
4\stackrel{o}{M}_\pi ^4L_6^r(\mu ) &=&\frac 14\frac{F_\pi ^2M_\pi ^2}{r+2}%
(1-X-\varepsilon (r))  \nonumber \\
&&-\frac{M_\pi ^4}{72(r-1)(r+2)}\left( \frac XZ\right)
^2[27rJ_{\pi \pi
}^r(0)+9(r+1)(r-2)J_{KK}^r(0)  \nonumber \\
&&+(2r+1)(r-4)J_{\eta \eta }^r(0)+\frac{11(r-1)(r+2)}{16\pi ^2}]
\\
&&-\frac{F_\pi ^2M_\pi ^2\delta _{F_\pi M_\pi
}[(r+1)^2]-4F_K^2M_K^2\delta
_{F_KM_K}}{4(r^2-1)(r+2)}  \nonumber \\
4\stackrel{o}{M}_\pi ^4L_7^r(\mu ) &=&-\frac 18F_\pi ^2M_\pi
^2\left(
\varepsilon (r)-\frac{\Delta _{GMO}}{(r-1)^2}\right)  \nonumber \\
&&-\frac{3(1+r)F_\eta ^2M_\eta ^2\delta _{F_\eta M_\eta
}+(2r^2+r-1)F_\pi ^2M_\pi ^2\delta _{F_\pi M_\pi
}-8rF_K^2M_K^2\delta _{F_KM_K}}{8(r-1)^2(r+1)}
\\
4\stackrel{o}{M}_\pi ^4L_8^r(\mu ) &=&\frac 14F_\pi ^2M_\pi
^2\varepsilon (r)
\nonumber \\
&&+\frac{M_\pi ^4}{24(r-1)}\left( \frac XZ\right) ^2[9J_{\pi \pi
}^r(0)-3(r+1)J_{KK}^r(0)-(2r+1)J_{\eta \eta
}^r(0)-\frac{5(r-1)}{16\pi ^2}]
\nonumber \\
&&-\frac{2F_K^2M_K^2\delta _{F_KM_K}-(r+1)F_\pi ^2M_\pi ^2\delta
_{F_\pi M_\pi }}{2(r^2-1)}.  \label{L_6_L_7_L8}
\end{eqnarray}

\section{\boldmath Lagrangian of $G\chi PT$ to $O(p^4)$ \label{appendix Lagrangian}}

Here we give the traditional form of the $G\chi PT$ Lagrangian. In
the following formulae
\begin{eqnarray}
\chi &=&\mathcal{M}+s+\mathrm{i}p \nonumber\\
\nabla U &=&\partial U-\mathrm{i}(v+a)U+\mathrm{i}U(v-a) \nonumber\\
\chi &=&\partial \chi -\mathrm{i}(v+a)\chi +\mathrm{i}\chi (v-a).
\end{eqnarray}
Up to the order $O(p^4)$, the Lagrangian can be split into the $O(p^2)$, $%
O(p^3)$ and $O(p^4)$ parts
\begin{equation}
\mathcal{L}=\mathcal{L}_2+\mathcal{L}_3+\mathcal{L}_4,
\end{equation}
where
\begin{equation}
\mathcal{L}_n=\sum_{i+j+k=n}\mathcal{L}^{(i,j,k)}
\end{equation}
and $(i,j,k)$ indicates the number of derivatives, $\chi $ sources
and powers of $B_0$ respectively. Then for $O(p^2)$ we get
\begin{eqnarray}
\mathcal{L}^{(2,0,0)} &=&\frac{{F}_0^2}4\langle \nabla _\mu
U^{+}\nabla ^\mu
U\rangle \nonumber\\
\mathcal{L}^{(0,1,1)} &=&\frac{F_0^2}2B_0\langle U^{+}\chi +\chi
^{+}U\rangle
\nonumber\\
\mathcal{L}^{(0,2,0)} &=&\frac{F_0^2}4\left( \overline{A}_0\langle
(U^{+}\chi )^2+(\chi ^{+}U)^2\rangle\right. \nonumber\\
&& +\left. \overline{Z}_0^S\langle U^{+}\chi +\chi ^{+}U\rangle ^2+\overline{%
Z}_0^P\langle U^{+}\chi -\chi ^{+}U\rangle ^2\right).
\end{eqnarray}
At the order $O(p^3)$ one has
\begin{eqnarray}
\mathcal{L}^{(2,1,0)} &=&\frac{F_0^2}4(\overline{\xi }\langle
\nabla _\mu
U^{+}\nabla ^\mu U(\chi ^{+}U+U^{+}\chi )\rangle +\overline{\widetilde{\xi }}%
\langle \nabla _\mu U^{+}\nabla ^\mu U\rangle \langle \chi
^{+}U+U^{+}\chi
\rangle ) \nonumber\\
\mathcal{L}^{(0,3,0)} &=&\frac{F_0^2}4(\overline{\rho }_1\langle
(\chi
^{+}U)^3+(U^{+}{\chi )}^3\rangle +\overline{\rho }_2\langle (\chi ^{+}U+U^{+}%
{\chi )\chi }^{+}\chi \rangle \nonumber\\
&&+\overline{\rho }_3\langle (\chi ^{+}U)^2-(U^{+}{\chi
)}^2\rangle \langle
\chi ^{+}U-U^{+}\chi \rangle \nonumber\\
&&+\overline{\rho }_4\langle (\chi ^{+}U)^2+(U^{+}{\chi
)}^2\rangle \langle
\chi ^{+}U+U^{+}\chi \nonumber\\
&&+\overline{\rho }_5\langle \chi ^{+}U+U^{+}\chi \rangle \langle {\chi }%
^{+}\chi \rangle +\overline{\rho }_6\langle \chi ^{+}U-U^{+}\chi
\rangle
^2\langle \chi ^{+}U+U^{+}\chi \rangle \nonumber\\
&&+\overline{\rho }_7\langle \chi ^{+}U+U^{+}\chi \rangle ^3) \nonumber\\
\mathcal{L}^{(2,0,1)} &=&\frac{{F}_0^2B_0}4\delta _d^{(1)}\langle
\nabla
_\mu U^{+}\nabla ^\mu U\rangle \nonumber\\
\mathcal{L}^{(0,2,1)} &=&\frac{F_0^2B_0}4\left( \delta ^{(1)}\overline{A}%
_0\langle (U^{+}\chi )^2+(\chi ^{+}U)^2\right. \nonumber\\
&&+\left. \delta ^{(1)}\overline{Z}_0^S\langle U^{+}\chi +\chi
^{+}U\rangle ^2+\delta ^{(1)}\overline{Z}_0^P\langle U^{+}\chi
-\chi ^{+}U\rangle
^2\right) \nonumber\\
\mathcal{L}^{(0,1,2)} &=&\frac{F_0^2}2B_0^2\delta _\chi
^{(1)}\langle U^{+}\chi +\chi ^{+}U\rangle .
\end{eqnarray}
For the $O(p^4)$ Lagrangian, the building blocks are

\begin{eqnarray}
\mathcal{L}^{(4,0,0)} &=&L_1\langle \nabla _\mu U^{+}\nabla ^\mu
U\rangle ^2+L_2\langle \nabla _\mu U^{+}\nabla _\nu U\rangle
\langle \nabla ^\mu
U^{+}\nabla ^\nu U\rangle \nonumber\\
&&+L_3\langle \nabla _\mu U^{+}\nabla ^\mu U\nabla _\nu
U^{+}\nabla ^\nu
U\rangle \nonumber\\
&&-iL_9\left\langle F_{\mu \nu }^R\nabla ^\mu U\nabla ^\nu
U^{+}+F_{\mu \nu
}^L\nabla ^\mu U^{+}\nabla ^\nu U\right\rangle \nonumber\\
&&+L_{10}\left\langle U^{+}F_{\mu \nu }^RUF_{\mu \nu
}^L\right\rangle +H_1\left\langle F_{\mu \nu }^RF^{R\mu \nu
}F_{\alpha \beta }^LF^{L\alpha
\beta }\right\rangle \nonumber\\
\mathcal{L}^{(2,1,1)} &=&\frac{F_0^2B_0}4(\delta ^{(1)}\overline{\xi }%
\langle \nabla _\mu U^{+}\nabla ^\mu U(\chi ^{+}U+U^{+}\chi
)\rangle +\delta ^{(1)}\overline{\widetilde{\xi }}\langle \nabla
_\mu U^{+}\nabla ^\mu
U\rangle \langle \chi ^{+}U+U^{+}\chi \rangle ) \nonumber\\
\mathcal{L}^{(2,0,2)} &=&\frac{{F}_0^2B_0^2}4\delta
_d^{(2)}\langle \nabla
_\mu U^{+}\nabla ^\mu U\rangle \nonumber\\
\mathcal{L}^{(2,2,0)} &=&\frac{F_0^2}4\{A_1\langle \nabla _\mu
U^{+}\nabla
^\mu U(\chi ^{+}{\chi +}U^{+}{\chi \chi }^{+}U)\rangle \nonumber\\
&&+A_2\langle (\nabla _\mu U^{+})U\chi ^{+}(\nabla ^\mu
U)U^{+}{\chi \rangle
} \nonumber\\
&&+A_3\left\langle \nabla _\mu U^{+}U(\chi ^{+}\nabla ^\mu \chi
-\nabla ^\mu \chi ^{+}\chi )+\nabla _\mu UU^{+}(\chi \nabla ^\mu
\chi ^{+}-\nabla ^\mu
\chi \chi ^{+})\right\rangle \nonumber\\
&&+A_4\langle \nabla _\mu U^{+}\nabla ^\mu U\rangle \langle \chi
^{+}\chi
\rangle \nonumber\\
&&+B_1\langle \nabla _\mu U^{+}\nabla ^\mu U(\chi ^{+}U\chi
^{+}U+U^{+}\chi
U^{+}\chi ){\rangle } \nonumber\\
&&+B_2\langle \nabla _\mu U^{+}\chi \nabla ^\mu U^{+}\chi +{\chi
^{+}}\nabla
^\mu U\chi ^{+}\nabla _\mu U\rangle \nonumber\\
&&+B_4\langle \nabla _\mu U^{+}\nabla ^\mu U\rangle \langle \chi
^{+}U\chi
^{+}U+U^{+}\chi U^{+}{\chi \rangle } \nonumber\\
&&+C_1^S\langle \nabla _\mu U\chi ^{+}+\chi \nabla _\mu
U^{+}\rangle \langle
\nabla ^\mu U\chi ^{+}+\chi \nabla ^\mu U^{+}\rangle \nonumber\\
&&+C_2^S\langle \nabla _\mu \chi ^{+}U+U^{+}\nabla _\mu \chi
\rangle \langle
\nabla ^\mu \chi ^{+}U+U^{+}\nabla ^\mu \chi \rangle \nonumber\\
&&+C_3^S\langle \nabla _\mu \chi ^{+}U+U^{+}\nabla _\mu \chi
\rangle \langle
\nabla ^\mu U^{+}\chi +\chi ^{+}\nabla ^\mu U\rangle \nonumber\\
&&+C_1^P\langle \nabla _\mu U\chi ^{+}-\chi \nabla _\mu
U^{+}\rangle \langle
\nabla ^\mu U\chi ^{+}-\chi \nabla ^\mu U^{+}\rangle \nonumber\\
&&+C_2^P\langle \nabla _\mu \chi ^{+}U-U^{+}\nabla _\mu \chi
\rangle \langle
\nabla ^\mu \chi ^{+}U-U^{+}\nabla ^\mu \chi \rangle \nonumber\\
&&+C_3^P\langle \nabla _\mu \chi ^{+}U-U^{+}\nabla _\mu \chi
\rangle \langle
\nabla ^\mu U^{+}\chi -\chi ^{+}\nabla ^\mu U\rangle \nonumber\\
&&+D^S\langle \nabla _\mu U^{+}\nabla ^\mu U(\chi ^{+}U+U^{+}\chi
)\rangle
\langle \chi ^{+}U+U^{+}\chi \rangle \nonumber\\
&&+D^P\langle \nabla _\mu U^{+}\nabla ^\mu U(\chi ^{+}U-U^{+}\chi
)\rangle
\langle \chi ^{+}U-U^{+}\chi \rangle \} \nonumber\\
&&+H_2\left\langle \nabla _\mu \chi \nabla ^\mu \chi ^{+}\right\rangle \nonumber\\
\mathcal{L}^{(0,4,0)} &=&\frac{F_0^2}4\{E_1\langle (\chi ^{+}U)^4+(U^{+}{%
\chi )}^4\rangle \nonumber\\
&&+E_2\langle \chi ^{+}{\chi (\chi }^{+}U\chi ^{+}U+U^{+}\chi
U^{+}\chi
)\rangle \nonumber\\
&&+E_3\langle \chi ^{+}\chi U^{+}{\chi \chi }^{+}U\rangle \nonumber\\
&&+F_1^S\langle \chi ^{+}U\chi ^{+}U+U^{+}\chi U^{+}\chi \rangle ^2 \nonumber\\
&&+F_2^S\langle (\chi ^{+}U)^3+(U^{+}{\chi )}^3\rangle \langle
\chi
^{+}U+U^{+}\chi \rangle \nonumber\\
&&+F_3^S\langle \chi ^{+}\chi (\chi ^{+}U+U^{+}\chi )\rangle
\langle \chi
^{+}U+U^{+}\chi \rangle \nonumber\\
&&+F_4^S\langle (\chi ^{+}U)^2+(U^{+}{\chi )}^2\rangle \langle
\chi ^{+}\chi
\rangle \nonumber\\
&&+F_1^P\langle \chi ^{+}U\chi ^{+}U-U^{+}\chi U^{+}\chi \rangle ^2 \nonumber\\
&&+F_2^P\langle (\chi ^{+}U)^3+(U^{+}{\chi )}^3\rangle \langle
\chi
^{+}U+U^{+}\chi \rangle \nonumber\\
&&+F_3^P\langle \chi ^{+}\chi (\chi ^{+}U-U^{+}\chi )\rangle
\langle \chi
^{+}U-U^{+}\chi \rangle \nonumber\\
&&+F_5^{SS}\langle (\chi ^{+}U)^2+(U^{+}{\chi )}^2\rangle \langle
\chi
^{+}U+U^{+}\chi \rangle ^2 \nonumber\\
&&+F_6^{SS}\langle \chi ^{+}\chi \rangle \langle \chi
^{+}U+U^{+}\chi
\rangle ^2 \nonumber\\
&&+F_5^{SP}\langle (\chi ^{+}U)^2+(U^{+}{\chi )}^2\rangle \langle
\chi
^{+}U-U^{+}\chi \rangle \nonumber\\
&&+F_6^{SP}\langle \chi ^{+}\chi \rangle \langle \chi
^{+}U-U^{+}\chi
\rangle ^2 \nonumber\\
&&+F_7^{SP}\langle (\chi ^{+}U)^2-(U^{+}{\chi )}^2\rangle \nonumber\\
&&\times \langle \chi ^{+}U-U^{+}\chi \rangle \langle \chi
^{+}U+U^{+}\chi
\rangle \nonumber\\
&&+H_3\left\langle \chi \chi ^{+}\chi \chi ^{+}\right\rangle
+H_4\left\langle \chi \chi ^{+}\right\rangle ^2 \nonumber\\
\mathcal{L}^{(0,3,1)} &=&\frac{F_0^2B_0}4(\delta ^{(1)}\overline{\rho }%
_1\langle (\chi ^{+}U)^3+(U^{+}{\chi )}^3\rangle +\delta ^{(1)}\overline{%
\rho }_2\langle (\chi ^{+}U+U^{+}{\chi )\chi }^{+}\chi \rangle \nonumber\\
&&+\delta ^{(1)}\overline{\rho }_3\langle (\chi ^{+}U)^2-(U^{+}{\chi )}%
^2\rangle \langle \chi ^{+}U-U^{+}\chi \rangle \nonumber\\
&&+\delta ^{(1)}\overline{\rho }_4\langle (\chi ^{+}U)^2+(U^{+}{\chi )}%
^2\rangle \langle \chi ^{+}U+U^{+}\chi \nonumber\\
&&+\delta ^{(1)}\overline{\rho }_5\langle \chi ^{+}U+U^{+}\chi
\rangle \langle {\chi }^{+}\chi \rangle +\delta
^{(1)}\overline{\rho }_6\langle \chi
^{+}U-U^{+}\chi \rangle ^2\langle \chi ^{+}U+U^{+}\chi \rangle \nonumber\\
&&+\delta ^{(1)}\overline{\rho }_7\langle \chi ^{+}U+U^{+}\chi \rangle ^3) \nonumber\\
\mathcal{L}^{(0,2,2)} &=&\frac{F_0^2B_0^2}4\left( \delta ^{(2)}\overline{A}%
_0\langle (U^{+}\chi )^2+(\chi ^{+}U)^2\rangle \right. \nonumber\\
&&+\left. \delta ^{(2)}\overline{Z}_0^S\langle U^{+}\chi +\chi
^{+}U\rangle ^2+\delta ^{(2)}\overline{Z}_0^P\langle U^{+}\chi
-\chi ^{+}U\rangle
^2\right) \nonumber\\
\mathcal{L}^{(0,1,3)} &=&\frac{F_0^2}2B_0^3\delta _\chi
^{(2)}\langle U^{+}\chi +\chi ^{+}U\rangle .
\end{eqnarray}
In fact, identifying $F_0$ with the Goldstone boson decay constant and $%
B_0=\Sigma /F_0^2$ where $\Sigma =-\langle \overline{u}u\rangle
_0$ (in the chiral limit), we have
\begin{equation}
\delta _d^{(i)}=\delta _\chi ^{(i)}=0.
\end{equation}
As usual, we can also resume the powers of $B_0$ already at the
Lagrangian level and write
\begin{equation}
\mathcal{L}_n=\sum_{i+j=n}\mathcal{L}^{(i,j)}
\end{equation}
with
\begin{equation}
\mathcal{L}^{(i,j)}=\sum_k\mathcal{L}^{(i,j,k)}
\end{equation}
and denote
\begin{eqnarray}
A_0 &=&\overline{A}_0+B_0\delta ^{(1)}\overline{A}_0+B_0^2\delta ^{(2)}%
\overline{A}_0+\ldots \nonumber\\
Z_0^{S,P} &=&\overline{Z}_0^{S,P}+B_0\delta ^{(1)}\overline{Z}%
_0^{S,P}+B_0^2\delta ^{(2)}\overline{Z}_0^{S,P}+\ldots \nonumber\\
\xi &=&\overline{\xi }+B_0\delta ^{(1)}\overline{\xi }+\ldots \nonumber\\
\widetilde{\xi } &=&\overline{\widetilde{\xi }}+B_0\delta ^{(1)}\overline{%
\widetilde{\xi }}+\ldots \nonumber\\
\rho _i &=&\overline{\rho }_i+B_0\delta ^{(1)}\overline{\rho
}_i+\ldots ,
\end{eqnarray}
\emph{etc.}. These LEC's without the bars are then used in the
main text. Note that while the $O(p^2)$ parameters $\overline{A}_0$, $\overline{Z}%
_0^{S,P}$ and the $O(p^3)$ LEC's $\overline{\widetilde{\xi }}$
$,\overline{\xi }$
are renormalization scale independent, the renormalized resumed parameters $%
Z_0^{S,P,r}$, $A_0^r$ and $\widetilde{\xi }^r$, $\xi ^r$ run with
$\mu $ in the same way as $16(B_0/F_0)^2L_{6-8}^r$ and
$8B_0/F_0^2L_{4,5}$ within the standard $\chi PT$.

\section{\boldmath Coefficients of the dispersive part of $G\chi PT$ amplitude \label%
{appendix coefficients}}

In these formulae as well as in the following two appendices, the
masses $\widetilde{M_P ^2}$ are the generalized $O(p^2)$ masses
given by (\ref{Op2_mass_G}).
\begin{eqnarray}
\alpha _{\pi \eta }\widetilde{M_\pi ^2} &=&2[\widehat{m}B_0+8\widehat{m}%
^2A_0+2\widehat{m}^2Z_0^S(5r+4)-8\widehat{m}^2Z_0^P(r-1)] \nonumber\\
\alpha _{\pi \eta K}\widetilde{M}_\pi ^2 &=&4\widehat{m}%
^2(r^2-1)(A_0+2Z_0^P) \nonumber\\
\alpha _{\pi \pi }\widetilde{M}_\pi ^2 &=&2\widehat{m}B_0+16\widehat{m}%
^2A_0+4\widehat{m}^2Z_0^S(r+8) \nonumber\\
\alpha _{\eta \eta }(4\widetilde{M}_\eta ^2-\widetilde{M}_\pi
^2)
&=&\frac 23\widehat{m}B_0(1+8r)+\frac{16}3\widehat{m}^2A_0(1+8r^2) \nonumber\\
&&+\frac 43\widehat{m}^2Z_0^S(8+41r+32r^2)+\frac{32}3\widehat{m}%
^2Z_0^P(r-1)(4r-1) \nonumber\\
(\alpha _{\pi K}-1)\widetilde{M}_K\widetilde{M}_\pi &=&6(A_0+2Z_0^S)%
\widehat{m}^2(r+1) \nonumber\\
\alpha _{\eta K}(2\widetilde{M}_\eta ^2-\frac
23\widetilde{M}_K^2)
&=&\frac 23[\widehat{m}B_0(1+3r)+\widehat{m}^2A_0(3+10r+19r^2) \nonumber\\
&&+2\widehat{m}^2Z_0^S(6+19r+11r^2)+16\widehat{m}^2Z_0^Pr(r-1)]
\end{eqnarray}

\section{\boldmath Parameters $\alpha -\omega $ within the generalized $\chi PT$ \label%
{appendix alpha}}

Here we summarize the formulae in terms of the decomposition of
the remainders. For the parameter $\alpha $ we write
\begin{equation}
\delta _\alpha =\delta _\alpha
^{loop}+3\frac{\widehat{m}^2F_0^2}{F_\pi ^2M_\pi ^2}\delta _\alpha
^{CT}(\mu )+\delta _\alpha ^{G\chi PT}.
\end{equation}
For the counterterm contribution we get
\begin{eqnarray}
\delta _\alpha ^{CT}(\mu ) &=&\frac 13\widehat{m}\left[ 81\,\,\rho
_1+\,\rho
_2+\left( 80-64\,\,r-16\,\,r^2\right) \,\,\rho _3\right. \nonumber\\
&&+\left. \left( 100+64\,\,r+34\,\,r^2\right) \,\rho _4+\left(
2\,+\,r^2\right) \,\rho _5\right. \nonumber\\
&&+\left. \left( 96-96\,r\right) \,\rho _6+\left(
144\,+288\,\,r+108\,\,r^2\right) \,\rho _7\right] \nonumber\\
&&+\frac 83\left[ -(B_1-B_2)\Sigma _{\pi \eta }+2D^PM_\pi
^2(r-1)-2C_1^PM_\eta ^2(r-1)-\frac 12D^S[\Sigma _{\pi \eta }(5r+4)]\right. \nonumber\\
&&-\left. 2B_4[3M_\eta ^2+M_\pi ^2(2r^2+1)]\right] \nonumber\\
&&+\frac 13\widehat{m}^2\left[
256E_1+16E_2+F_1^P(256-256r^2)+F_4^S(32+16r^2)\right. \nonumber\\
&&+\left. F_1^S(256+320r^2)+F_5^{SP}(192-320r+160r^2-32r^3)\right. \nonumber\\
&&+\left. F_2^P(240-216r-24r^3)+F_6^{SP}(32-32r+16r^2-16r^3)\right. \nonumber\\
&&+\left. F_3^P(16-8r-8r^3)+F_3^S(16+10r+10r^3)\right. \nonumber\\
&&+\left.
F_6^{SS}(32+40r+16r^2+20r^3)+F_7^{SP}(384-160r-256r^2+32r^3)\right.
\nonumber\\
&&+\left.
F_2^S(400+234r+74r^3)+F_5^{SS}(576+720r+480r^2+168r^3)\right]
\end{eqnarray}
and the loops contribute as

\begin{eqnarray}
\frac 13F_\pi ^2M_\pi ^2\delta _\alpha ^{loop} &=&\frac 13\left\{ \widetilde{[M}_\pi ^2\left( 3B_0\widehat{m}+64A_0\widehat{m}^2+2Z_0^S\widehat{m}%
^2(15r+32)-8Z_0^P\widehat{m}^2(3r-8)\right) ]\right. \nonumber\\
&&\left. -6B_0^2\widehat{m}^2\right\} \left( J_{\pi \pi
}^r(0)+\frac 1{16\pi
^2}\right) \nonumber\\
&&+\frac 23\left\{ \widetilde{[M}_K^2\left( B_0\widehat{m}+2A_0\widehat{m}%
^2(r+8)+2Z_0^S\widehat{m}^2(15r+8)-8Z_0^P\widehat{m}^2(3r-2)\right)
]\right.
\nonumber\\
&&\left. -B_0^2\widehat{m}^2(r+1)\right\} \left( J_{KK}^r(0)+\frac
1{16\pi
^2}\right) \nonumber\\
&&\frac 19\left\{ [\widetilde{M}_\eta ^2(B_0\widehat{m}+32A_0\widehat{m}%
^2-16Z_0^P\widehat{m}^2(5r-2)+2Z_0^S\widehat{m}^2(41r+16)]\right. \nonumber\\
&&\left. -\frac 23B_0^2\widehat{m}^2(2r+1)\right\} \left( J_{\eta
\eta
}^r(0)+\frac 1{16\pi ^2}\right) \nonumber\\
&&+\frac 29\left\{ \widetilde{[M}_\pi ^2+4\widehat{m}%
^2(3A_0-4(r-1)Z_0^P+2(2r+1)Z_0^S)]^2-4B_0^2\widehat{m}^2\right\}
J_{\pi \eta
}^r(0) \nonumber\\
&&+\frac 34\left\{ [\frac 23\widetilde{M}_\pi ^2-\frac 83(r-1)\widehat{m}%
^2(A_0+2Z_0^P)]^2-\frac{16}9B_0^2\widehat{m}^2\right\} J_{KK}^r(0) \nonumber\\
&&+\frac 13\left\{ \widetilde{[M}_\pi ^2+4\widehat{m}%
^2(3A_0-4(r-1)Z_0^P+2(2r+1)Z_0^S)]\right. \nonumber\\
&&\left. \times [-2M_\pi ^2+\frac 32\widetilde{M}_\pi ^2+10\widehat{m}%
^2(A_0+2Z_0^S)]-2B_0\widehat{m}(3B_0\widehat{m}-2M_\pi ^2)\right\}
J_{\pi
\pi }^r(0) \nonumber\\
&&+\frac 29\left\{ \widetilde{[M}_\pi ^2+4\widehat{m}%
^2(3A_0-4(r-1)Z_0^P+2(2r+1)Z_0^S)]\right. \nonumber\\
&&\left. \times [\widetilde{M}_\eta ^2-\frac 14\widetilde{M}_\pi ^2+%
\widehat{m}^2((8r^2+1)A_0+8r(r-1)Z_0^P+2(2r+1)^2Z_0^S)]\right. \nonumber\\
&&\left. -\frac 13B_0^2\widehat{m}^2(8r+1)\right\} J_{\eta \eta }^r(0) \nonumber\\
&&+\frac 18\left\{ [-2M_\pi ^2+2\widetilde{M}_\pi ^2+8(r+1)\widehat{m}%
^2(A_0+2Z_0^S))]\right. \nonumber\\
&&\left. \times [-6M_\eta ^2+6\widetilde{M}_\eta ^2-\frac 83\widetilde{M}%
_K^2+\frac 83(r+1)\widehat{m}^2(3rA_0+2(r-1)Z_0^P+2(2r+1)Z_0^S]\right. \nonumber\\
&&\left. -2(2B_0\widehat{m}-M_\pi ^2)(\frac
43B_0\widehat{m}(4r+1)-6M_\eta ^2)\right\} J_{KK}^r(0).
\end{eqnarray}

\noindent In the same way we have for $\beta$
\begin{equation}
\beta \delta _\beta =\beta \delta _\beta
^{loop}+\widehat{m}^2F_0^2\delta _\beta ^{CT}(\mu )+\beta \delta
_\beta ^{G\chi PT},
\end{equation}
where
\begin{eqnarray}
\delta _\beta ^{CT}(\mu ) &=&\frac 83[(C_1^S+D^S)(2r+1)+2B_4(r^2+2)] \\
\beta \delta _\beta ^{loop} &=&-\frac 34\left\{ [\frac
23\widetilde{M}_\pi ^2-\frac
83(r-1)\widehat{m}^2(A_0+2Z_0^P)]-\frac 43B_0\widehat{m}\right\}
J_{KK}^r(0) \nonumber\\
&&+\frac 13\left\{ \widetilde{[M}_\pi ^2+4\widehat{m}%
^2(3A_0-4(r-1)Z_0^P+2(2r+1)Z_0^S)]-2B_0\widehat{m}\right\} J_{\pi
\pi }^r(0)
\nonumber\\
&&+\frac 18\left\{ [6(\widetilde{M}_\eta ^2-M_\eta
^2+\widetilde{M}_\pi
^2-M_\pi ^2)-\frac 83\widetilde{M}_K^2\right. \nonumber\\
&&\left. +\frac 83(r+1)\widehat{m}^2(3A_0(r+3)+4Z_0^S(r+5)+2(r-1)Z_0^P)]%
\right. \nonumber\\
&&\left. -[\frac 83B_0\widehat{m}(2r+5)-6M_\eta ^2-6M_\pi
^2]\right\} J_{KK}^r(0).
\end{eqnarray}
For the remaining two parameters the corresponding decomposition
of the remainders
\begin{eqnarray}
\gamma \delta _\beta (\mu ) &=&\gamma \delta _\gamma {}^{loops}(\mu )+%
\widehat{m}^2F_0^2\delta _\gamma ^{CT}(\mu )+\gamma \delta _\gamma
^{G\chi
PT} \\
\omega \delta _\omega (\mu ) &=&\omega \delta _\omega {}^{loops}(\mu )+%
\widehat{m}^2F_0^2\delta _\omega ^{CT}(\mu )+\omega \delta _\omega
^{G\chi PT}
\end{eqnarray}
is trivial, \emph{i.e.}
\begin{equation}
\delta _\gamma {}^{loops}(\mu )=\delta _\gamma ^{CT}(\mu )=\delta
_\omega {}^{loops}(\mu )=\delta _\omega ^{CT}(\mu )=0.
\end{equation}

\section{\boldmath Generalized $\chi PT$ contributions to the bare expansion remainders for the masses 
\label{appendix generalized Li}}

The expressions for $\xi $, $\widetilde{\xi }$ can be
obtained from the exact algebraic identities (\ref{L_5_L_6}) after
identification (\ref {identification}) and using the
representation (\ref{Fpi_G_rem}) and (\ref {Fpi_G_rem_explicit})
for the remainder of $F_\pi ^2$ and (\ref{Fk_G_rem}) and
(\ref{Fk_G_rem_explicit}) for the remainder of $F_K^2$. In the
same spirit, $A_0$, $Z_0^S$ and $Z_0^P$ can be expressed using
identities (\ref {L_6_L_7_L8}) and the following remainders

\begin{eqnarray}
F_\pi ^2M_\pi ^2\delta _{F_\pi M_\pi } &=&F_\pi ^2M_\pi ^2\delta
_{F_\pi M_\pi }^{loop}(\mu )+F_0^2\widehat{m}^2\delta _{F_\pi
M_\pi }^{CT}(\mu
)+F_\pi ^2M_\pi ^2\delta _{F_\pi M_\pi }^{G\chi PT} \nonumber\\
F_K^2M_K^2\delta _{F_KM_K} &=&F_K^2M_K^2\delta _{F_KM_K}^{loop}(\mu )+F_0^2%
\widehat{m}^2\delta _{F_KM_K}^{CT}(\mu )+F_K^2M_K^2\delta
_{F_KM_K}^{G\chi
PT} \nonumber\\
F_\eta ^2M_\eta ^2\delta _{F_\eta M_\eta } &=&F_\eta ^2M_\eta
^2\delta _{F_\eta M_\eta }^{loop}(\mu )+F_0^2\widehat{m}^2\delta
_{F_\eta M_\eta }^{CT}(\mu )+F_\eta ^2M_\eta ^2\delta _{F_\eta
M_\eta }^{G\chi PT},
\end{eqnarray}
where
\begin{eqnarray}
F_\pi ^2M_\pi ^2\delta _{F_\pi M_\pi }^{loop}(\mu ) &=&\left[ \widetilde{M}%
_\pi ^2\left( 3B_0\widehat{m}+16A_0\widehat{m}^2+2Z_0^S\widehat{m}%
^2(3r+16)\right) -6B_0^2\widehat{m}^2\right] \nonumber\\
&&\times \left( J_{\pi \pi }^r(0)+\frac 1{16\pi ^2}\right) \nonumber\\
&&+2\left[ \widetilde{M}_K^2\left( B_0\widehat{m}+2A_0\widehat{m}%
^2(r+2)+2Z_0^S\widehat{m}^2(3r+4)\right) -B_0^2\widehat{m}^2(r+1)\right] \nonumber\\
&&\times \left( J_{KK}^r(0)+\frac 1{16\pi ^2}\right) \nonumber\\
&&+\frac 13\left[ \widetilde{M}_\eta ^2(B_0\widehat{m}+8A_0\widehat{m}%
^2-8Z_0^P\widehat{m}^2(r-1)+2Z_0^S\widehat{m}^2(5r+4)\right. \nonumber\\
&&\left. -\frac 23B_0^2\widehat{m}^2(2r+1)\right] \left( J_{\eta
\eta
}^r(0)+\frac 1{16\pi ^2}\right) \\
\delta _{F_\pi M_\pi }^{CT}(\mu ) &=&\widehat{m}[9\rho _1+\rho
_2+2\rho
_4(10+4r+r^2) \nonumber\\
&&+\rho _5(2+r^2)+12\rho _7(4+4r+r^2)] \nonumber\\
&&+2\widehat{m}^2[8E_1+2E_2+8F_1^S\left( 2+r^2\right) +F_2^S\left(
9r+r^3+20\right) +F_3^S\left( 4+r+r^3\right) \nonumber\\
&&+2F_4^S\left( 2+r^2\right) +4F_5^{SS}\left( r+2\right) \left(
r^2+2r+6\right) +2F_6^{SS}\left( r+2\right) \left( 2+r^2\right) ],
\end{eqnarray}
then
\begin{eqnarray}
F_K^2M_K^2\delta _{F_KM_K}^{loop}(\mu ) &=&\left\{ \frac 34\left[ \widetilde{M}_\pi ^2\left( B_0\widehat{m}+A_0\widehat{m}^2(r+5)+2Z_0^S\widehat{m}%
^2(r+6)\right) -2B_0^2\widehat{m}^2\right] \right. \nonumber\\
&&\left. \times \left( J_{\pi \pi }^r(0)+\frac 1{16\pi ^2}\right) \right. \nonumber\\
&&\left. +\frac 32\left[ \widetilde{M}_K^2\left( B_0\widehat{m}+3A_0%
\widehat{m}^2(r+1)+2Z_0^S\widehat{m}^2(3r+4)\right) -B_0^2\widehat{m}%
^2(r+1)\right] \right. \nonumber\\
&&\left. \times \left( J_{KK}^r(0)+\frac 1{16\pi ^2}\right) \right. \nonumber\\
&&\left. +\frac 1{12}\left[ \widetilde{M}_\eta ^2\left( 5B_0\widehat{m}+A_0%
\widehat{m}^2(17r+5)+8Z_0^P\widehat{m}^2(r-1)+2Z_0^S\widehat{m}%
^2(13r+14)\right) \right. \right. \nonumber\\
&&\left. \left. -\frac{10}3B_0^2\widehat{m}^2\left( 2r+1\right)
\right]
\left( J_{\eta \eta }^r(0)+\frac 1{16\pi ^2}\right) \right\} (r+1) \\
\delta _{F_KM_K}^{CT}(\mu ) &=&\frac 12\widehat{m}[3\rho
_1(1+r)(1+r+r^2)+\rho _2(1+r^3)+6\rho _4\left( r+1\right) \left(
2+2r+r^2\right) \nonumber\\
&&+\rho _5\left( r+1\right) \left( 2+r^2\right) +12\rho _7\left(
r+1\right)
\left( r+2\right) ^2] \nonumber\\
&&+\widehat{m}^2[2E_1\left( 1+r\right) ^2\left( 1+r^2\right)
+E_2\left(
1+r\right) ^2\left( 1-r+r^2\right) \nonumber\\
&&+\frac 12E_3\left( r^2-1\right) ^2 \nonumber\\
&&+4F_1^S\left( 1+r\right) ^2\left( 2+r^2\right) +F_2^S\left(
1+r\right)
\left( 8+9r+9r^2+4r^3\right) \nonumber\\
&&-F_3^S\left( 1+r\right) \left( 4-r+r^2+2r^3\right) +F_4^S\left(
1+r\right)
^2\left( 2+r^2\right) \nonumber\\
&&+4F_5^{SS}\left( 1+r\right) \left( 2+r\right) \left( 4+3r+2r^2\right) \nonumber\\
&&+2F_6^{SS}\left( 1+r\right) \left( 2+r\right) \left(
2+r^2\right) ]
\end{eqnarray}
and
\begin{eqnarray}
F_\eta ^2M_\eta ^2\delta _{F_\eta M_\eta }^{loop}(\mu )
&=&\,\,\,\left[
\widetilde{M}_\pi ^2\left( B_0\widehat{m}+8\,A_0\,\widehat{m}^2-8\,%
\widehat{m}^2\,Z_0^P\left( r-1\right)
\,+2\,\widehat{m}^2\,Z_0^S\left(
4+5\,r\right) \,\right) -2B_0^2\widehat{m}^2\right] \nonumber\\
&&\times \left( J_{\pi \pi }^r(0)+\frac 1{16\pi ^2}\right) \nonumber\\
&&+\frac{2\,}3\,\left[ \widetilde{M}_K^2\,\left(
B_0\widehat{m}\,\left(
1+4\,r\right) +2\,A_0\,\widehat{m}^2\,\left( 2+r+8\,r^2\right) +8\,\widehat{m%
}^2\,Z_0^P\,\left( r-1\right) \,\left( 2\,r-1\right) \right. \,\right. \nonumber\\
&&\left. \left. +2\,\,\widehat{m}^2\,Z_0^S\,\left(
4+15\,r+8\,r^2\right)
\right) -B_0^2\widehat{m}^2\left( 4\,r+1\right) (r+1)\right] \nonumber\\
&&\times \left( J_{KK}^r(0)+\frac 1{16\pi ^2}\right) \nonumber\\
&&+\frac{1\,}9\,\left[ \widetilde{M}_\eta ^2\left(
B_0\widehat{m}\,\left(
1+8\,r\right) +8A_0\,\widehat{m}^2\,\left( 1+8\,r^2\right) +16\,\widehat{m}%
^2\,Z_0^P\,\left( r-1\right) \,\left( 4\,r-1\right) \right. \,\right. \nonumber\\
&&\left. \left. +2\,\,\widehat{m}^2\,Z_0^S\,\left(
8+41\,r+32\,r^2\right) \right) -\frac 23B_0^2\widehat{m}^2\left(
8\,r+1\right) \left( 2r+1\right)
\right] \nonumber\\
&&\times \left( J_{\eta \eta }^r(0)+\frac 1{16\pi ^2}\right) \\
\delta _{F_\eta M_\eta }^{CT}(\mu ) &=&\frac 13\widehat{m}[9\rho
_1\left( 1+2r^3\right) +\rho _2\left( 1+2r^3\right) +16\rho
_3\left( r-1\right)
^2\left( 1+r\right) \nonumber\\
&&+2\rho _4\left( 10+8r+17r^2+10r^3\right) \nonumber\\
&&+\rho _5\left( 1+2r\right) \left( 2+r^2\right) +16\rho _6\left(
2-3r+r^3\right) +12\rho _7\left( 2+r\right) ^2\left( 1+2r\right) ] \nonumber\\
&&+\frac 23\widehat{m}^2[8E_1\left( 1+2r^4\right) +2E_2\left(
1+2r^4\right)
\nonumber\\
&&+8F_1^S\left( r^2+2\right) \left( 2r^2+1\right) +16F_1^P\left(
r^2-1\right) ^2 \nonumber\\
&&+F_2^S\left( 20+13r+37r^3+20r^4\right) +12F_2^P\left(
r^2+r+1\right)
\left( r-1\right) ^2 \nonumber\\
&&+F_3^S\left( 4+5r+5r^3+4r^4\right) +4F_3^P\left( r^2+r+1\right)
\left(
r-1\right) ^2 \nonumber\\
&&+2F_4^S\left( r^2+2\right) \left( 2r^2+1\right) \nonumber\\
&&+12F_5^{SS}\left( r+2\right) \left( 2r^3+3r^2+2r+2\right)
+8F_5^{SP}\left(
r^2+2\right) \left( r-1\right) ^2 \nonumber\\
&&+2F_6^{SS}\left( 2r+1\right) \left( r+2\right) \left(
r^2+2\right)
+4F_6^{SP}\left( r^2+2\right) \left( r-1\right) ^2 \nonumber\\
&&+16F_7^{SP}\left( r+2\right) \left( r+1\right) \left( r-1\right)
^2].
\end{eqnarray}

\section{Resonance amplitude and remainders estimates\label%
{Appendix resonances}}

Here we give the contribution to the amplitude
\cite{Bernard:1991xb} related to the resonance exchange,
derived from the leading order Lagrangian of $R\chi T$ (we have
confirmed this expression by independent calculation)
\begin{eqnarray}
G_R(s,t;u) &=&4\frac 1{M_{S_1}^2-t}\left( \widetilde{c}_d(t-2M_\pi ^2)+2%
\widetilde{c}_m\stackrel{o}{M}_\pi ^2\right) \left( \widetilde{c}%
_d(t-2M_\eta ^2)+2\widetilde{c}_m\stackrel{o}{M}_\eta ^2\right) \nonumber\\
&&+4\frac{\widetilde{c}_m^2}{M_{S_1}^2}\stackrel{o}{M}_\pi
^2\left(
\stackrel{o}{M}_\pi ^2+\stackrel{o}{M}_\eta ^2\right) +4\frac{c_m^2}{3M_S^2}%
\stackrel{o}{M}_\pi ^2\left( \stackrel{o}{M}_\pi
^2-\stackrel{o}{M}_\eta
^2\right) \nonumber\\
&&+\frac 23\frac 1{M_S^2-s}\left( c_d(s-M_\pi ^2-M_\eta ^2)+2c_m\stackrel{o}{%
M}_\pi ^2\right) ^2 \nonumber\\
&&+\frac 23\frac 1{M_S^2-u}\left( c_d(u-M_\pi ^2-M_\eta ^2)+2c_m\stackrel{o}{%
M}_\pi ^2\right) ^2 \nonumber\\
&&-\frac 23\frac 1{M_S^2-t}\left( c_d(t-2M_\pi
^2)+2c_m\stackrel{o}{M}_\pi
^2\right) \left( c_d(t-2M_\eta ^2)+2c_m(2\stackrel{o}{M}_\eta ^2-\stackrel{o%
}{M}_\pi ^2)\right) \nonumber\\
&&-16\frac{\widetilde{d}_m^2}{M_{\eta _1}^2}\stackrel{o}{M}_\pi
^2\left( \stackrel{o}{M}_\pi ^2-\stackrel{o}{M}_\eta ^2\right).
\end{eqnarray}
The resonance estimate of the remainders $\delta _\gamma ^R$ and
$\delta _\omega ^R$ are
\begin{eqnarray}
\gamma \delta _\gamma ^R &=&-\frac 8{3M_S^6}\left( c_dM_\pi ^2-c_m\stackrel{o%
}{M}_\pi ^2\right) \left( c_dM_\eta ^2-c_m(2\stackrel{o}{M}_\eta ^2-%
\stackrel{o}{M}_\pi ^2)\right) \nonumber\\
&&+\frac 43\frac{c_d}{M_S^4}\left( c_dM_\eta ^2-c_m(2\stackrel{o}{M}_\eta ^2-%
\stackrel{o}{M}_\pi ^2)\right) +\frac 13\frac{c_d^2\Sigma _{\pi \eta }}{%
M_S^2(M_S^2-\Sigma _{\pi \eta })} \nonumber\\
&&+\frac 43\frac{c_dc_m}{(M_S^2-\Sigma _{\pi \eta
})^2}\stackrel{o}{M}_\pi ^2+\frac 43\frac{c_m^2}{(M_S^2-\Sigma
_{\pi \eta })^3}\stackrel{o}{M}_\pi ^4
\nonumber\\
&&+\frac{16}{M_{S_1}^6}\left( \widetilde{c}_dM_\pi ^2-\widetilde{c}_m%
\stackrel{o}{M}_\pi ^2\right) \left( \widetilde{c}_dM_\eta ^2-\widetilde{c}_m%
\stackrel{o}{M}_\eta ^2\right) \nonumber\\
&&-\frac{8\widetilde{c}_d}{M_{S_1}^4}\left( \widetilde{c}_d\Sigma
_{\pi \eta
}-\widetilde{c}_m(\stackrel{o}{M}_\pi ^2+\stackrel{o}{M}_\eta ^2)\right) \\
\omega \delta _\omega ^R &=&-\frac 13\frac{c_d^2\Sigma _{\pi \eta }}{%
M_S^2(M_S^2-\Sigma _{\pi \eta })}+\frac 43\frac{c_m^2\stackrel{o}{M}_\pi ^4}{%
(M_S^2-\Sigma _{\pi \eta })^3}+\frac 43\frac{c_dc_m}{(M_S^2-\Sigma
_{\pi \eta })^2}\stackrel{o}{M}_\pi ^2.
\end{eqnarray}

\end{document}